\documentclass[useAMS,usenatbib, graphicx]{mn2e}

\usepackage{eucal}%prettier mathcal fonts
\usepackage{aas_macros}
\usepackage{amsmath}
\usepackage{amssymb}
\usepackage{color}
\usepackage{epsfig}

\usepackage{graphicx}
\usepackage{ifthen}
\usepackage{latexsym}
\usepackage{placeins}
\usepackage{multicol}
\usepackage{ogonek}
\usepackage{overpic}
\usepackage{rotating}
\usepackage{subfigure}
\usepackage[absolute,overlay]{textpos} 
\usepackage{times,epsf}
\usepackage{txfonts}
\usepackage{varioref}
\usepackage{verbatim}
\usepackage{url}
\newcommand{\beq}{\begin{equation}}
\newcommand{\eeq}{\end{equation}}

\usepackage{float}

\begin{document}

\title[Simulations of Blandford-Znajek Jets]{General Relativistic Magnetohydrodynamic Simulations of Blandford-Znajek Jets and the Membrane Paradigm}

\author[R.~F. Penna, R. Narayan, \&  A.~S. S\k{a}dowski]
{Robert F. Penna$^{1,2}$
\thanks{E-mail: rpenna@cfa.harvard.edu~(RFP),
rnarayan@cfa.harvard.edu~(RN), asadowski@cfa.harvard.edu~(AS)},
Ramesh Narayan$^2$\footnotemark[1],
Aleksander S\k{a}dowski$^2$\footnotemark[1]\\
$^1$Department of Physics, and Kavli Institute for Astrophysics and Space Research,
Massachusetts Institute of Technology, Cambridge, MA 02139, USA\\
$^2$Harvard-Smithsonian Center for Astrophysics, 60 Garden Street,
Cambridge, MA 02138, USA \\ }

\date{\today}

\maketitle

\begin{abstract}

Recently it has been observed that the scaling of jet power with black hole spin in galactic X-ray binaries is consistent with the predictions of the Blandford-Znajek (BZ) jet model.  These observations motivate us to revisit the BZ model using general relativistic magnetohydrodynamic simulations of magnetized jets from accreting ($h/r\sim 0.3$), spinning ($0<a_*<0.98$) black holes.   We have three main results.  First, we quantify the discrepancies between the BZ jet power and our simulations: assuming maximum efficiency and uniform fields on the horizon leads to a $\sim10\%$ overestimate of jet power, while ignoring the accretion disk leads to a further $\sim50\%$ overestimate.  Simply reducing the standard BZ jet power prediction by $60\%$ gives a good fit to our simulation data.  Our second result is to show that the membrane formulation of the BZ model correctly describes the physics underlying simulated jets: torques, dissipation, and electromagnetic fields on the horizon.  This provides intuitive yet rigorous pictures for the black hole energy extraction process.  Third, we compute the effective resistance of the load region and show that the load and the black hole achieve near perfect impedance matching. Taken together, these results increase our confidence in the BZ model as the correct description of jets observed from astrophysical black holes.

\end{abstract}


\section{Introduction}
\label{sec:intro}

The spins of ten stellar-mass black holes have been measured using the continuum fitting method (see \citealt{2013arXiv1303.1583M} for a recent review of the details of the continuum fitting method and the uncertainties in the derived spin estimates).  Seven of these black holes are so-called ``transient'' systems which have large amplitude outbursts.  During outburst they reach close to the Eddington luminosity limit and near peak luminosity they eject blobs of plasma. The blobs move ballistically outward at relativistic speeds (Lorentz factor $\gamma>2$).   \citet{2012MNRAS.419L..69N} and \citet{2013ApJ...762..104S} measured the peak radio luminosities of ballistic jet blobs, a proxy for jet power, from five transient systems.  They find that the jet power is correlated with black hole spin, increasing by a factor of 1000 as the spin varies from 0.1 to 1.  This suggests the blobs may be powered by black hole rotational energy.

The observed scaling of jet power with black hole spin is consistent with the predictions of the Blandford-Znajek (BZ) jet model \citep{bz77,1982MNRAS.198..345M,1986bhmp.book.....T}.  This model describes how magnetic fields drain a black hole of its rotational energy and drive powerful jets.   It builds on earlier proposals for tapping black hole rotational energy using particles \citep{1969NCimR...1..252P} and magnetic fields \citep{1975PhRvD..12.2959R}, and is a close cousin of the pulsar magnetosphere model of \citet{1969ApJ...157..869G}.  The similarity between black hole and pulsar jets is particularly transparent in the membrane formulation of the BZ model \citep{1982MNRAS.198..345M,1986bhmp.book.....T}.

The BZ model has three free parameters: the angular velocity of the 
event horizon $\Omega_H$, the angular velocity of magnetic field lines, $\Omega_F$, and the magnetic flux threading the jet, $\Phi$.  Early attempts to determine $\Omega_F$ \citep{1979AIPC...56..399L,1982MNRAS.198..345M,phinney1983} found (up to factors of order unity) $\Omega_F/\Omega_H\approx 1/2$.  If one assumes $\Omega_F/\Omega_H=1/2$, then the jet power predicted by the BZ model is \citep{bz77,1986bhmp.book.....T,2000PhR...325...83L,2010ApJ...711...50T}:
%\begin{align}\label{eq:BZ1}
%P_{\rm jet} &\sim \frac{1}{128}\left(\frac{a}{M}\right)^2 B_n^2 r_H^2 \\
%&\sim \left(10^{45}\frac{\text{erg}}{\text{sec}}\right)\left(\frac{a}{M}\right)^2
%\left(\frac{M}{10^9M_\odot}\right)^2\left(\frac{B_n}{10^4\text{G}}\right)^2,
%\end{align}
%where $M$ and $a$ are the black hole mass and spin, and $r_H=M+\sqrt{M^2-a^2}$ is the radius of the horizon (we set $G=c=1$).  This equation is valid for all spins, provided  $\Omega_F/\Omega_H=1/2$ (maximum efficiency).  
\begin{equation}\label{eq:BZ2}
P^{\rm BZ} \approx \frac{1}{6\pi}\Omega_H^2 \Phi^2.
\end{equation}
We re-derive this equation in Appendix \ref{sec:BZpjet}.  The assumption $\Omega_F/\Omega_H=1/2$ gives maximum jet efficiency.  We refer to the BZ model with this assumption as the standard BZ model.   Equation \eqref{eq:BZ2} is consistent with the jet power scaling observed from astrophysical black holes by \citet{2012MNRAS.419L..69N} and \citet{2013ApJ...762..104S}.

There is now an extensive, decade-old literature on general relativistic magnetohydrodynamic (GRMHD) simulations of black hole jets, which further support and extend the BZ model.   \citet{2001MNRAS.326L..41K} presented the first time-dependent simulations of BZ jets and demonstrated that the model is stable.  In these simulations, magnetic fields were imposed on the black hole at the outset.  

Later simulations (including those in this paper), embed the black hole in a turbulent accretion disk, which then deposits magnetic fields onto the hole self-consistently.  A funnel-shaped region develops along the black hole spin axis where the field geometry resembles a split monopole \citep{2004ApJ...606.1083H}.
In this region, the flux of magnetic energy at the horizon is sometimes directed outwards, away from the black hole \citep{2004ApJ...611..977M,2005ApJ...620..878D}. The distribution of electromagnetic fields and the angular momentum flux at the horizon are consistent with the BZ model
\citep{2004ApJ...611..977M,2012MNRAS.423.3083M}.  The flux of energy carried by gas is always directed into the black hole \citep{2005MNRAS.359..801K}.  The strength of the jet is an increasing function of black hole spin \citep{2005ApJ...622.1008K,2005ApJ...630L...5M,2006ApJ...641..103H} and  the energy in the jet can be comparable to the energy in the accretion flow \citep{2006ApJ...641..103H}.  The magnetic field geometry is intermediate between the split-monopole and paraboloidal geometries considered by BZ  \citep{2007MNRAS.375..513M,2007MNRAS.375..531M}.  The strength of the simulated jet depends on the field geometry in the initial conditions, because this affects the final field strength of the black hole and disk \citep{2008ApJ...678.1180B,2012MNRAS.423.3083M}.  
%The jets appear to be stable \citep{2009MNRAS.394L.126M}.  
The simulated scaling of jet power with black hole spin agrees with the BZ prediction \eqref{eq:BZ2} \citep{2010ApJ...711...50T,2012JPhCS.372a2040T}.     If the magnetic field is very strong, the jet can carry off more energy from the black hole than the accretion flow puts in \citep{2011MNRAS.418L..79T}.   If the magnetic field is very weak, gas accretion can quench jet formation 
\citep{2012MNRAS.423.3083M}.  Prograde black holes drive more powerful jets than retrograde holes \citep{2012MNRAS.423L..55T}.

In this paper, we revisit our GRMHD simulations of jets from accreting, spinning black holes.  
We have three main results. First, we quantify the error introduced into the BZ jet power prediction \eqref{eq:BZ2} by the standard approximations (maximum efficiency, uniform magnetic fields on the horizon, no disk thickness, and no gas accretion).  Second, we check that the underlying physics generating simulated jets (torques, dissipation, and electromagnetic fields at the horizon) is correctly described by the membrane formulation of the BZ model.  Third, we compute the effective resistance of the load region, where magnetic energy is converted into bulk gas motion.  This analysis supports the prediction $\Omega_F/\Omega_H\approx 1/2$ of simple load region models  \citep{1979AIPC...56..399L,1982MNRAS.198..345M,phinney1983}.

The paper is organized as follows.  In \S\ref{sec:prelim}, we give an overview of the membrane formalism and our GRMHD simulations.
In \S\ref{sec:jets}, we show that the simulated jet power is consistent with the BZ formula \eqref{eq:BZ2} and quantify the main sources of discrepancy.  In \S\ref{sec:horizon} and \S\ref{sec:fields}, we show that the torques, dissipation, and electromagnetic fields at the horizon in the GRMHD simulations are in excellent agreement with the BZ model.  In \S\ref{sec:load}, we discuss the conversion of magnetic energy into bulk gas motion in the load region and its relationship to $\Omega_F/\Omega_H$.   We summarize our results in \S\ref{sec:conc}.  Appendix A re-derives the BZ jet power prediction \eqref{eq:BZ2}.  Appendix B gives 3D visualizations of the torques, dissipation, and electromagnetic fields on the black hole membranes of our GRMHD simulations.

\section{Preliminaries}
\label{sec:prelim}
\subsection{The membrane formalism}
\label{sec:membrane}

\citet{1982MNRAS.198..345M} recast the BZ model in the membrane formalism (see \citealt{1986bhmp.book.....T} for an overview).   We will use this formulation for much of our analyses.  It allows a local description of the conversion of gravitational energy into magnetic energy at the horizon.   Another advantage is that it does not require any mathematics beyond three-dimensional vector algebra, so relations to non-relativistic mechanics are particularly transparent.   

There are two pieces to the membrane formalism: we introduce fiducial observers (the ZAMOs), and we switch to a dual description of black holes that treats the horizon as a viscous membrane (but is mathematically equivalent to the usual description of black holes).

Understanding the flow of energy at the horizon presents a conceptual challenge: it is always possible to change the metric at a point to the flat, zero-energy Minkowski metric by a change of reference frame (equivalence principle), so there is no observer-independent way of defining the energy of the gravitational field at a point.\footnote{A ``quasi-local'' energy, defined on surfaces rather than at points, can be defined in at least some cases.  See \citet{2009PhRvL.102b1101W} for a recent approach and a summary of earlier work.}  The only way to give a local description of black hole energy extraction is to fix an observer.  The fiducial choice in the Kerr metric is the zero angular momentum observer (ZAMO) \citep{1972ApJ...178..347B}.  Quantities measured at infinity do not depend on the choice of local observer.  However, introducing the ZAMO is useful because it gives a concrete picture for the intermediate interactions between black hole and jet that result in black hole energy extraction.  So we will often work in the ZAMO frame.

It is well-known that to an observer outside a black hole (such as a ZAMO), matter falling into the hole appears to freeze just outside as a result of gravitational redshift.  This applies equally well to the matter which first formed the black hole.  So the black hole's energy, $M$, and angular momentum, $J$, appear spread out in a membrane covering the horizon.  The ``membrane paradigm'' treats this membrane as a surrogate for the black hole.  Every interaction of the external world with the black hole becomes concretely realized as an interaction with the membrane.   It would be difficult to give a complete discussion of black hole energy extraction without the membrane formalism because we would not be able to say where the black hole energy is located to begin with.  Of course, an observer falling into the black hole does not find the hole's energy and angular momentum at the horizon (equivalence principle), but infalling observers are not relevant for astrophysics.  The fact that different observers see the energy of the gravitational field in different places has been called ``black hole complementarity'' to highlight its similarity with wave-particle duality in quantum theory \citep{1993PhRvD..48.3743S}.

\subsection{GRMHD simulations}
\label{sec:thesims}

Our GRMHD simulations have been described in detail elsewhere \citep{2012MNRAS.426.3241N,2013MNRAS.428.2255P,olek2013}, so we can be brief.  We use the GRMHD code HARM \citep{2003ApJ...589..444G,2006MNRAS.367.1797M} to evolve a magnetized, turbulent accretion disk in the Kerr metric.  The code conserves energy to machine precision, so any energy lost at the grid scale by e.g. turbulent dissipation or numerical magnetic reconnection is returned to the fluid, increasing its entropy.  The stress-energy tensor of the fluid is
\begin{align}
T_{\mu\nu}&=T^{\rm gas}_{\mu\nu}+T^{\rm mag}_{\mu\nu},
\end{align}
where
\begin{align}
T^{\rm gas}_{\mu\nu}&=(\rho_0+u)u_\mu u_\nu + p h_{\mu\nu},\\
T^{\rm mag}_{\mu\nu}&=\frac{1}{2}\left(b^2 u_\mu u_\nu +b^2 h_{\mu\nu}-2b_\mu b_\nu\right).
\end{align}
The notation is standard: $\rho_0$, $u$, $p$, and $u_\mu$ are the fluid rest mass density, internal energy, pressure, and four-velocity. The field $b^\mu$ is the fluid frame magnetic field and $h_{\mu\nu}=g_{\mu\nu}+u_\mu u_\nu$ is the projection tensor.  The equation of state for the gas is $p=(\Gamma-1)u$, where $\Gamma=5/3$.  There is no radiative cooling, so the disk becomes thick and hot.  It is similar to an advection dominated accretion flow \citep{1994ApJ...428L..13N,1995ApJ...444..231N,2012MNRAS.426.3241N}.

The Kerr metric in Boyer-Lindquist ($t$, $r$, $\theta$, $\phi$) coordinates is:
\begin{align}
ds^2 = 
&- (1-2Mr/\rho^2) dt^2 - (4 Mar \sin^2 \theta/\rho^2) dtd\phi +\rho^2 d\theta^2\\
&+(\rho^2/\Delta) dr^2 + (r^2+a^2+2Ma^2 r\sin^2\theta/\rho^2)\sin^2 \theta d\phi^2. 
\end{align}
We have defined the metric functions
\begin{align}
\Delta &= r^2 -2Mr+a^2,\\
\rho^2 &= r^2 + a^2 \cos^2\theta,\\
\Sigma^2 &= (r^2+a^2)^2-a^2\Delta \sin^2 \theta.
\end{align}
Our notation follows \citet{1986bhmp.book.....T}.
The simulations do not evolve the metric (self-gravity is ignored).  We consider four black hole spins: $a_* = a/M = 0, 0.7, 0.9$, and $0.98$.  

HARM uses Kerr-Schild coordinates and Boyer-Lindquist coordinates.  These coordinates have a couple of advantages over the membrane formulation for numerical computations.  First, unlike the membrane formulation, they make Lorentz invariance manifest, so it is easier to impose energy and momentum conservation numerically.   Second, computing in Kerr-Schild coordinates makes it trivial to impose boundary conditions at the inner edge of the numerical grid; the coordinates are horizon penetrating, so we simply  place the inner boundary of the grid between the inner and outer black hole horizons and the black hole behaves as an event horizon.   A major part of our work in this paper is converting the simulation results from Kerr-Schild and Boyer-Lindquist coordinates to the membrane formalism.  It is only possible to give a local description of black hole energy extraction in the membrane formalism.

We use a logarithmically spaced radial grid and put the outer boundary of the grid at $r\sim 10^5M$: far enough away that nothing reaches it over the course of the simulation (so we do not need to worry about boundary conditions there).  The $\theta$ coordinate runs from 0 to $\pi$ and the $\phi$ coordinate runs from 0 to $2\pi$. The duration of the simulations varies from as short as $t=25,000M$ to as long as $t=200,000M$.  The typical resolution is $256\times 128\times 64$ in $(r, \theta, \phi)$.    The simulations are summarized in Table \ref{tab:sims}.

\begin{table*}
 \begin{minipage}{100mm}
  \caption{GRMHD simulations}
  \begin{tabular}{@{}ccccccc@{}}
  \hline
  Simulation
  & Initial field
  & $a/M$
  & $h/r$
  & Resolution ($r$, $\theta$, $\phi$)
  & Duration\\
%  & $\Delta t_{\rm steady}$\\
  \hline
%                                                                                                  
% \thin   cd /n/home08/rpenna/blackhole/thindisks/hr1a0_V/alpha/                                   
% \thinA  cd /n/home08/rpenna/blackhole/thindisks/hr05a7_IV/alpha/                                 
% \thinLR cd /n/home08/rpenna/blackhole/thindisks/hr05a0_IV/alpha/                                 
% \sane   cd /n/home08/rpenna/blackhole/limotorus_calc/pol-trans-bc/a0-hr03/alpha/                 
% \saneA  cd /n/home08/rpenna/blackhole/limotorus_calc/pol-trans-bc/a0-hr03/alpha/                 
% \mad    cd /n/home08/rpenna/blackhole/limotorus_calc/mad/a0/alpha/                               
%                                                                                                  
 1.  &  SANE & 0     &  0.3 & $256\times128\times64$ & 200,000M \\% &  $10,000M$ \\
 2.  &  SANE & 0.7  &  0.3 & $256\times128\times64$ & 100,000M \\% & $30,000M$ \\
 3.  &  SANE & 0.9  &  0.3 & $256\times128\times64$ & 50,000M  \\%& $10,000M$ \\
 4.  &  SANE & 0.98&  0.3 & $256\times128\times64$ & 25,000M \\% & $10,000M$ \\
 \quad\\
 5.  & MAD    & 0      &  0.3 & $264\times126\times60$ & 100,000M \\% & $10,000M$ \\
 6.  & MAD    & 0.7   &  0.3 & $264\times126\times60$ & 91,500M  \\%& $80,000M$ \\
 7.  & MAD    & 0.9   &  0.3 & $264\times126\times60$ & 44,000M\\% & $10,000M$ \\
\hline \label{tab:sims}
\end{tabular}
\end{minipage}
\end{table*}

Initially, the fluid is in a hydrostatic equilibrium torus outside $r=20M$.  The fluid is threaded with a weak poloidal ($\beta=p^{\rm gas}/p^{\rm mag}=100$) magnetic field.  We consider two initial field geometries: multiple, smaller poloidal loops (we call these runs SANE, for Standard and Normal Evolution) and a single, large poloidal loop (we call these runs MAD, for Magnetically Arrested Disk).  During the first few orbits of the fluid torus, the magnetic field is sheared and the magnetorotational instability is triggered \citep{velikhov1959,chandrasekhar1960,balbus1991,balbus1998}.  This causes the fluid to become turbulent, leading to outward angular momentum transport and allowing fluid to accrete 
inwards and form an accretion disk.  The accretion disk feeds the black hole and threads it with magnetic field lines.  These magnetic fields tap the rotational energy of the black hole and drive jets.  The remainder of this paper is an analysis of these jets.

\section{Simulation jets and jet power}
\label{sec:jets}

In this section, we first verify that the simulations described in the previous section are generating jets.  Then we show that the jet power matches the BZ prediction \eqref{eq:BZ2}.  In subsequent sections, we will show that the underlying physics producing jets in GRMHD simulations is indeed described by the BZ model in its membrane formulation.  

\subsection{Jet Lorentz factor}

\begin{figure*} 
\begin{overpic}[scale=0.12,unit=1mm]{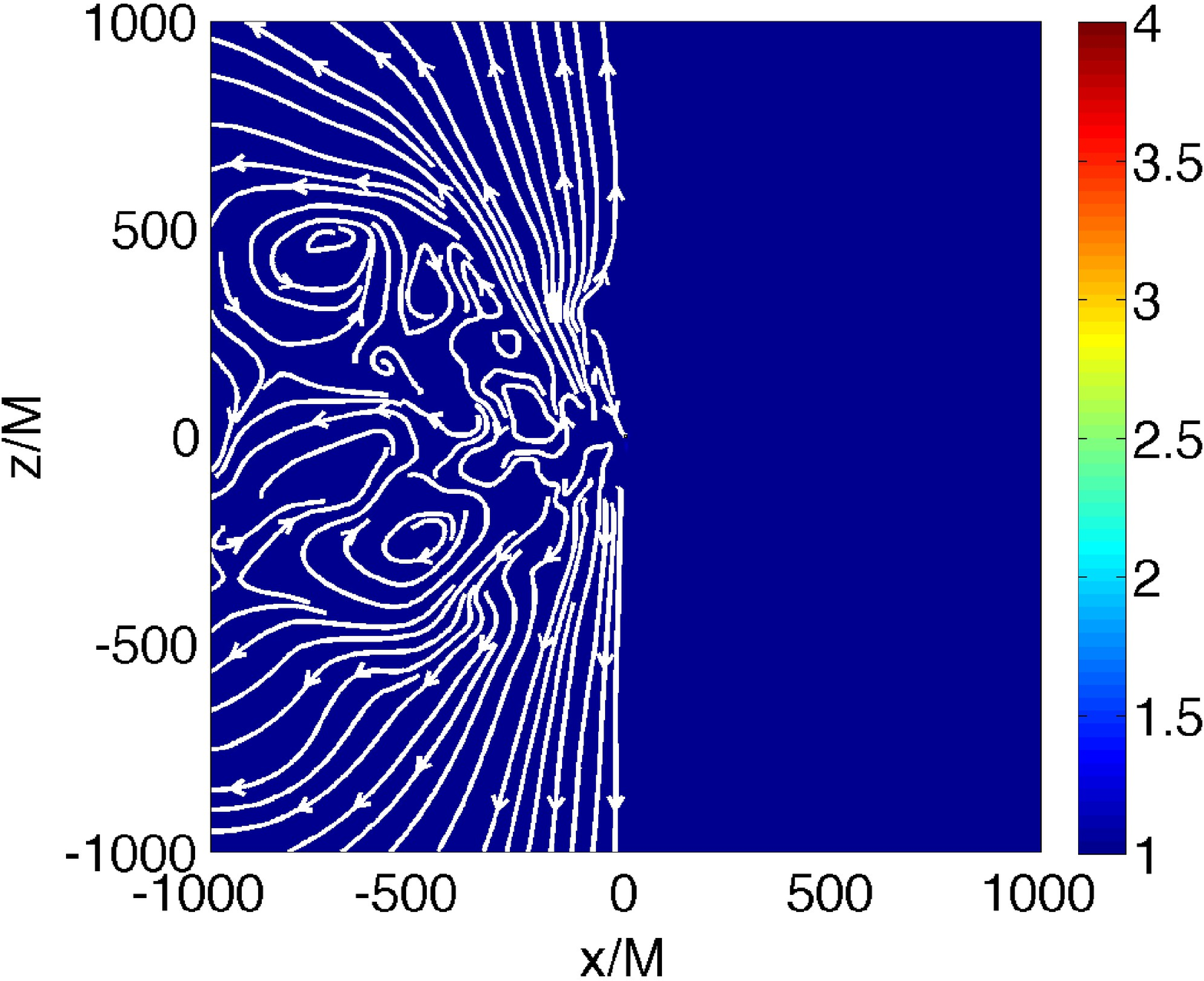}
	\put(57,53){\LARGE\color{white}$a_*=0$}
	\put(57,45){\LARGE\color{white}SANE}
\end{overpic}
\begin{overpic}[scale=0.12,unit=1mm]{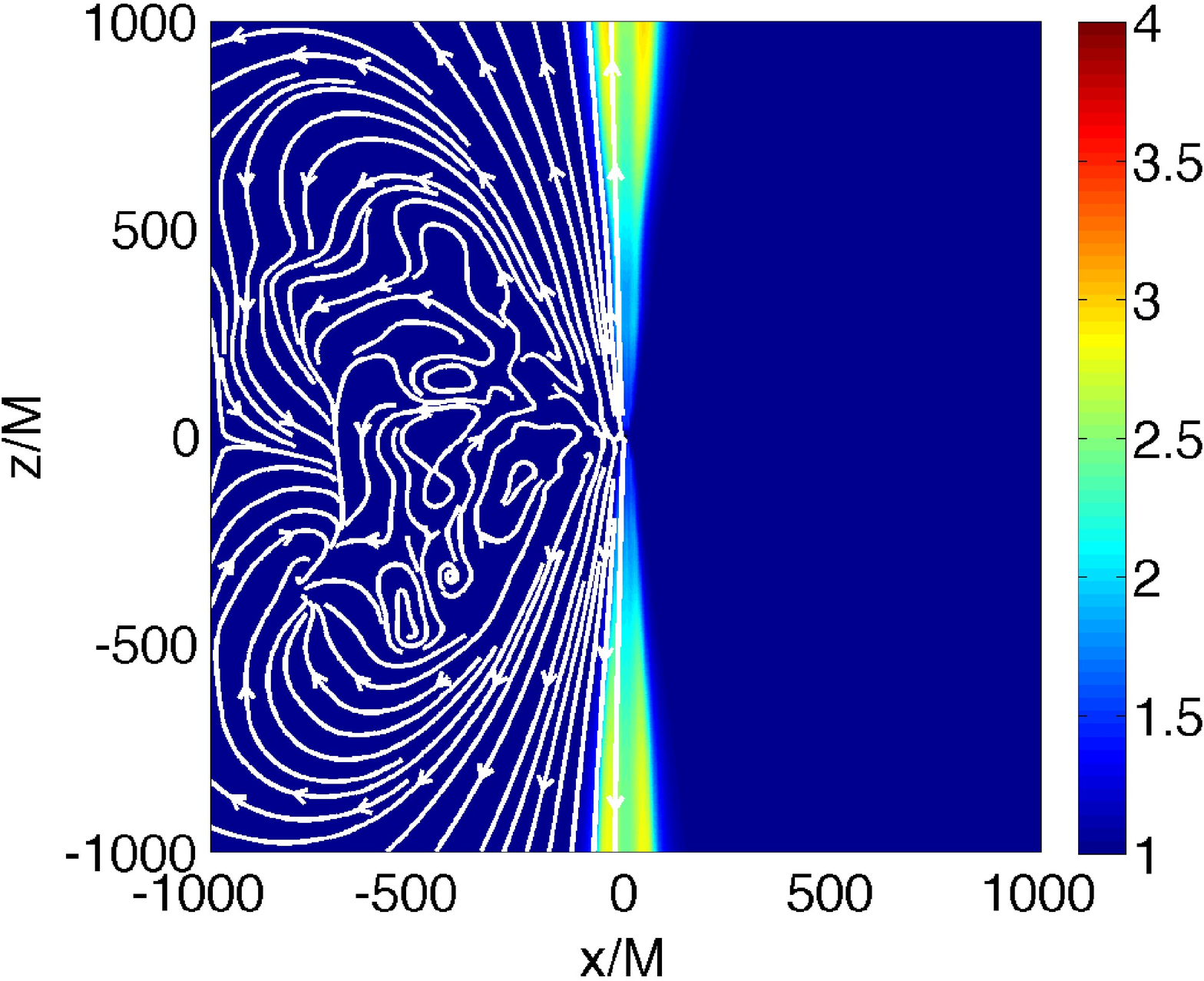}
	\put(57,53){\LARGE\color{white}$a_*=0.7$}
	\put(57,45){\LARGE\color{white}SANE}
\end{overpic}
\\
\begin{overpic}[scale=0.12,unit=1mm]{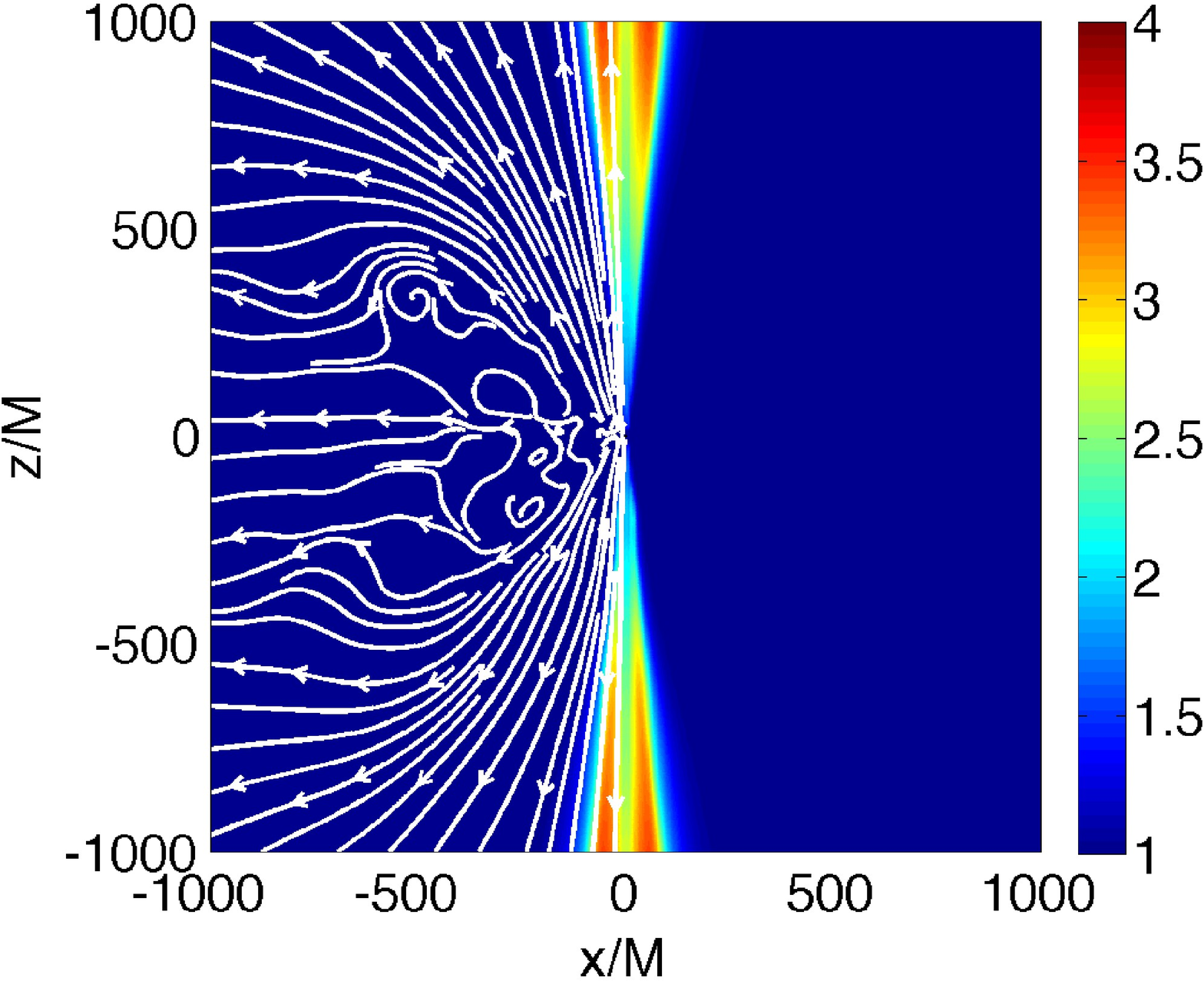}
	\put(57,53){\LARGE\color{white}$a_*=0.9$}
	\put(57,45){\LARGE\color{white}SANE}
\end{overpic}
\begin{overpic}[scale=0.12,unit=1mm]{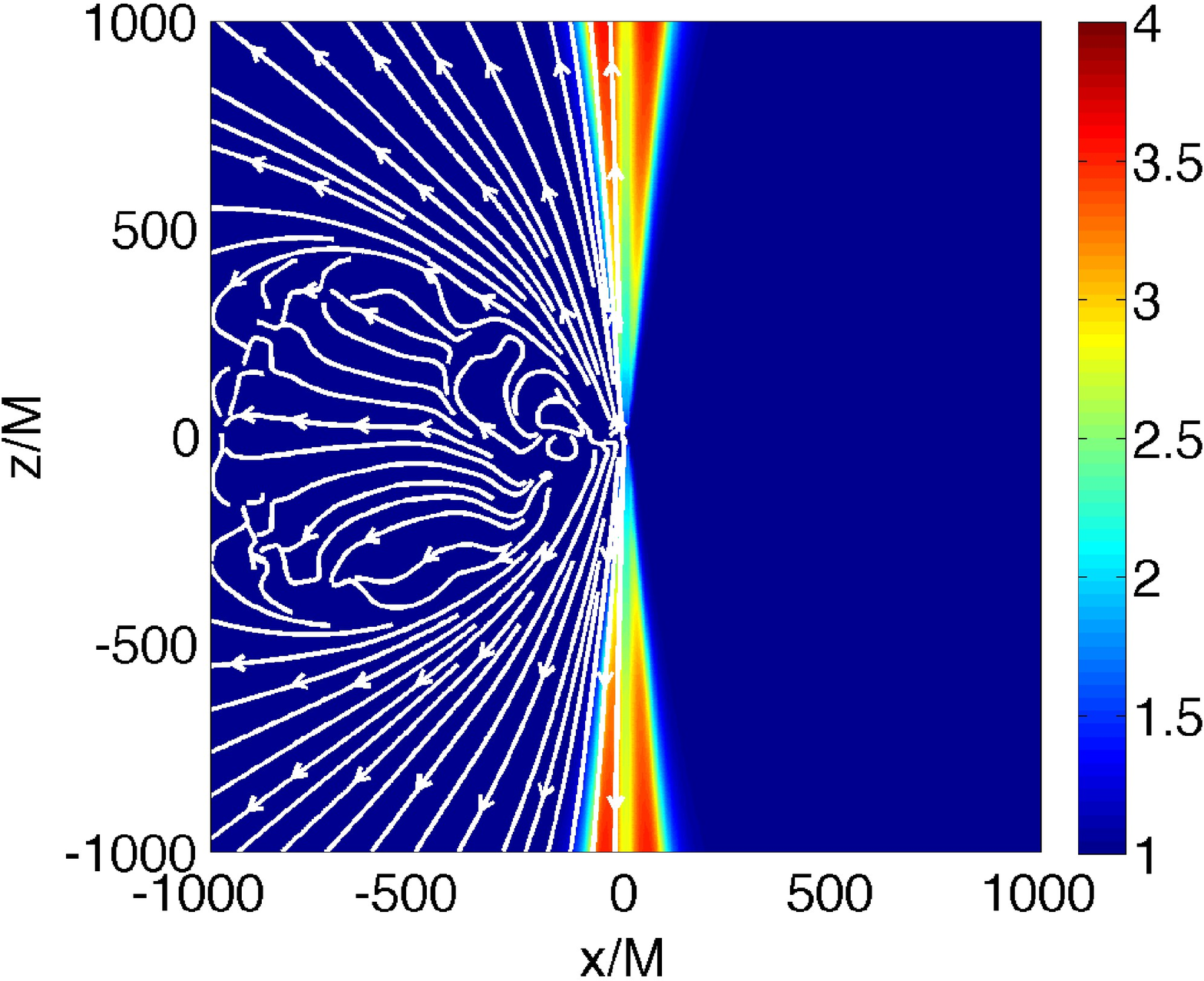}
	\put(57,53){\LARGE\color{white}$a_*=0.98$}
	\put(57,45){\LARGE\color{white}SANE}
\end{overpic}
\caption{\label{fig:lorentz} GRMHD simulation Lorentz factor and velocity streamlines for SANE runs.  The spinning black holes power jets.}
\end{figure*}

\begin{figure} 
\begin{overpic}[scale=0.12,unit=1mm]{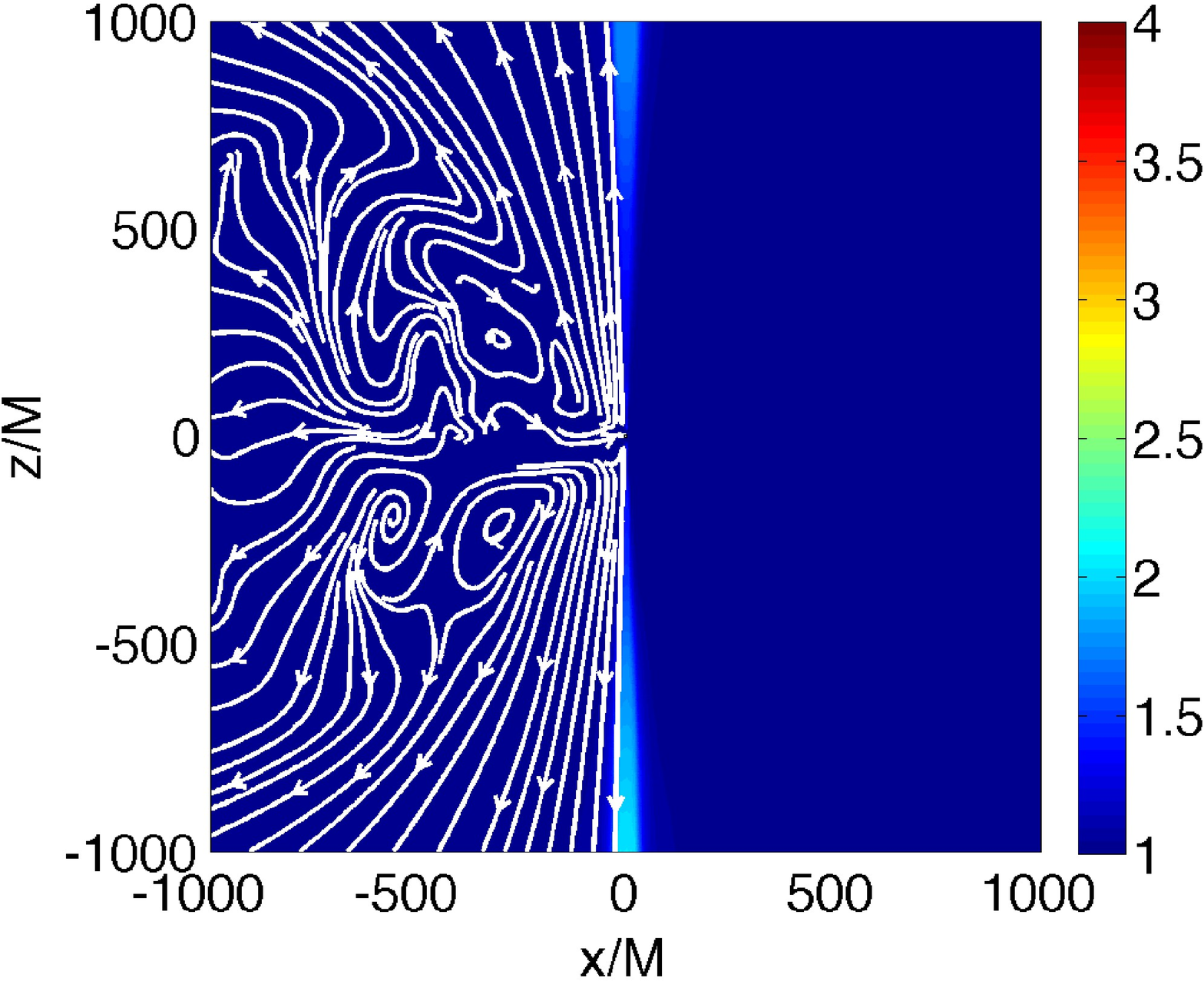}
	\put(55,53){\LARGE\color{white} $a_*=0$}
	\put(55,43){\LARGE\color{white} MAD}
\end{overpic}
\\
\begin{overpic}[scale=0.12,unit=1mm]{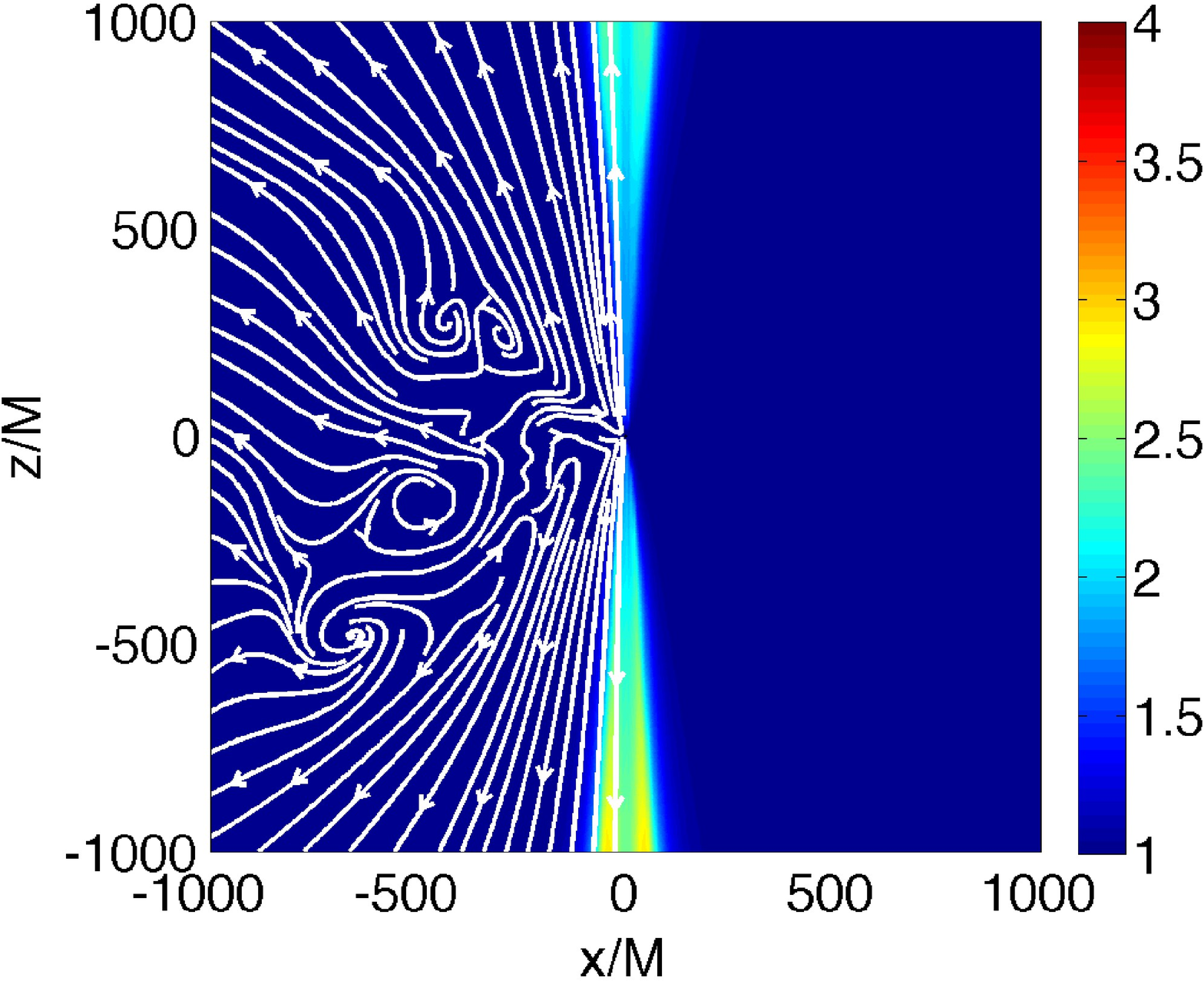}
	\put(55,53){\LARGE\color{white} $a_*=0.7$}
	\put(55,43){\LARGE\color{white} MAD}
\end{overpic}
\\
\begin{overpic}[scale=0.12,unit=1mm]{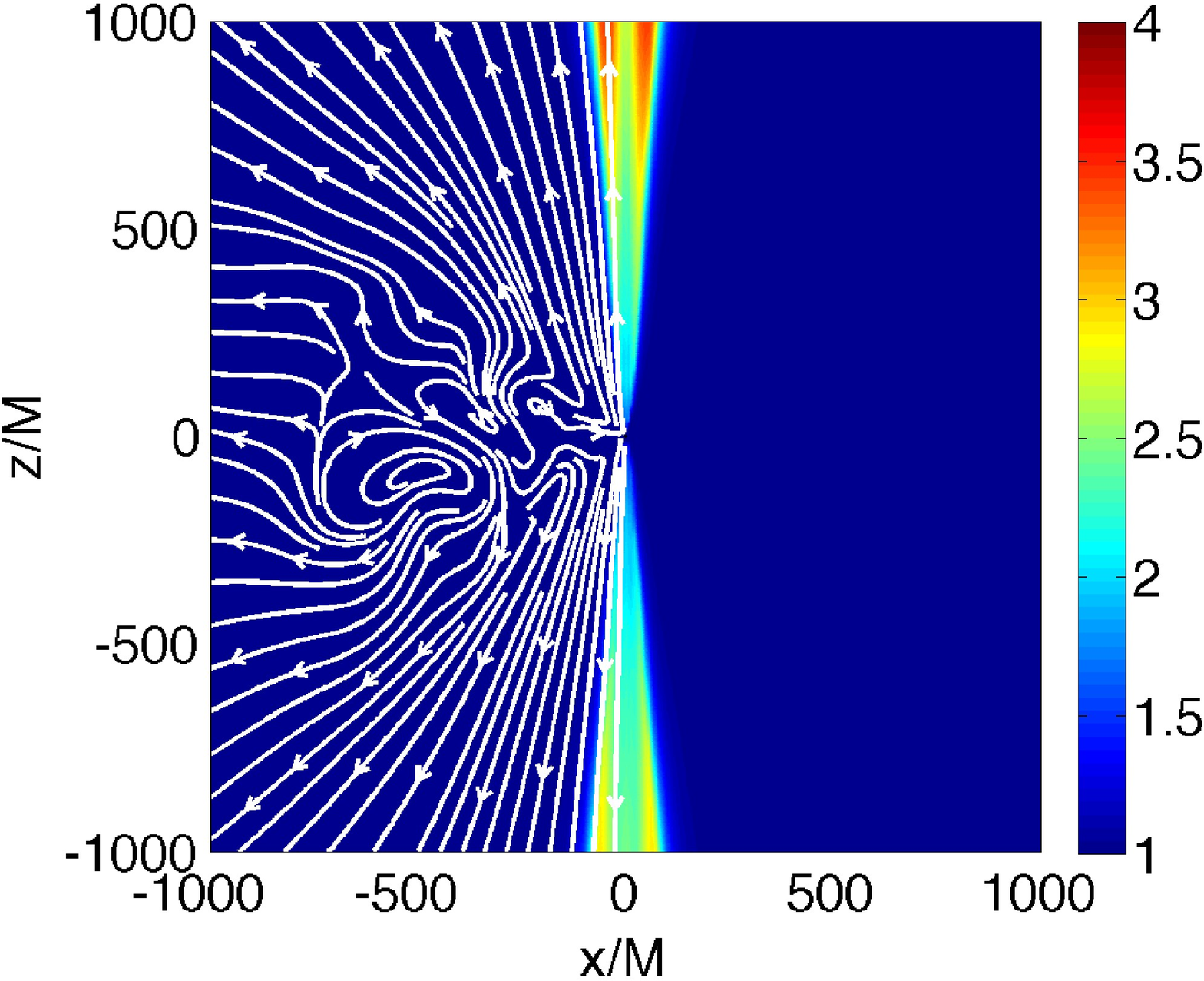}
	\put(55,53){\LARGE\color{white} $a_*=0.9$}
	\put(55,43){\LARGE\color{white} MAD}
\end{overpic}
\caption{\label{fig:lorentz_mad} Same as Figure \ref{fig:lorentz} but for MAD runs.} 
\end{figure}

The easiest way to detect jets in the simulations is to look at the gas Lorentz factor, $\gamma$, in the $x$-$z$ plane (where $z=r \cos\theta$ is along the black hole spin axis, and $x=r\sin\theta$ is along the equatorial plane).  This is plotted in Figures \ref{fig:lorentz} (for the SANE runs) and \ref{fig:lorentz_mad} (for the MAD runs).  We have time-averaged the simulation data over the last $10,000M$ of each run.  The jets show up as bright, collimated outflows along the black hole spin axes.   Collimation of the jet is provided by the accretion disk.  The accretion disk is invisible in these images because accreting gas has $\gamma \sim 1$.

% ------- Lorentz factor ----------

%----------------------------------------

There are already several hints of a connection between jet power and black hole spin in these images.  Jets only appear when the black hole is spinning; the two non-spinning black hole simulations have no jets.  The Lorentz factor (a proxy for jet power) increases slightly with spin, from $\gamma\sim3$ at low spins to $\gamma\sim 4$ at high spins.  These Lorentz factors should be interpreted with some caution.  HARM uses density floors to avoid the high magnetizations and low densities that lead to inversion failures.  The floors are mostly activated in the highly magnetized, low density regions along the jet axes, and could affect $\gamma$ \citep{2011MNRAS.418L..79T,2012MNRAS.423.3083M}.

The Lorentz factor peaks near the edges of the jets rather than along the axes: there is a low Lorentz factor core.  This is consistent with the BZ model.   Magnetic torques extracting black hole rotational energy scale with the lever arm radius, $\varpi$ (roughly the cylindrical distance from the black hole spin axis to the black hole membrane) as $\mathbf{\tau} = \mathbf{\varpi} \times \mathbf{F}$.  Near the axes, $\varpi$ (and so also torque) goes to zero, so the cores of the jets are less accelerated than the edges. 

\subsection{Jet power vs. time}

We introduce two proxies for jet power.  The first is
\beq\label{eq:mypjet}
P^{\rm mag} =  -\int_{\mathcal{H}} \alpha T^{\rm mag}_{nt} dA,
\quad (r=r_H),
\eeq
where $dA=(r_H^2+a^2)\sin\theta d\theta d\phi$ is an area element on the membrane and $\alpha=\rho/\Sigma\sqrt{\Delta}$ is the lapse function.  The outward normal vector is $e_n=\sqrt{\Delta}/\rho(\partial/\partial r)$ and $t$ is Boyer-Lindquist time.   The integral is over the entire black hole horizon, $\mathcal{H}$, so it includes the jet and the accretion disk in the integral over $\theta$.

The accretion disk might extract energy from the black hole, but only the energy extracted into the jet is directly relevant for observations of black hole jets.  (The jets observed by \citealt{2012MNRAS.419L..69N} and \citealt{2013ApJ...762..104S} are at $r\sim 10^{10}M$ so can be observationally distinguished from the accretion disk.) 
We thus introduce a second proxy for jet power:
\beq\label{eq:mypjet2nd}
P^{\rm tot}_{\rm jet} = -\frac{dM}{dt}\bigg\vert_{\mathcal{A}_{\rm jet}} = -\int_{\mathcal{A}_{\rm jet}} \alpha T_{nt} dA,
\quad (r=r_H).
\eeq
Now we have restricted the integral to the \emph{jet region}, $\mathcal{A}_{\rm jet}$, defined to be the region of the horizon where $-T_{nt}>0$ (net energy leaving the hole).  In other words, we do not include the accretion disk in the integral over $\theta$.  We have also switched from including only the magnetic energy flux, $T^{\rm mag}_{nt}$, to including the combined magnetic and gas energy flux, $T_{nt}$.  

We expect $P^{\rm mag}$ to give the best fit to the BZ model, because the BZ prediction \eqref{eq:BZ2} includes the entire horizon and considers only magnetic torques.  We expect $P^{\rm tot}_{\rm jet}$ to be more relevant for jet observations.

Our simulations with non-spinning black holes always have $-T_{nt}<0$ across the entire horizon, meaning the black hole never loses energy.  So $\mathcal{A}_{\rm jet}$ is empty and $P^{\rm tot}_{\rm jet}=0$, always.

\begin{figure} 
\includegraphics[width=\columnwidth]{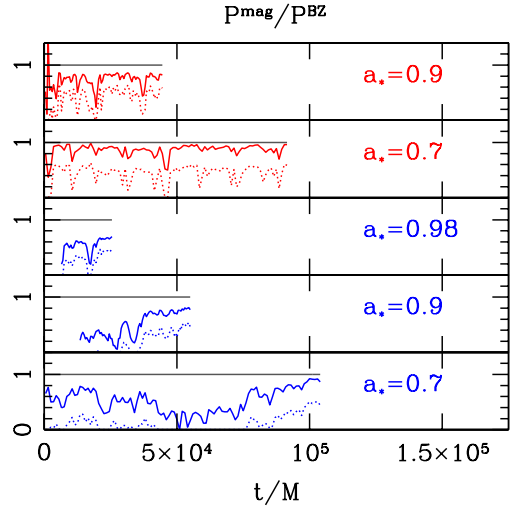}
\caption{\label{fig:vst} Jet power vs. time for our five GRMHD simulations with spinning black holes.  $P^{\rm mag}$ (solid) and $P^{\rm tot}_{\rm jet}$ (dotted) are shown separately.  The MAD runs (red) converge more quickly than the SANE runs (blue).   The standard BZ prediction \eqref{eq:BZ2} is also shown (solid black lines).}
\end{figure}

The simulations with spinning black holes are more lively.  We show $P^{\rm mag}$ and $P^{\rm tot}_{\rm jet}$ vs. time for these runs in Figure \ref{fig:vst}.   Initially, none of the simulations have jets.  The MAD simulations develop jets quickly, after just a few $1000M$, and soon thereafter the jet power saturates around a quasi-steady value.  Remaining fluctuations in the jet power arise from turbulent fluctuations in the accretion disk feeding the black hole.  The SANE simulations develop jets much more slowly.  Even by the end of this set of runs the jet power has probably not converged to a quasi-steady value.  The different onset time of jets in the SANE and MAD runs is easily understood.  MAD runs begin with a single magnetic loop.  This loop is so big that less than half of it is dragged onto the hole over the duration of the simulation.  So the accretion flow is continually depositing magnetic flux of the same polarity on the horizon and a large mean field builds up.  In the SANE runs on the other hand, the initial magnetic field is a series of small poloidal loops.  The sign of the magnetic flux arriving on the hole is alternating with time and it is difficult for the hole to build up a large mean field.  Nonetheless, at late times the SANE runs appear to be converging to what is perhaps the same quasi-steady jet power as the MAD runs.

% ----------- Jet power vs. time --------------

It is interesting to compare $P^{\rm mag}$ (solid red and blue curves) and $P^{\rm tot}_{\rm jet}$ (dotted red and blue curves) in Figure \ref{fig:vst}.  The former is always $\sim 2-3$ times larger than the latter.  There are two reasons for this.  First,  $P^{\rm tot}_{\rm jet}$  does not include the accretion disk, so it is limited to a smaller region of the horizon.  Second, even the jet regions, $\mathcal{A}_{\rm jet}$, are not devoid of gas.  In the polar regions of the flow, gas within a few gravitational radii of the horizon is falling onto the black hole.  This gas torques the hole in the opposite sense as the magnetic fields, so the total jet power is lower than the magnetic jet power.  The BZ prediction \eqref{eq:BZ2} (black lines) gives an acceptable fit to the magnetic jet power, $P^{\rm mag}$, but overestimates the total jet power, $P^{\rm tot}_{\rm jet}$.  The neglect of gas accretion is the main shortcoming of the BZ model identified in this paper.  To quantify this effect, we first need to time-average the data.

\subsection{Time averaging}

Time averaging eliminates (or at least reduces) the imprint of the accretion disk's turbulent variability on the jets.  The duration of the time interval is limited by the run's duration and the jet onset time.  We have time averaged the simulation data over the last $10,000M$ of each run.  Inspection of Figure \ref{fig:vst} shows that $P_{\rm jet}(t)$ has a quasi-steady value over this period.

%and we continue time averaging until the end of the run.  The smallest time averaging interval is $\Delta t_{\rm steady}=10,000M$.  Table \ref{tab:sims} gives $\Delta t_{\rm steady}$ for each of our simulations.

\subsection{Jet power vs. black hole spin}
\label{sec:pjetvsspin}
\begin{figure} 
\includegraphics[width=\columnwidth]{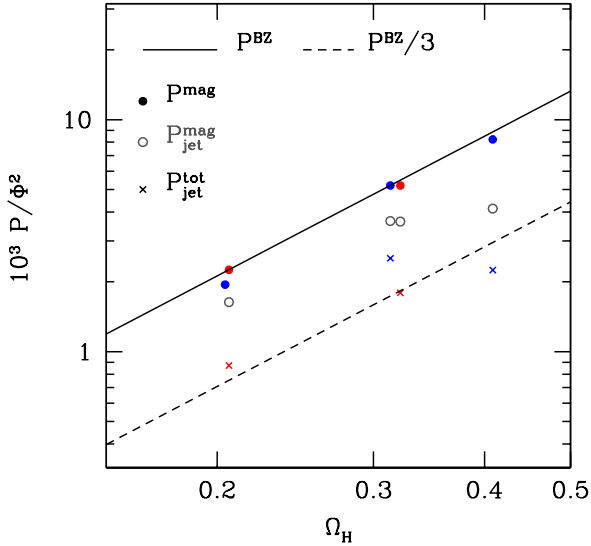}
\caption{\label{fig:bz} Jet power versus black hole spin for SANE (blue) and MAD (red) GRMHD simulations.   $P^{\rm mag}$ (filled circles), $P^{\rm mag}_{\rm jet}$ (open circles), and $P^{\rm tot}_{\rm jet}$ (crosses) are shown separately.  The BZ model (solid line) gives a good fit to $P^{\rm mag}$.  Reducing the BZ jet power by a factor of 3 gives a good fit to $P^{\rm tot}_{\rm jet}$. Data for the MAD runs has been shifted slightly to the right for readability.  We have normalized by $\Phi^2$ because this varies across the simulations according to the magnetic field geometry in the initial conditions.}
\end{figure}

We now come to the main result of this section, the dependence of jet power on black hole spin.  This is shown in Figure \ref{fig:bz} for the five simulations with spinning black holes.  
The magnetic power, $P^{\rm mag}$, (filled circles) agrees with the BZ prediction \eqref{eq:BZ2} to within $10\%$.  This is consistent with earlier simulation results \citep{2010ApJ...711...50T,2012JPhCS.372a2040T}.  (Our $P^{\rm mag}$ is similar to the $\eta$ parameter of \citealt{2010ApJ...711...50T} and \citealt{2012JPhCS.372a2040T}.)  

% ----------- Jet power vs. spin --------------

Discrepancies between $P^{\rm mag}$ and the BZ model can be traced to  approximations in the BZ prediction \eqref{eq:BZ2}.  First, the standard BZ model assumes $\Omega_F/\Omega_H=1/2$.  As we will show in \S\ref{sec:load}, our simulations have $\Omega_F/\Omega_H\approx 0.35$.  For general $\Omega_F/\Omega_H$, the BZ model predicts \citep{bz77,1986bhmp.book.....T}:
\beq\label{eq:pjeteff}
P^{\rm BZ} = \frac{1}{6\pi}4\Omega_F/\Omega_H(1-\Omega_F/\Omega_H) \Omega_H^2 \Phi^2,
\eeq
where $\Omega_F$ is the field line angular velocity, $\Omega_H$ is the horizon angular velocity, and $\Phi$ is the magnetic flux threading the horizon.  One obtains maximum efficiency for $\Omega_F\Omega_H=1/2$.  Switching from $\Omega_F/\Omega_H=1/2$ to $\Omega_F/\Omega_H=0.35$ lowers the jet power by $8\%$.  In other words, assuming $\Omega_F/\Omega_H=1/2$ introduces an $8\%$ error into the jet power estimate for our simulations.  We conclude that $\Omega_F/\Omega_H=1/2$ is a good assumption for analytical applications of the BZ model.

The BZ jet power \eqref{eq:BZ2} also assumes uniform  $B_n$ over the horizon (see Appendix A).   This assumption  breaks down at high black hole spins, for which the magnetic field tends to bunch up near the polar axes \citep{2010ApJ...711...50T}.  However, Figure \ref{fig:bz} shows $P^{\rm BZ}$ and $P^{\rm mag}$ continue to agree to within $10\%$ even at $a_*=0.98$.   We conclude that the assumptions $\Omega_F/\Omega_H=1/2$ and uniform $B_n$ together create a $\sim10\%$ discrepancy between the BZ prediction \eqref{eq:BZ2} and our simulations.

The jet power proxy $P^{\rm mag}$ measures the electromagnetic energy extracted across the  entire black hole horizon.  Not all of this energy is extracted into the jet: some of it is extracted into the accretion disk.  We expect $P^{\rm tot}_{\rm jet}$ to be more relevant for jet observations than $P^{\rm mag}$ because the former is restricted to the jet region.  Figure \ref{fig:bz} shows $P^{\rm tot}_{\rm jet}$ (crosses) is about a factor of three lower than $P^{\rm mag}$ (filled circles).  There are two reasons for this reduction.  First, $P^{\rm tot}_{\rm jet}$ is restricted to a smaller region of the horizon, the jet region.  Second, even within the jet region, there are gas torques counteracting the magnetic torques, further lowering $P^{\rm tot}_{\rm jet}$.  

To isolate the relative importance of these two effects, we introduce a third proxy for jet power
\beq\label{eq:mypjet3}
P^{\rm mag}_{\rm jet} = -\int_{\mathcal{A}_{\rm jet}} \alpha T^{\rm mag}_{nt} dA,
\quad (r=r_H),
\eeq
which is intermediate between $P^{\rm mag}$ and $P^{\rm tot}_{\rm jet}$ (Figure \ref{fig:bz}, open circles).  It is limited to the jet region, but it does not include gas torques.  So the discrepancy between $P^{\rm mag}$ and $P^{\rm mag}_{\rm jet}$ measures the effect of restricting the jet power to the jet region.  The discrepancy between $P^{\rm mag}_{\rm jet}$ and $P^{\rm tot}_{\rm jet}$ measures the importance of gas torques within the jet region.  Figure \ref{fig:bz} shows that these two effects are comparable: each contributes about $25\%$  to the discrepancy between $P^{\rm mag}$ and $P^{\rm tot}_{\rm jet}$.

To summarize, the jet power proxy which is probably most relevant for jet observations, $P^{\rm tot}_{\rm jet}$,  is roughly $60\%$ lower than the BZ prediction \eqref{eq:BZ2}.   Simply lowering $P^{\rm BZ}$ by $60\%$ gives a good fit to our simulations.  The main sources of discrepancy between the simulations and the BZ model are that the simulated jets do not cover the whole horizon and, within the jet region, there are gas torques partially counteracting the electromagnetic torques.    \citet{2010ApJ...711...50T}  have previously considered restricting the BZ model to a subregion of the full horizon and our power estimates are consistent with theirs.  Our estimate for the importance of gas torques in the jet region should be considered an upper limit, as numerical floor activations in the polar regions introduce more gas there than should otherwise be present.   More work is needed to isolate the effect of the numerical floors and better determine the importance of gas torques in the jet region.

We have found that assuming $\Omega_F/\Omega_H=1/2$ and uniform horizon $B_n$ introduces order $10\%$ discrepancies between the BZ jet power estimate and our simulations.    It is worth noting that simulations of thicker ($h/r\sim 1$), more magnetized accretion flows have found $\Omega_F/\Omega_H\approx 0.2$ \citep{2012MNRAS.423.3083M,2013arXiv1303.1644B}.  Reducing $\Omega_F/\Omega_H$ from 0.5 to 0.2 lowers the jet power by $40\%$ (equation \ref{eq:pjeteff}).  This example serves to emphasize that details of the accretion flow can change jet power estimates by order unity factors.  This suggests it will be difficult in practice to predict the power of astrophysical jets to better than a factor of order unity.

\section{Energy Extraction At The Horizon}
\label{sec:horizon}

In the previous section, we found good agreement between the simulated and BZ jet power.  In this section and the next, we turn to the underlying physics.  We show that the torques and  electromagnetic fields acting on the black hole membrane in our GRMHD simulations are in excellent agreement with the BZ model in its membrane formulation.   All of the numbered equations in this section can be found in \citet{1986bhmp.book.....T}.

\subsection{First law of black hole thermodynamics}

The black hole membrane has mass $M$, angular momentum $J$, angular velocity $\Omega_H$, Bekenstein-Hawking temperature $T_H$, and Bekenstein-Hawking entropy $S_H$.  These are related by the first law of black hole thermodynamics:
\beq\label{eq:firstlaw}
dM = \Omega_H dJ + T_H dS_H.
\eeq
The left hand side is the total energy entering ($dM>0$) or leaving ($dM<0$) the membrane.  It is related to jet power by $P_{\rm jet}=-dM/dt$.  The right hand side splits $dM$ into contributions from torques on the membrane, $\Omega_H dJ$, and dissipation in the membrane $T_H dS_H$.  
It is impossible to extract energy from a nonspinning black hole because $\Omega_H=0$.   The torque is negative when black hole rotational energy is being extracted and positive otherwise.  The dissipation is always positive, as demanded by the second law of black hole thermodynamics.  

In the BZ model, torques and dissipation have similar magnitudes.  This might seem surprising: the Bekenstein-Hawking temperature  $T_H\sim \hbar$ is miniscule.  However, the Bekenstein Hawking entropy $S_H \sim 1/\hbar$ is huge, so the product $T_H dS_H$ is finite and astrophysically relevant.  The importance of the dissipation term is a key distinction between black holes and pulsars.  Pulsars are perfect conductors, so $dM_{\rm pulsar}=\Omega_{\rm pulsar} dJ_{\rm pulsar}$.  Black holes are not perfect conductors, so jet power is a combination of torques and dissipation on the membrane. This explains why a pulsar magnetosphere has $\Omega_F = \Omega_{\rm pulsar}$ while a black hole magnetosphere has $\Omega_F < \Omega_H$: pulsar field lines are frozen into the star, but black hole field lines slip with respect to the membrane because the latter is not a perfect conductor.

\subsection{Torques and dissipation on the membrane}

We have shown that GRMHD simulations of jets and the BZ model have the same $P^{\rm mag}$.  Now we will show they produce the same torques and dissipation on the membrane.  Energy flow, torques, and dissipation on the membrane are related to the GRMHD stress energy tensor by:
\begin{align}\label{eq:1stlawgrmhd}
\frac{dM}{dt} 
&= -\int \alpha T_{nt} dA,\quad (r=r_H), \\
\Omega_H\frac{dJ}{dt} 
&= \Omega_H \int \alpha T_{n\phi} dA,\quad (r=r_H),\label{eq:torque} \\
T_H\frac{dS_H}{dt} 
&= -\int \alpha^2 T_{n\hat{t}} dA,\quad (r=r_H),\label{eq:dissip}
\end{align}
where $\hat{t}$ is the proper time of a ZAMO.  The mathematical distinction between $dM/dt$ and $T_H dS_H/dt$ is that the former is related to the Boyer-Lindquist energy flux, $T_{nt}$, and the latter is related to the ZAMO energy flux,  $T_{n\hat{t}}$.   Boyer-Lindquist and ZAMO time are related at the membrane by
\beq\label{eq:zamot}
\vec{e}_{\hat{t}} = \frac{1}{\alpha}\left(\frac{\partial}{\partial t}+\Omega_H \frac{\partial}{\partial \phi}\right).
\eeq

The first law of black hole thermodynamics is equivalent to equations \eqref{eq:1stlawgrmhd}-\eqref{eq:zamot}, as an easy calculation shows:
\begin{align}
T_H\frac{dS_H}{dt}  &= -\int \alpha^2 T_{n\hat{t}} dA\\
&= -\int \alpha T_{nt} dA -\Omega_H \int \alpha T_{n\phi} dA\\
&=\frac{dM}{dt} - \Omega_H \frac{dJ}{dt}.
\end{align}
The second law of black hole thermodynamics, $T_H dS_H\geq 0$,  is equivalent to the right hand side of equation \eqref{eq:dissip} by the weak energy condition:  $-T_{n\hat{t}}\geq0$ because ZAMOs are orthonormal observers.

The energy flow, torques, and dissipation on the membrane in the GRMHD simulations are shown in Figure \ref{fig:horizonedot} (for the SANE runs) and Figure \ref{fig:horizonedot_mad} (for the MAD runs).  We have normalized each time-averaged quantity to the time-averaged accretion rate, $\dot{m}$.   The membrane is depicted as a spherical surface with coordinates on the sphere corresponding to Boyer-Lindquist coordinates $(\theta,\phi)$ in the usual way.  The membrane is gaining energy at blue regions and losing energy at red regions.  

The blue band near the equator is the accretion disk and the thickness of this band is set by the thickness of the disk.  The accretion disk adds energy to the black hole.  The yellow and red bands wrapping around the poles are the jet regions, $\mathcal{A}_{\rm jet}$, where the black hole is losing rotational energy.  
In the BZ model there is no accretion disk (so there would be no blue band) and the jet regions  extend to the equator.  In the GRMHD simulations the jet region is limited to areas outside the disk.  The disk thickness in our simulations is $h/r\sim 0.3$ and this reduction in $\mathcal{A}_{\rm jet}$ lowers the jet power by a factor of $2-3$.  Physically this is not a particularly interesting distinction between GRMHD simulations and BZ jets, as it is easily absorbed into the BZ model by simply restricting the BZ jet power to $\mathcal{A}_{\rm jet}$. 

The torques and dissipation on the membrane in the GRMHD simulations pass basic consistency checks.  The torque on nonspinning black holes is zero everywhere.  Dissipation is always positive as demanded by the second law of black hole thermodynamics (except possibly at the last couple grid cells near the poles where HARM has trouble with floors on magnetization).  The net energy flux is always smaller than the torque because there is dissipation.  Torques decrease near the poles as the lever arm radius $\varpi$ goes to zero.   This causes the dissipation and total energy flux to drop as well.  Everything is amplified by black hole spin.

Figures \ref{fig:horizonsplit} and \ref{fig:horizonsplit_mad}  provide a more fine grained look at the membrane energy flux, torques, and dissipation in the GRMHD simulations.  Each quantity is split into electromagnetic and hydrodynamic components.   We find that the hydrodynamic torques are everywhere positive, so the extraction of black hole energy is a purely electromagnetic process.  We also find that electromagnetic torques become negligible in the accretion disk region.  The accretion disk's magnetic fields are extracting energy from the spin of the black hole, but it is small compared to the energy in the accretion flow.

\begin{figure*} 
\includegraphics[width=\linewidth]{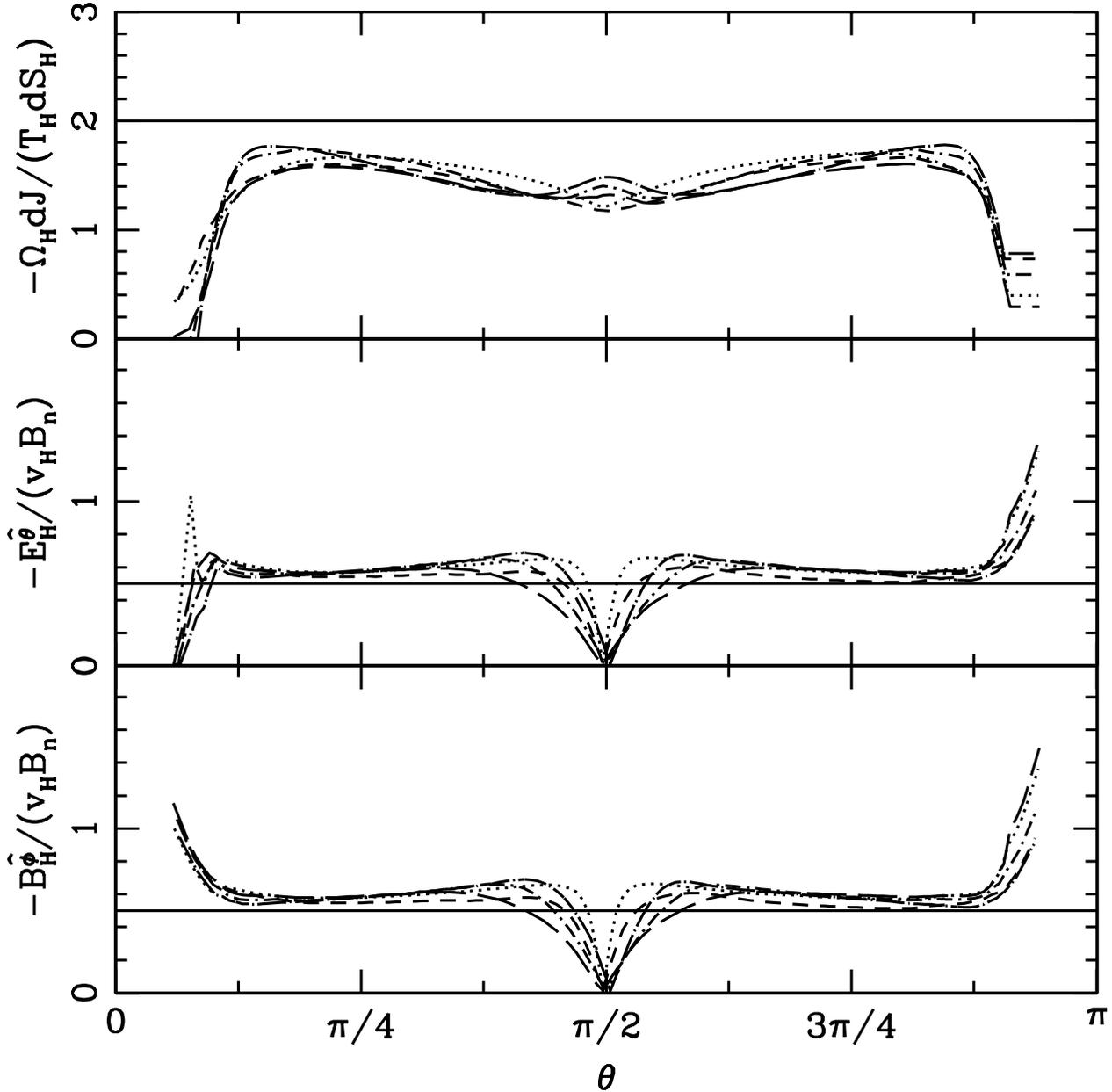}
\caption{\label{fig:ratios} Torques and electromagnetic fields at the membrane as a function of $\theta$.  The line types are as follows.   SANE runs: $a_*=0.7$  (long dashed), $a_*=0.9$ (dot-dashed), and $a_*=0.98$ (dot-long dashed).  MAD runs: $a_*=0.7$ (dotted) and $a_*=0.9$ (dashed). Solid lines indicate the standard BZ prediction.  The simulation data should not be trusted near $\theta=0$ and $\theta=\pi$, where low gas densities require numerical floor activations.  The simulations differ from the standard BZ model near $\theta=\pi/2$ because the simulations have accretion disks and the BZ model does not.  Over the remaining range of polar angles, the simulations and BZ model are in good agreement.  The discrepancies that do appear can be traced to the field line angular velocity: the simulations have $\Omega_F/\Omega_H \approx 0.3$ and the standard BZ model assumes $\Omega_F/\Omega_H=1/2$.}
\end{figure*}

The crucial test of  the BZ model is that it makes a specific prediction for the relative strength of torques and dissipation.  The standard BZ model predicts:
\begin{equation}\label{eq:BZtorquediss}
\Omega_H dJ = -2T_H dS_H, \quad \text{(for $\Omega_F/\Omega_H=1/2$)}.
\end{equation}
Figure \ref{fig:ratios} shows the ratio $-\Omega_HdJ/(T_HdS_H)$ as a function of $\theta$ for our GRMHD simulations.  The ratio is roughly independent of $\theta$ except near the polar axes, where numerical floor activations make the simulation results untrustworthy.  The simulations have $\Omega_HdJ\approx -1.5 T_HdS_H$, meaning they produce about $25\%$ less torque per unit dissipation than the standard BZ model (equation \ref{eq:BZtorquediss}).  This can be traced back to the fact that the standard BZ model assumes maximum efficiency ($\Omega_F/\Omega_H= 1/2$).  For general $\Omega_F/\Omega_H$, the BZ model predicts \citep{1986bhmp.book.....T}
\beq\label{eq:torquegeneral}
-\frac{\Omega_H dJ}{T_H dS_H}=\left(1-\frac{\Omega_F}{\Omega_H}\right)^{-1}.
\eeq
Our simulations have $\Omega_F/\Omega_H\approx 0.35$ (see \S\ref{sec:load}), for which equation \eqref{eq:torquegeneral} gives $-\Omega_H dJ/(T_H dS_H)\approx 1.5$, just as we observe in Figure \ref{fig:ratios}.   Note that while assuming maximum efficiency introduces a $25\%$ error into $-\Omega_H dJ/(T_H dS_H)$, it only introduces an $8\%$ error into $P^{\rm tot}_{\rm jet}$, as discussed in \S\ref{sec:pjetvsspin}.

%-----------------Ratios---------------------------

\section{Electromagnetic fields at the horizon}
\label{sec:fields}

In the previous section, we showed that membrane torques and dissipation in the GRMHD simulations are related as the BZ model predicts: $\Omega_H dJ \approx -2T_H dS_H$.  We also showed that extraction of black hole energy is a purely electromagnetic process mediated by electric and magnetic fields and currents acting on the black hole membrane.  We now turn to the study of these fields and currents.  

\subsection{Membrane formalism}

First, we review the necessary pieces of the membrane formalism \citep{1986bhmp.book.....T}.

The electric and magnetic fields measured by ZAMOs are computed from the GRMHD Faraday tensor according to
\begin{align}
\vec{E} = F^{i\hat{t}}, \quad
\vec{B} =F^{*i\hat{t}},
\end{align}
where ZAMO time, $\hat{t}$, is given by equation \eqref{eq:zamot}.  These electric and magnetic fields are  three-dimensional vectors.  

At the black hole membrane, we split these electric and magnetic fields into their components perpendicular and parallel to the membrane.  The perpendicular component of the magnetic field is $B^{n}$, 
where the outward normal vector is related to $\partial/\partial r$ by $e_{n}=\sqrt{\Delta}/\rho (\partial/\partial r)$.

The perpendicular component of the electric field  is
\beq
\sigma_H = E^{n}, \quad (r=r_H)
\eeq
and called the membrane's charge density, for the following reason.  An observer outside a black hole never sees a field line cross the membrane, because gravitational redshifting causes everything falling in to appear to freeze just outside.  Thus field lines appear to terminate on the membrane.  As a result, ZAMOs infer a charge distribution on the membrane, $\sigma_H$, sourcing the radial electric fields.  The inferred distribution of positive and negative charges on the membrane sums to $Q$, the charge of the black hole.  Our simulations use the Kerr metric, so $Q=0$ even when $\sigma_H$ is nonzero.  \footnote{Reasoning by analogy, $B_n$ is sometimes called the membrane's magnetic monopole distribution.}

For the same reason, an observer outside the black hole believes the parallel components of $\vec{E}$ and $\vec{B}$ are terminated at the membrane.  To ZAMOs, the membrane is a conductor endowed with just the right fields to terminate $\vec{E}^\parallel$ and $\vec{B}^\parallel$.  This defines the membrane  fields:
\beq
\vec{E}_H = \alpha \vec{E}^\parallel, \quad 
\vec{B}_H = \alpha \vec{B}^\parallel, \quad (r=r_H),
\eeq
where $\alpha=\rho/\Sigma \sqrt{\Delta}$ is the lapse function.  These electric and magnetic fields are two-dimensional vectors.  

We have now packaged the six degrees of freedom of $F_{\mu\nu}$ into the set $B_n, \sigma_H, \vec{E}_H$, and $\vec{B}_H$.  From this set, we can compute the horizon current, $\vec{J}_H$.  The horizon behaves as a resistor with $R_H=1\thinspace(=377 \text{\thinspace ohms in physical units})$ and the current is 
\beq
\vec{J}_H = \vec{E}_H/R_H.
\eeq

In these variables, the power generated by electromagnetic torques and dissipation on the horizon \eqref{eq:torque}-\eqref{eq:dissip} are
\begin{align}
\Omega_H \frac{dJ}{dt} &= \Omega_H\int 
\varpi\left[\sigma_H E^\phi +(\vec{J}_H \times \vec{B}_n)^\phi\right]  dA,\\
T_H \frac{dS_H}{dt} &= \int R_H J_H^2 dA.
\end{align}
These are familiar expressions from ordinary three-dimensional mechanics for the electromagnetic torque on a membrane, $\vec{\varpi}\times \vec{F}_L$, and dissipation in a resistor, $R_HJ_H^2$.  This simplicity is an advantage of the membrane formalism.

\subsection{Comparison of GRMHD and BZ electromagnetic fields}

In \S\ref{sec:horizon}, we showed that the BZ model and GRMHD simulations produce the same torques and dissipation on the membrane.  We now show that the underlying electromagnetic fields are also the same.  

There are six degrees of freedom: $B_n$, $\sigma_H$, $\vec{E}_H$, and $\vec{B}_H$.   The BZ model leaves $B_n$ a free parameter.  The remaining degrees of freedom are predictions we can test against our GRMHD simulations.  The ($t$,$\phi$)-averaged electromagnetic fields of our simulations are shown in Figures \ref{fig:horizonfields} (SANE runs) and \ref{fig:horizonfields_mad} (MAD runs).  The accretion rate varies from run to run, so we plot the fields in units of $\sqrt{\dot{m}}$, where $\dot{m}$ is the ($t$,$\theta$,$\phi$)-averaged accretion rate.

First consider the radial magnetic field, $B_n$.  This is a free parameter of the BZ model.  In our GRMHD simulations, it is spontaneously generated by the accretion disk.  The relaxed field geometry has a simple structure.  It is roughly uniform over the jet regions and the sign of the field is reversed in the northern and southern hemispheres.   So it is similar to a split monopole, the simplest possible field.

Now consider the membrane electric field, $\vec{E}_H$.  The BZ model assumes axisymmetry, which implies \citep{1982MNRAS.198..339T}:
\beq
E_H^{\hat{\phi}}=0.
\eeq  
Our GRMHD simulations do not enforce axisymmetry, but the time-averaged fields become roughly axisymmetric at the membrane.  Figures \ref{fig:horizonfields} and \ref{fig:horizonfields_mad} show $E_H^{\hat{\phi}}\approx0$ and the membrane electric field runs north-south.  

The standard BZ model fixes the $\hat{\theta}$-component of the electric field by winding up $B_n$:
\beq\label{eq:ehstandard}
E_H^{\hat{\theta}} = -\frac{1}{2}v_H B_n, \quad (\Omega_F/\Omega_H=1/2),
\eeq
where $v_H=\varpi\Omega_H$ is the velocity of the membrane (note $v_H=1$ at the equator when $a_*=1$).  Figure \ref{fig:ratios} (middle panel) shows $-E_H^{\hat{\theta}}/(v_H B_n)$ as a function of $\theta$ for our GRMHD simulations.  The simulations have $E_H^{\hat{\theta}} \approx -0.65 v_H B_n$, so the simulated electric field is about $30\%$ larger than predicted by the standard BZ model (equation \ref{eq:ehstandard}).   This is because the simulations do not achieve perfect efficiency.   For general $\Omega_F/\Omega_H$, the BZ model predicts  \citep{1986bhmp.book.....T}
\beq\label{eq:eh}
E_H/(v_H B_n) = -(1-\Omega_F/\Omega_H).
\eeq
Our simulations have $\Omega_F/\Omega_H\approx 0.35$ (see \S\ref{sec:load}), for which equation \eqref{eq:eh} predicts $E_H/(v_H B_n)\approx -0.65$, in excellent agreement with our results in Figure \ref{fig:ratios}.

The magnetic field,  $\vec{B}_H$, is not really an independent variable, because the black hole metric enforces 
\beq
\vec{B}_H=\hat{n}\times\vec{E}_H
\eeq
at the membrane.  Figures \ref{fig:horizonfields} and \ref{fig:horizonfields_mad} show that our GRMHD simulations pass this basic consistency test: the magnetic field runs east-west around the membrane and $|B_H|=|E_H|$.

The final degree of freedom is $\sigma_H=E_n$.  The BZ model assumes force-free fields, $\vec{E}\cdot\vec{B}=0$, which implies 
\beq
\sigma_H=0,
\eeq 
because $E_H^{\hat{\phi}}=B_H^{\hat{\theta}}=0$.  Our GRMHD simulations enforce $\vec{E}\cdot\vec{B}=0$, so they do not give an independent test of this assumption.   Figures \ref{fig:horizonfields} and \ref{fig:horizonfields_mad} show $\sigma_H\approx 0$ in our simulations, as expected.

In summary, the GRMHD simulations' membrane fields, $B_n$, $\sigma_H$, $\vec{E}_H$, and $\vec{B}_H$, are correctly described by the BZ model.  The BZ solution with $\Omega_F/\Omega_H=1/2$ differs from the simulations' membrane fields by as much as $30\%$.  The BZ solution with $\Omega_F/\Omega_H\approx 0.35$ gives an excellent fit to the simulations.

\section{The Load Region}
\label{sec:load}

We have compared the BZ predictions for jet power and torques, dissipation, and electromagnetic fields on the membrane with our GRMHD simulations.  A crucial assumption of the standard BZ model is $\Omega_F/\Omega_H=1/2$.  In this section we compute $\Omega_F/\Omega_H$ from our simulations and relate it to the physics of gas acceleration in the load region.

In the absence of gas, many field geometries are possible, each with their own angular velocity, $\Omega_F/\Omega_H$.  For example, a slowly rotating split monopole has $\Omega_F/\Omega_H=1/2$, while a slowly rotating paraboloidal field has $\Omega_F/\Omega_H\approx 0.4$ \citep{bz77}.  \citet{2013arXiv1303.1644B} have found solutions with $\Omega_F/\Omega_H\approx 0.2$.  

In simulations with gas (such as ours),  $\Omega_F/\Omega_H$ is not a free parameter, it is determined self-consistently by the dynamics of the MHD flow.    
%The region in the jet where magnetic energy is converted into bulk gas motion is called the load region.  Simple models for the load region suggest (to within factors of order unity) $\Omega_F/\Omega_H\approx 1/2$ \citep{1979AIPC...56..399L,1982MNRAS.198..345M,phinney1983}.  In this section, we connect these load region models to our GRMHD simulations. 
A current, $I$, flowing along the black hole membrane, draws energy from the hole's rotation.  The black hole acts as a battery with EMF, $V$, given by \citep{1978MNRAS.185..833Z,1979agn..book..241B,1982MNRAS.198..345M,phinney1983,1986bhmp.book.....T}
\beq
%\Delta V = \Omega_H B_n (r_H^2 +a^2)\sin\theta \Delta \theta,
IV = \Omega_H\frac{dJ}{dt}.
\eeq
Less than half of this energy is available for powering the jet.  The remainder is dissipated by the membrane's internal resistance, $R_H$.  In physical units, $R_H=377 \text{ ohms}$, while in the dimensionless units of this paper, $R_H=1$. 
So the membrane's internal resistance creates a potential drop, $V_H$, given by
\beq
IV_H = T_H \frac{dS_H}{dt}. 
\eeq
As the current circulates through the magnetosphere, the magnetic energy extracted at the membrane is converted into bulk gas motion in the load region.  The potential drop through the load region, $V_L$, is related to $V$ and $V_H$ by energy conservation:
\beq
V=V_H + V_L.
\eeq
The effective resistance of the load region is
\beq
R_L = V_L/I.
\eeq
In the BZ model, the effective resistance of the load region and $\Omega_F/\Omega_H$   are related by \citep{1986bhmp.book.....T}:
\beq\label{eq:rlomega}
\Omega_F/\Omega_H=\frac{R_L/R_H}{1+R_L/R_H}.
\eeq
If the load and black hole achieve perfect impedance matching, $R_L/R_H= 1$, then equation \eqref{eq:rlomega} gives $\Omega_F/\Omega_H=1/2$ .

\begin{figure} 
\includegraphics[width=\columnwidth]{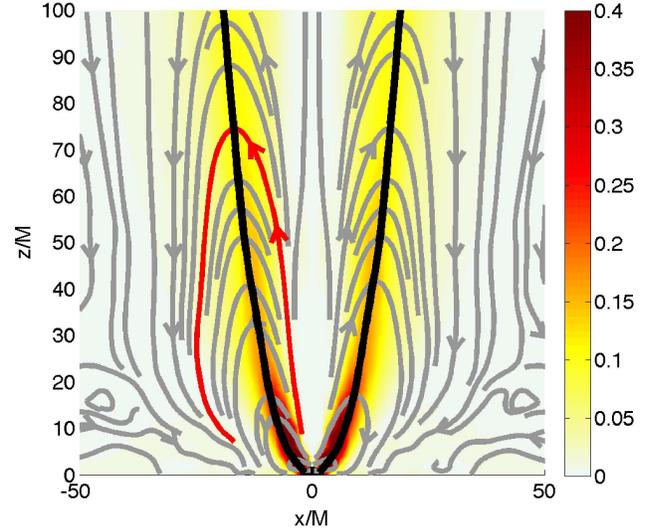}
\caption{\label{fig:bulkJ} ZAMO frame currents (silver streamlines) and the effective resistance, $dV_L/dl$, (yellow-red) of the current carrying ``wires.'' The jet is indicated with heavy black lines.}
\end{figure}

This analysis can be applied to our simulations.  Figure \ref{fig:bulkJ} shows the structure of the magnetosphere currents of the $a_*=0.7$ MAD simulation.  Currents flow out from the black hole into the jet, loop around the boundary of the jet, and return to the black hole through the accretion disk's corona.   The boundary of the jet is marked in Figure \ref{fig:bulkJ} with heavy black lines.  It is defined by following streamlines of energy flux, $-T^\mu_t$, outward from $\mathcal{A}_{\rm jet}$ (the region of the horizon where $-T_{nt}>0$).

%----------------Load currents

The voltage drop along a current streamline, $\mathcal{L}$, through the load region, is 
\beq\label{eq:vl}
V_L=\int_{\mathcal{L}} \vec{E}\cdot d\vec{l}.
\eeq
In the load region, the electric field is predominantly along $\theta$.  So $R_L=V_L/I$ peaks at turning points of the current, where $\vec{j}$ is along $\theta$.  Figure \ref{fig:bulkJ} shows $dV/dl\sim R_L$ is concentrated at turning points of the current, as expected.  The integral curves of $\vec{j}$ through the magnetosphere behave like wires attached to resistors, $R_L$, at their turning points.

Numerically evaluating the line integral \eqref{eq:vl} for the loop $\mathcal{L}$ highlighted in red in Figure \ref{fig:bulkJ}, gives $V_L\approx 0.9$.  Connecting the footpoints of this current loop across the membrane, we find $V_H\approx 2$. So
\beq
\frac{R_L}{R_H} = \frac{V_L}{V_H} \approx 0.45.
\eeq
Plugging $R_L/R_H=0.45$ into equation \eqref{eq:rlomega} gives $\Omega_F/\Omega_H=0.31$.  The load resistance does not appear to depend strongly on the choice of current loop or black hole spin, but we save a detailed investigation for a future paper.

\begin{figure} 
\centering
\includegraphics[width=0.8\columnwidth]{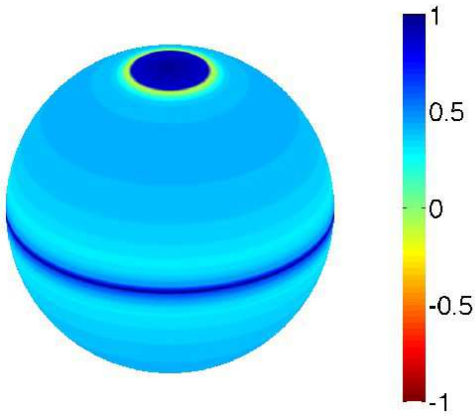}
\caption{\label{fig:omegaF2d} Angular velocity of magnetic fields lines, $\Omega_F/\Omega_H$, at the membrane of our $a_*=0.7$ MAD simulation.} 
\end{figure}

We can test this analysis by computing the field line angular velocity directly:
\beq
\Omega_F = \frac{\alpha}{\varpi}\left(v^{\hat{\phi}} - v^{n}\frac{B^{\hat{\phi}}}{B^{n}}\right),
\eeq
where $\vec{v}$ is the gas velocity in the ZAMO frame.  
Figure  \ref{fig:omegaF2d} shows that the simulated $\Omega_F/\Omega_H$ is roughly constant over the horizon (except near the poles, where numerical floor activations make the simulations unreliable).

% ----------- Omega_F vs. spin; Horizon Omega_F --------------

\begin{figure} 
\includegraphics[width=\columnwidth]{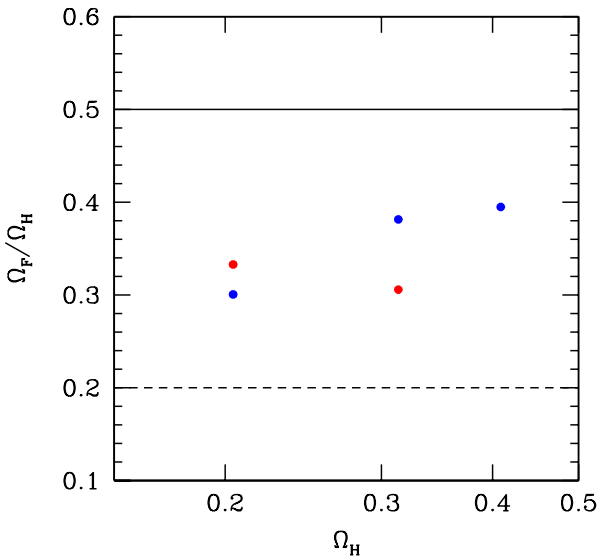}
\caption{\label{fig:omegaF} \ ($t,\theta,\phi$)-averaged field line angular velocity, $\Omega_F/\Omega_H$, as a function of $\Omega_H$ for SANE (blue) and MAD (red) simulations. } 
\end{figure}

For all five of our simulations with spinning black holes, the ($t$,$\theta$,$\phi$)-averaged field line velocity is $\Omega_F/\Omega_H\approx 0.35$  (see Figure \ref{fig:omegaF}), close to the value estimated from $R_L/R_H$.  This supports the idea that the angular velocity of magnetic field lines is tied to the effective resistance of the load region, $R_L/R_H$.  Furthermore, the simulated load region and black hole achieve near perfect impedance matching, as anticipated \citep{1979AIPC...56..399L,phinney1983}.  

More work is needed to understand how the simulated black hole and load achieve $\Omega_F/\Omega_H \approx 0.3$.
\citet{1982MNRAS.198..345M} have give an intuitive argument for nearly perfect impendance matching.  The minimum velocity of particles sliding along magnetic field lines is \citep{1982MNRAS.198..345M}
\beq
v_{\rm min} = \frac{\Omega_F/\Omega_H}{1-\Omega_F/\Omega_H}.
\eeq
  If $\Omega_F/\Omega_H \ll 1/2$, then matter sliding along field lines has little inertia and the field tends to spin up.  If $\Omega_F/\Omega_H \gg 1/2$, then matter is flung off of field lines and the resulting backreaction tends to spin down the field.  This suggests the field is driven towards $\Omega_F/\Omega_H \approx 1/2$ in equilibrium.   Perhaps this argument, or a variant thereof, underlies the simulated physics.  We save a detailed study of this question for the future.

\begin{figure} 
\includegraphics[width=\columnwidth]{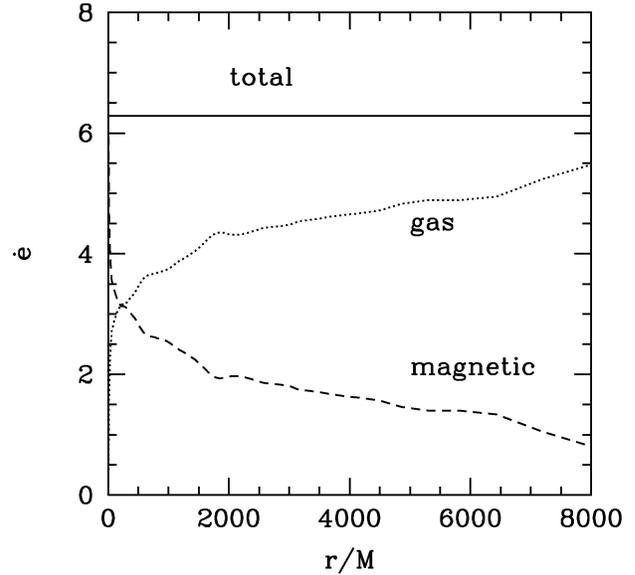}
\caption{\label{fig:load} Gas and magnetic energy fluxes in the jet.}
\end{figure}

Figure \ref{fig:load} shows the gas and magnetic energy fluxes in the
jet for the $a_*=0.7$ MAD simulation. The conversion of magnetic
energy into gas energy is a gradual process and the load is a broad
region, beginning a few gravitational radii from the black hole and
continuing to $r\sim 10,000M$. The gas energy flux first exceeds the
magnetic energy flux around $r\sim 300M$, at which point roughly half of the magnetic energy has been converted into gas energy.  By $r\sim 10,000M$ the energy in the jet is carried almost entirely by the gas.  

The fact that the load region is concentrated far from the black hole, where frame dragging is unimportant, may go some way towards explaining why $\Omega_F/\Omega_H$ is roughly independent of black hole spin (see Figure \ref{fig:omegaF}).

% ----------- The load region--------------

\section{Summary and Conclusions}
\label{sec:conc}

We have discussed GRMHD simulations of jets from accreting ($h/r\sim 0.3$), spinning ($0\leq a_* \leq 0.98$) black holes, and their relationship with the BZ model in its membrane formulation. We showed that the simulated magnetic jet power, $P^{\rm mag}$, integrated over the entire horizon, agrees with the standard BZ prediction \eqref{eq:BZ2} to within $10\%$. This is consistent with earlier simulations \citep{2010ApJ...711...50T,2012JPhCS.372a2040T}.  We traced the order $10\%$ discrepancies between the BZ model and our simulated $P^{\rm mag}$ to the standard BZ assumptions $\Omega_F/\Omega_H=1/2$ and uniform $B_n$.

The jet power proxy $P^{\rm mag}$ is integrated over the entire horizon, so it also includes power extracted into the accretion disk.  We have separately considered the power extracted into the jet alone, $P^{\rm tot}_{\rm jet}$, and found this to be roughly $50\%$ lower than $P^{\rm mag}$.  This quantity is probably more relevant for jet observations.   Simply lowering the standard BZ prediction \eqref{eq:BZ2} by $60\%$ gives a good fit to the simulated $P^{\rm tot}_{\rm jet}$.  

We then turned to the physics underlying jet power.  We showed that the torques, dissipation, and electromagnetic fields at the horizon are correctly described by the BZ model in its membrane formulation.  This extends earlier GRMHD tests of the BZ model in its Boyer-Lindquist formulation \citep{2004ApJ...611..977M,2010ApJ...711...50T,2012JPhCS.372a2040T,2012MNRAS.423.3083M}.  We showed that the BZ model with $\Omega_F/\Omega_H=1/2$ correctly describes our simulations to within factors of order unity.  The BZ solution with 
$\Omega_F/\Omega_H=0.35$ gives the best fit to our simulations.

Finally, we computed the effective resistance of the load region for the $a_*=0.7$ MAD simulation.  We found $R_L/R_H\approx 0.45$, for which the BZ model implies $\Omega_F/\Omega_H\approx 0.31$.  This is close to the actual field line angular velocity of this simulation: $\Omega_F/\Omega_H\approx 0.33$.  This supports the idea that $\Omega_F/\Omega_H$ is connected to the physics of gas acceleration in the load region.  Near perfect impedance matching between black hole and load was anticipated long ago by \citet{1979AIPC...56..399L} and \citet{phinney1983}.   The load region extends to $r\sim 10,000M$, at which point the energy in the jet is almost entirely in the gas (rather than the magnetic fields).  The fact that the load region is far from the black hole, where frame dragging is unimportant, may be connected to the fact that $\Omega_F/\Omega_H$ is found to be roughly independent of spin.

\section*{Acknowledgments}

We thank Jon McKinney and Sasha Tchekhovskoy for discussions.  R.F.P was supported in part by a Pappalardo Fellowship in Physics at MIT.  R.N. and A.S. were supported in part by NASA grant NNX11AE16G.

\appendix

\section{Blandford-Znajek jet power}
\label{sec:BZpjet}

In this Appendix, we derive the BZ jet power equation \eqref{eq:BZ2}.  
The magnetic flux, $d\Phi$, through a circular ribbon on the horizon between $\theta$ and $\theta+d\theta$ is \citep{bz77,1986bhmp.book.....T}
\beq\label{eq:dflux} 
d\Phi = 2\pi B_n (r_H^2 +a^2)\sin\theta d\theta.
\eeq
The jet power from the ribbon is \citep{bz77,1986bhmp.book.....T} 
\beq\label{eq:dpjet}
dP_{\rm jet} =2\pi \lambda \Omega_H^2 B_n^2 (r_H^2+a^2)^3\frac{\sin^3\theta}{r_H^2+a^2 \cos^2\theta} d\theta,
\eeq 
where
\beq
\lambda = \Omega_F/\Omega_H (1-\Omega_F/\Omega_H)
\eeq
is the efficiency factor.  

Now we assume $\Omega_F/\Omega_H=1/2$ and uniform $B_n$.
Integrating \eqref{eq:dflux} gives the total flux threading the jet:
\beq
\Phi =  2\pi B_n (r_H^2 +a^2).
\eeq
Plugging this into \eqref{eq:dpjet} and integrating gives the total jet power:
\beq\label{eq:mypjetphi}
P = \frac{1}{8\pi}\Omega_H^2\Phi^2
(1+4\Omega_H^2)
\int_0^{\pi}\frac{\sin^3\theta}{1+4\Omega_H^2\cos^2\theta}d\theta.
\eeq
For all spins, equation \eqref{eq:mypjetphi} is approximately
\beq\label{eq:mypjet2}
P^{\rm BZ} \approx \frac{1}{6\pi} \Omega_H^2 \Phi^2. 
\eeq
This is the BZ model's prediction for jet power.   We compare it with the jet power of GRMHD simulations in \S\ref{sec:pjetvsspin}.  

%For completeness, we give a closed form expression for the integral appearing in equation \eqref{eq:mypjetphi}:
%\begin{align}
%&\int_0^{\pi}\frac{\sin^3\theta}{1+4\Omega_H^2\cos^2\theta}d\theta = \notag\\
%&\frac{
%\left(1+4 \Omega_H ^2\right) \cot ^{-1}(2 \Omega_H )
%-\left(1+4 \Omega_H ^2\right) \tan ^{-1}\left(\frac{1}{4 \Omega_H }-\Omega_H \right)
%-2 \Omega_H }{4\Omega_H ^3}
%\end{align}

\vspace{-2ex}\section{3D Visualizations of GRMHD Membranes}

On the following pages we give 3D visualizations of our GRMHD simulation data.  We show the simulated torques, dissipation, and electromagnetic fields on the black hole membranes.

\clearpage

\onecolumn

\label{sec:3d}
\FloatBarrier
\vspace{-1in}

%---------- Horizon 1st Law

\begin{figure}
\hspace{1.3in}
{\LARGE $\frac{dM}{dmdA}$}
\hspace{1.4in}
{\LARGE $\Omega_H \frac{dJ}{dmdA}$}
\hspace{1.3in}
{\LARGE $T_H \frac{dS_H}{dmdA}$}
\\
{\LARGE $a_*=0.98$}
\includegraphics[width=0.28\textwidth]{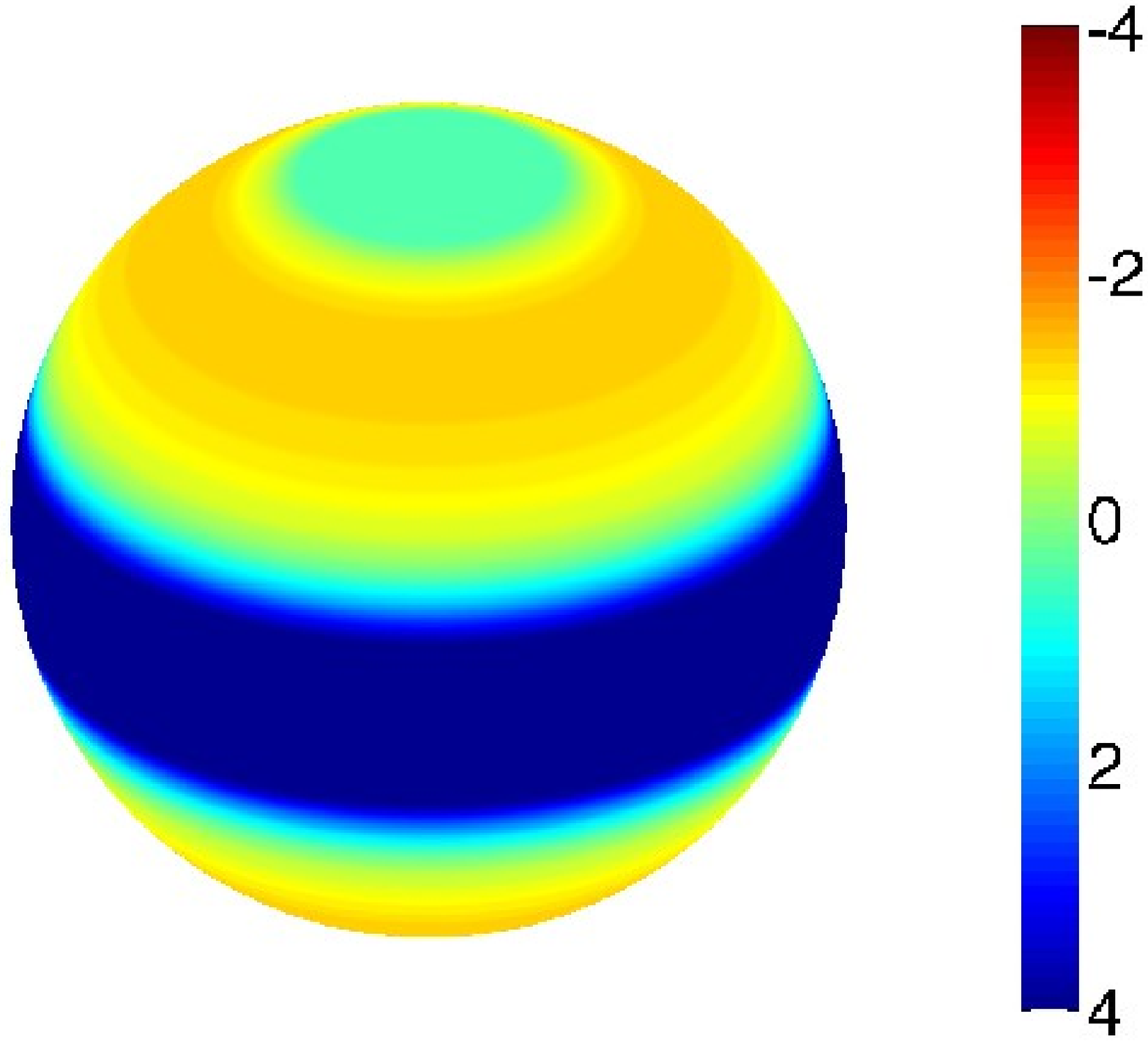} 
\includegraphics[width=0.28\textwidth]{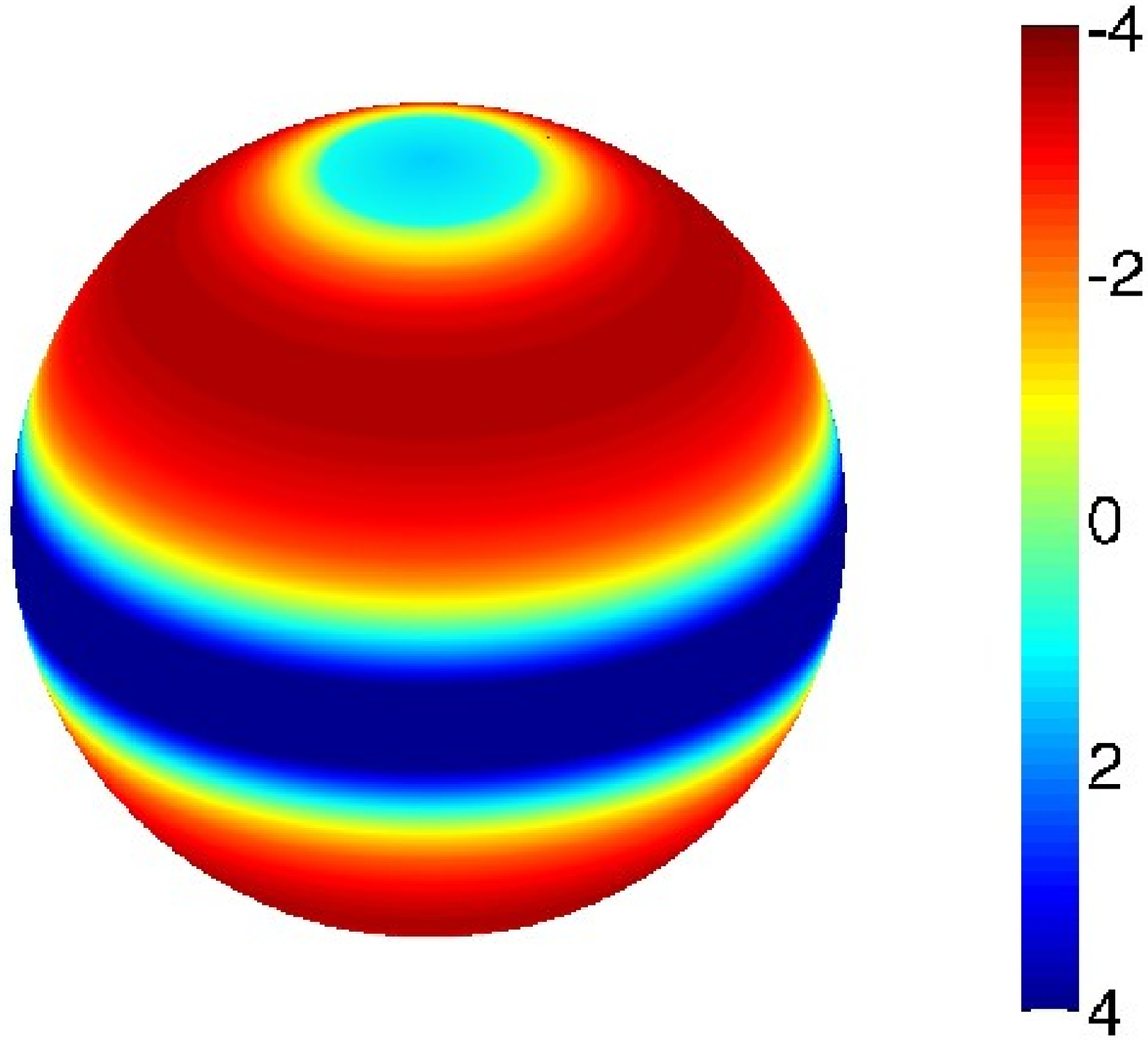} 
\includegraphics[width=0.28\textwidth]{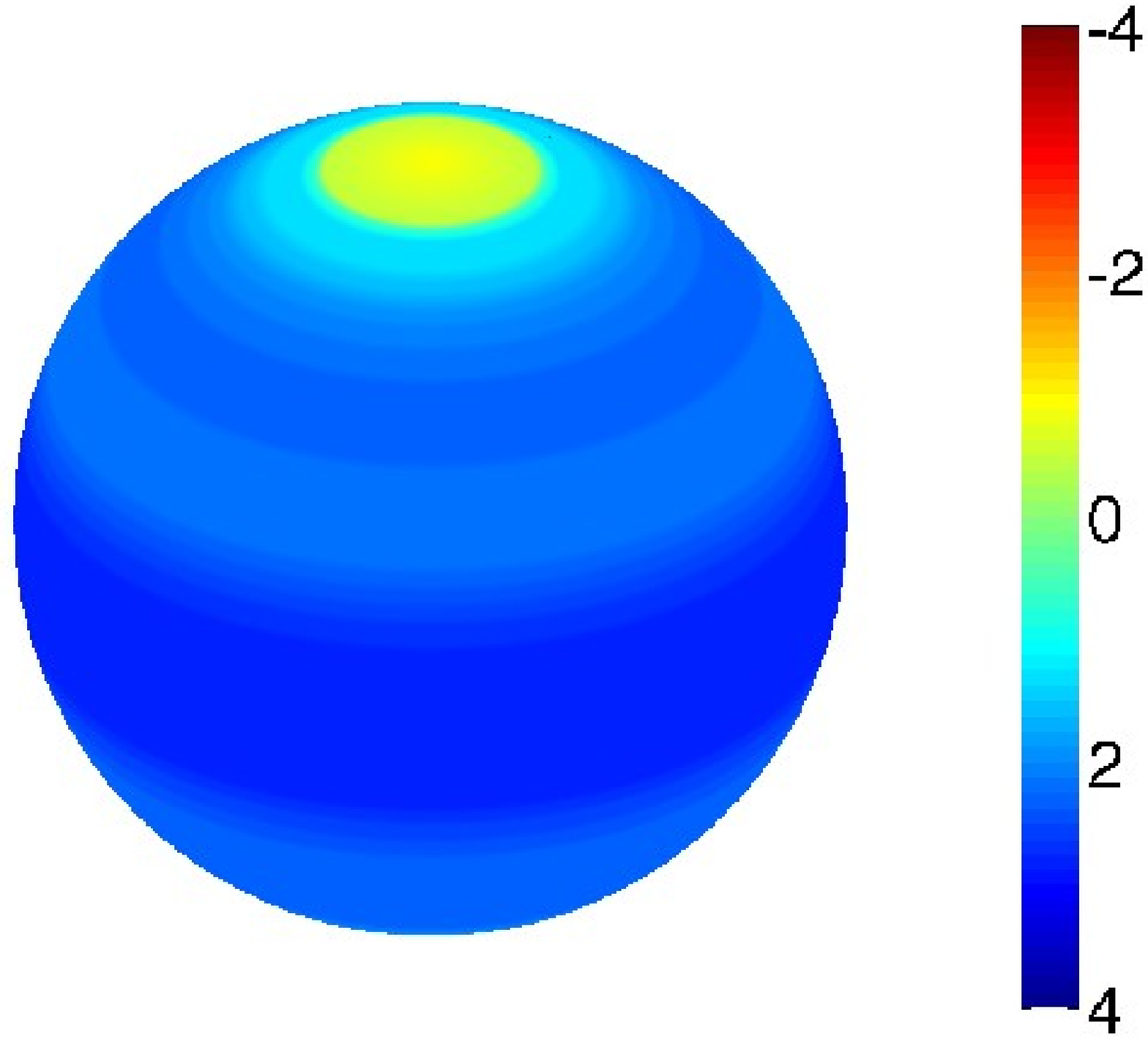}
\\
{\LARGE $a_*=0.90$}
\includegraphics[width=0.28\textwidth]{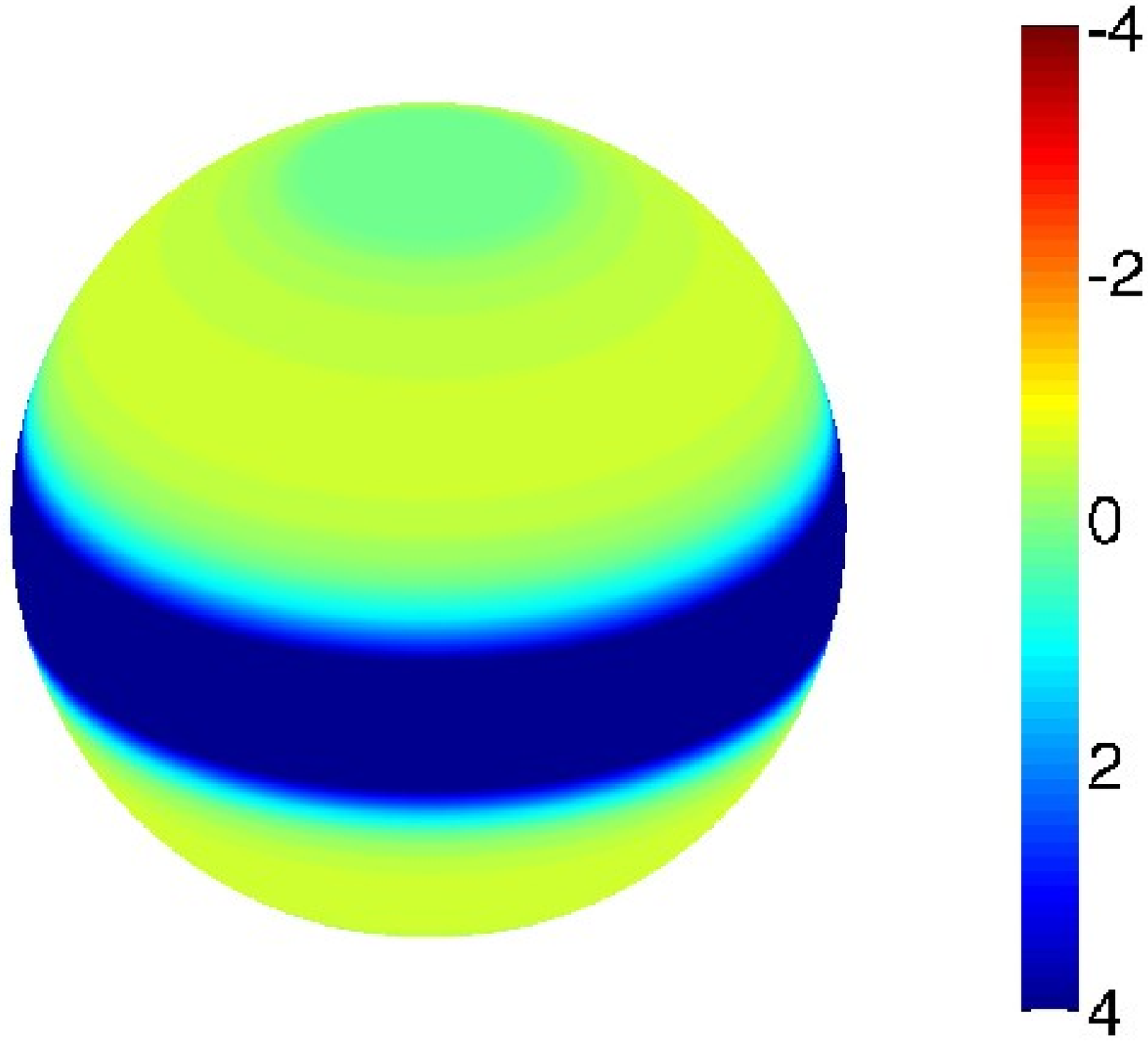}
\includegraphics[width=0.28\textwidth]{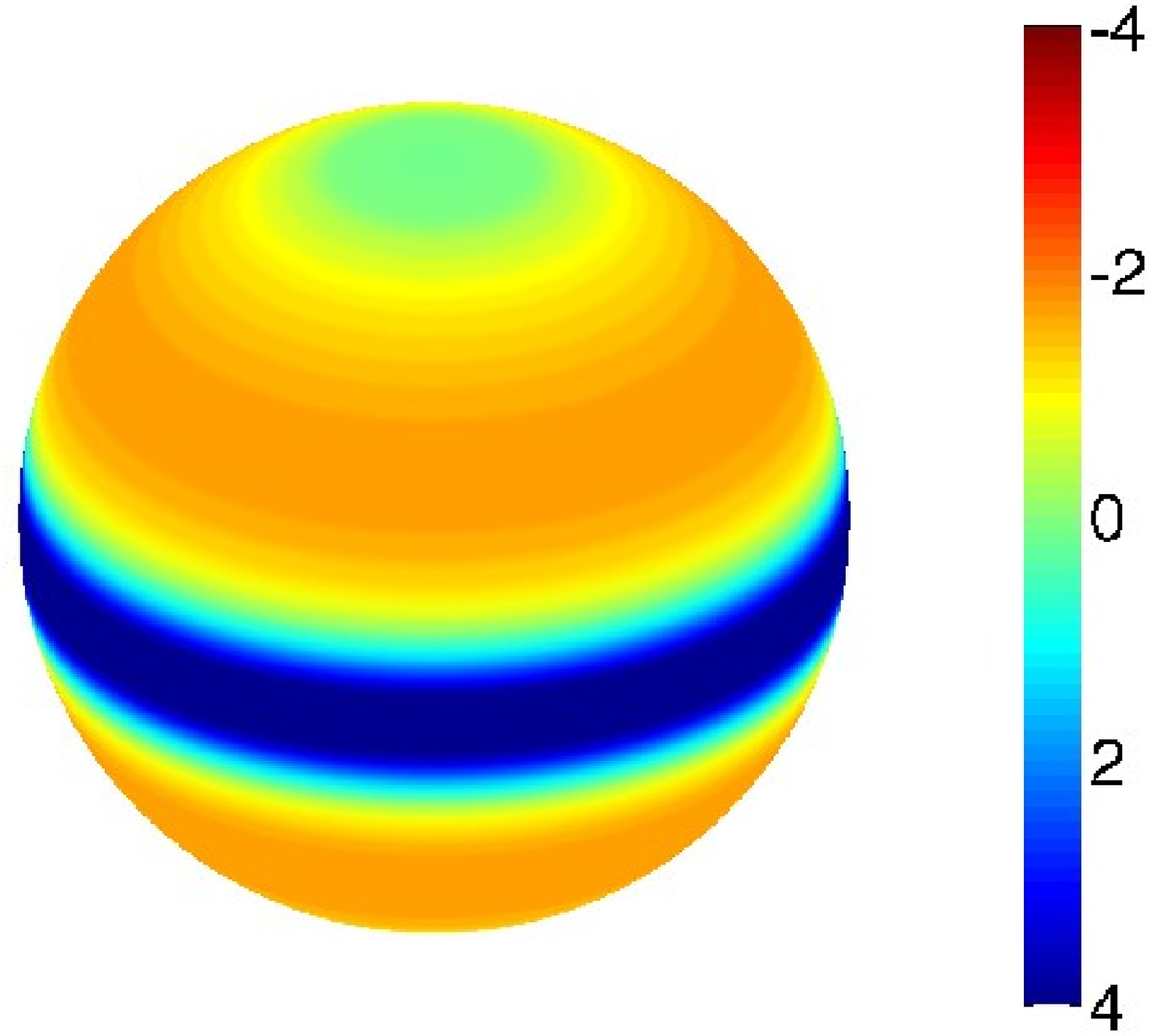}
\includegraphics[width=0.28\textwidth]{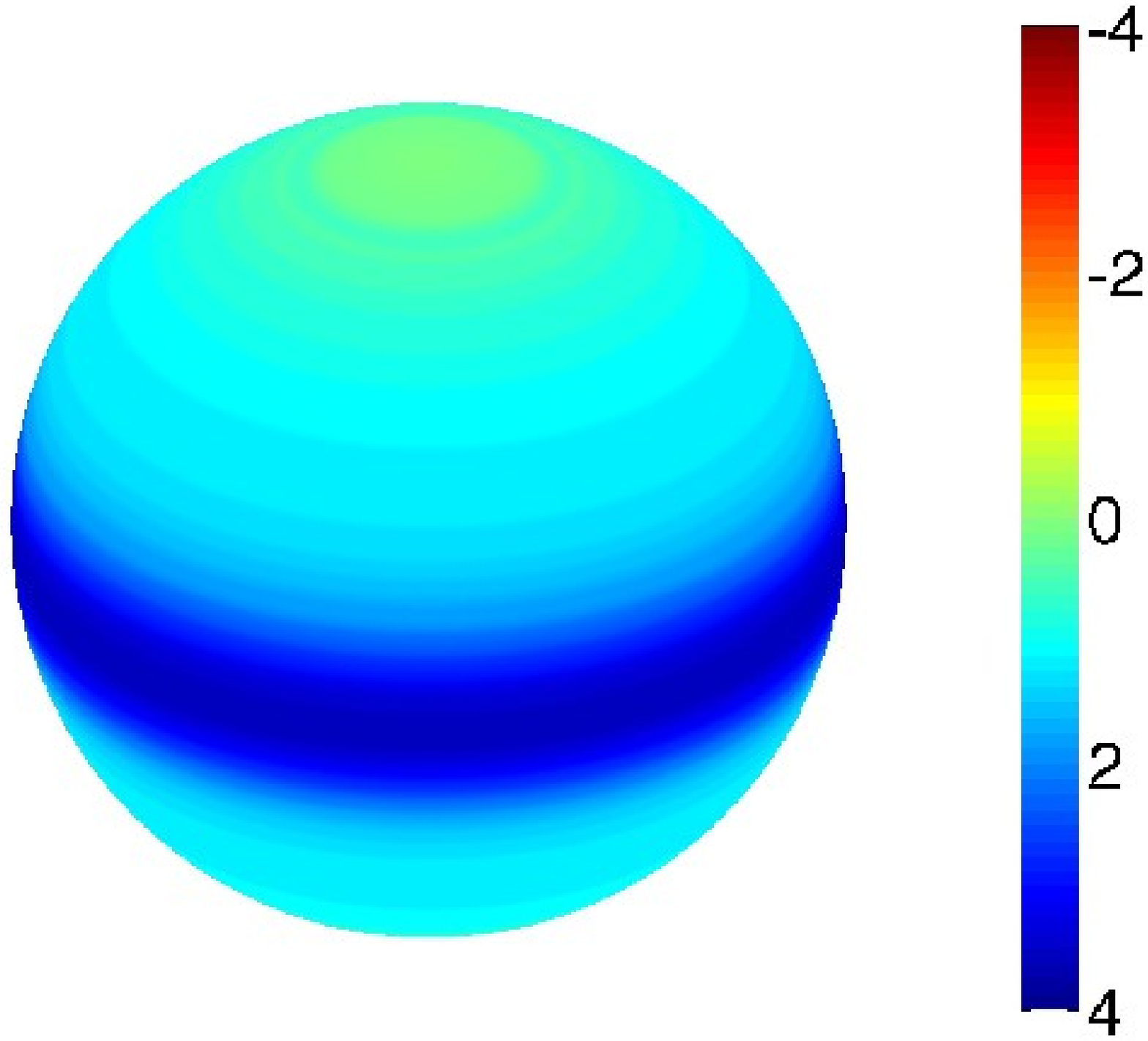}
\\
{\LARGE $a_*=0.70$}
\includegraphics[width=0.28\textwidth]{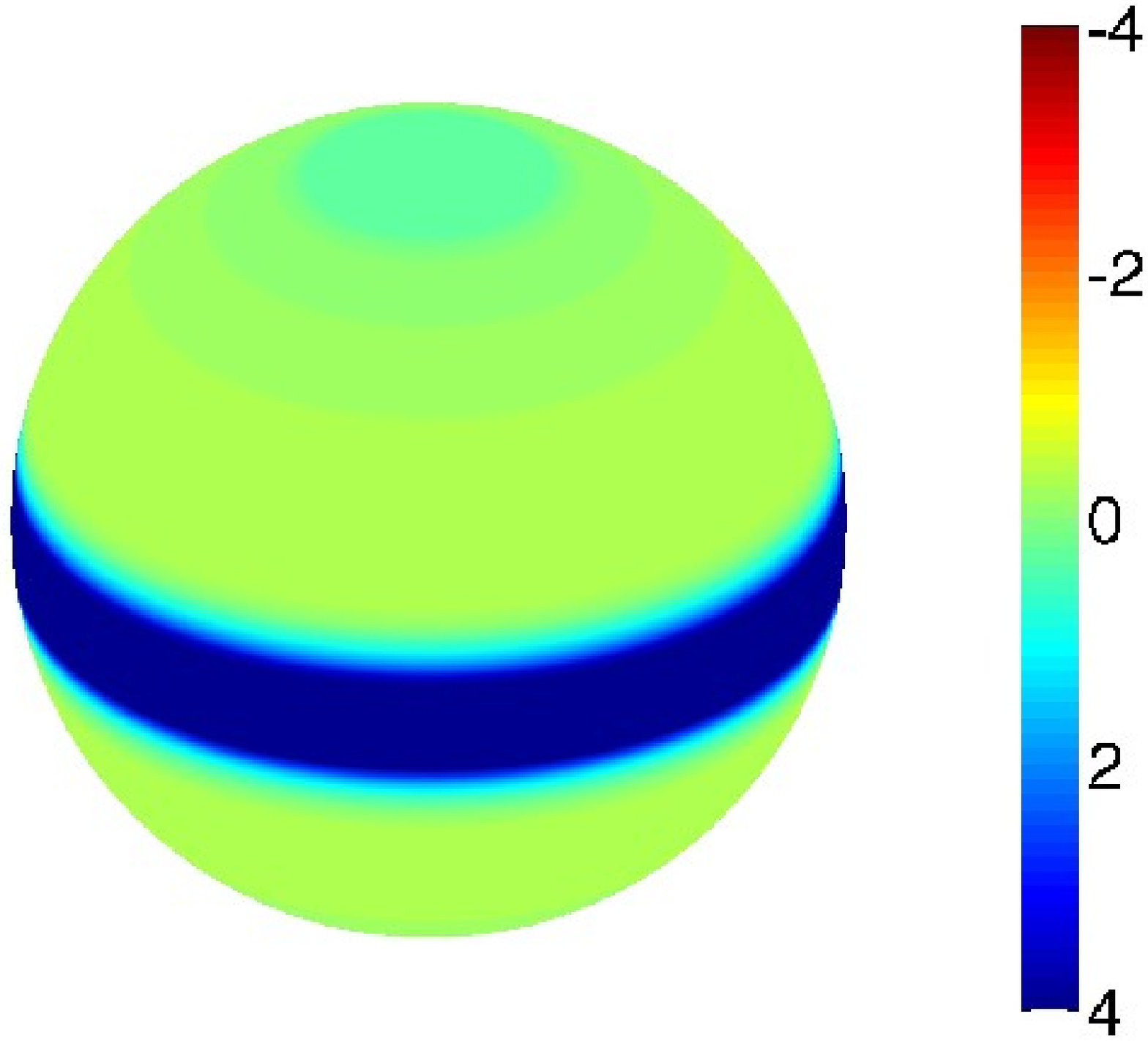}
\includegraphics[width=0.28\textwidth]{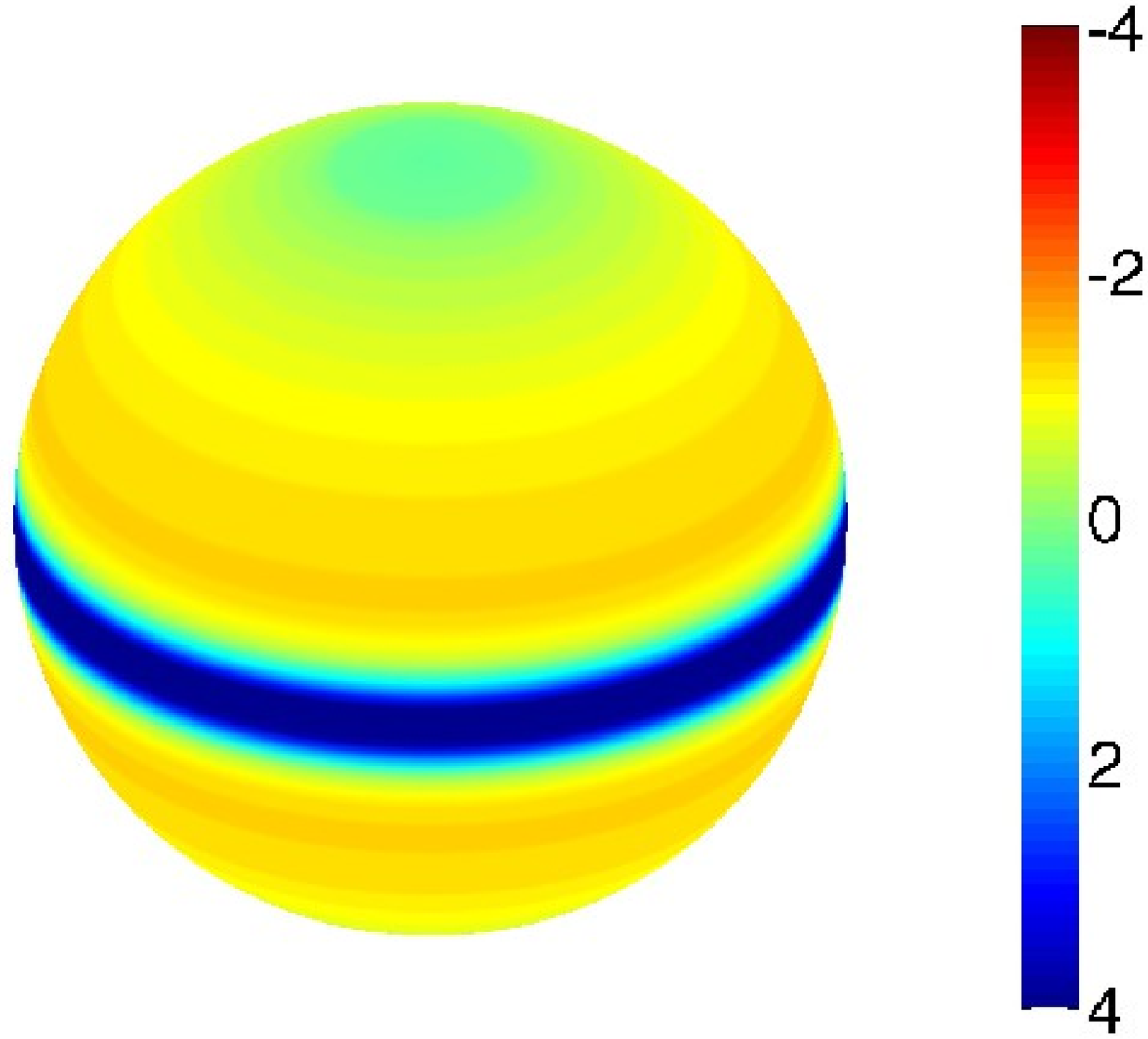}
\includegraphics[width=0.28\textwidth]{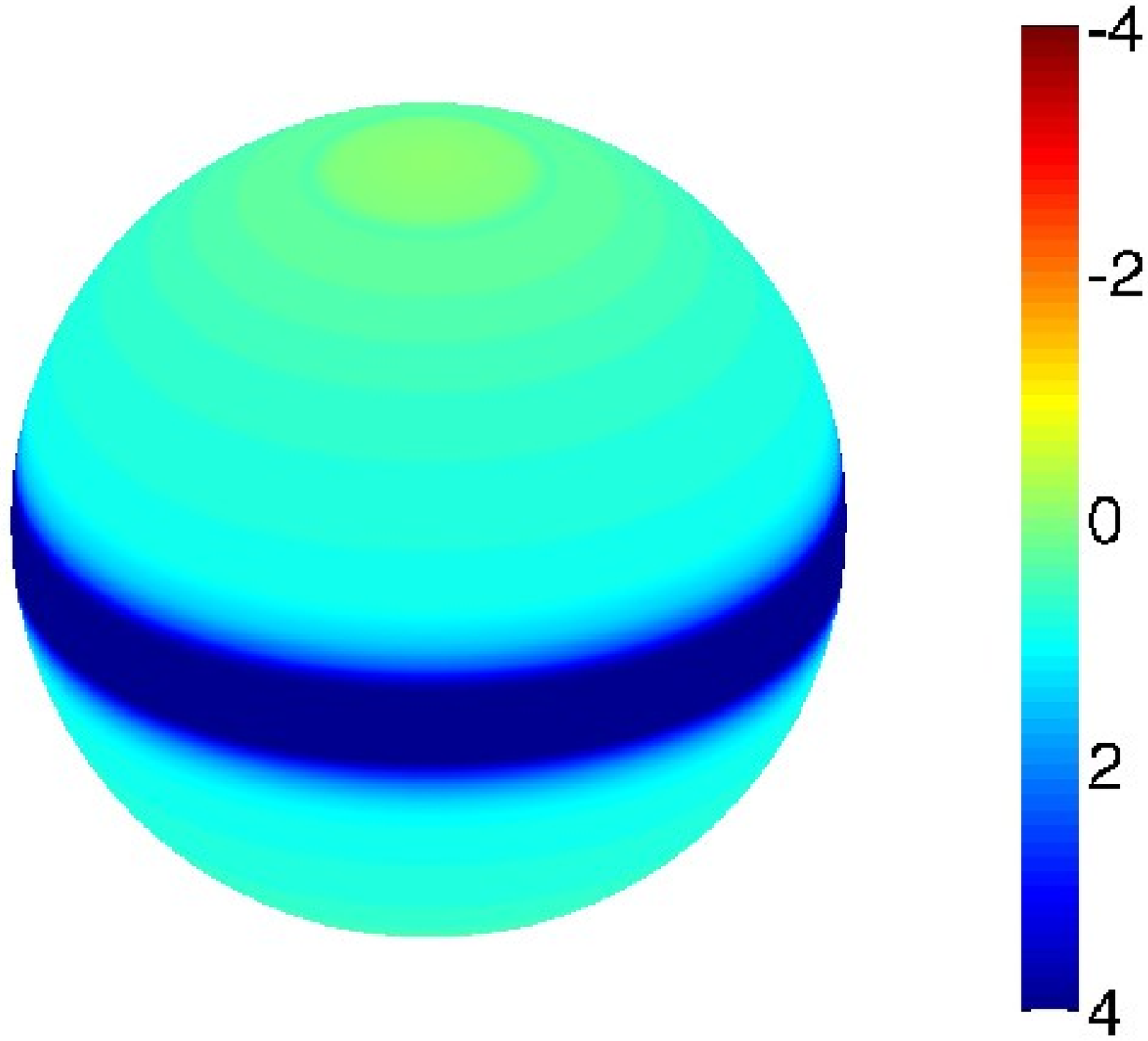}
\\
{\LARGE $a_*=0.00$}
\includegraphics[width=0.28\textwidth]{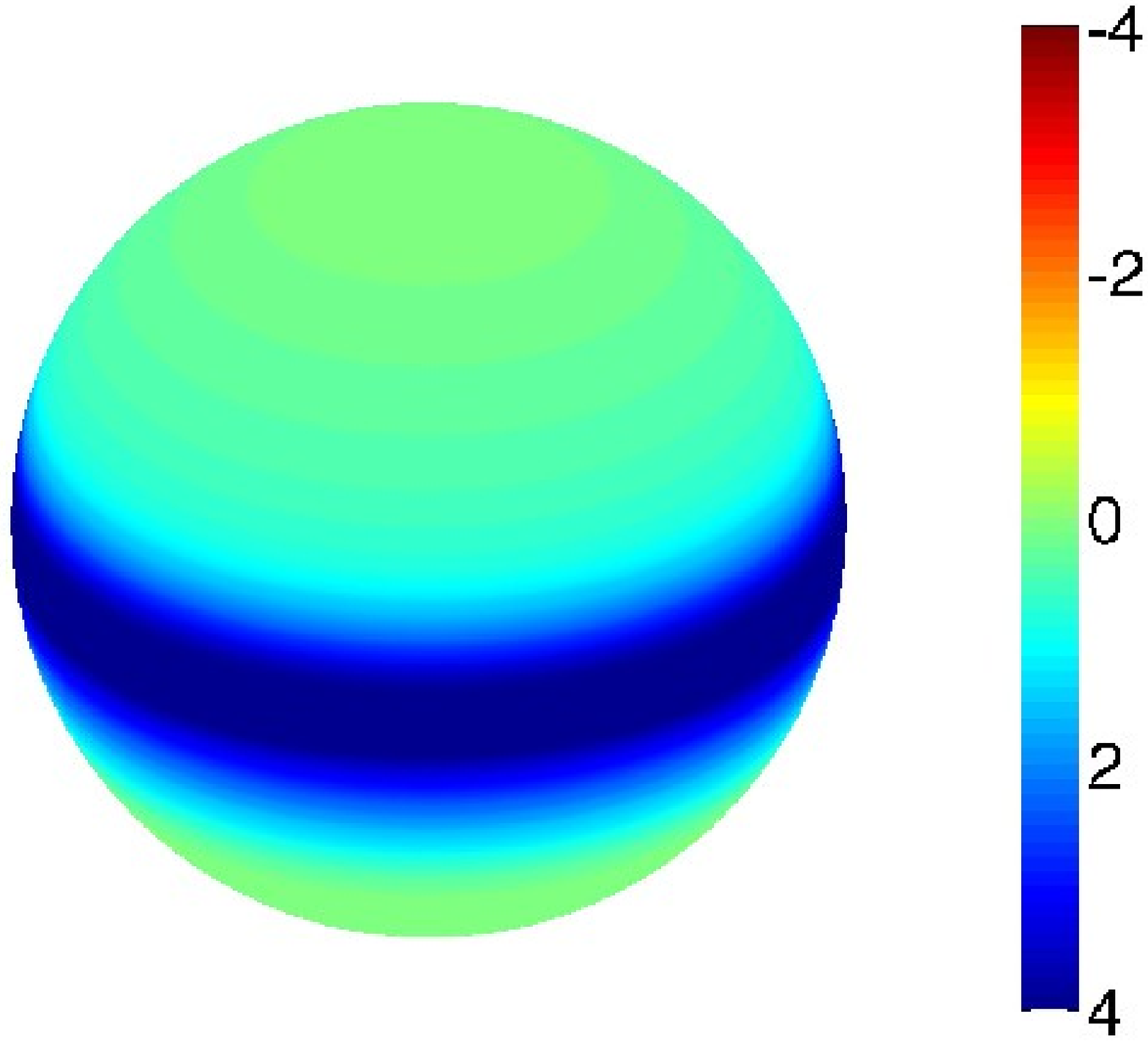}
\includegraphics[width=0.28\textwidth]{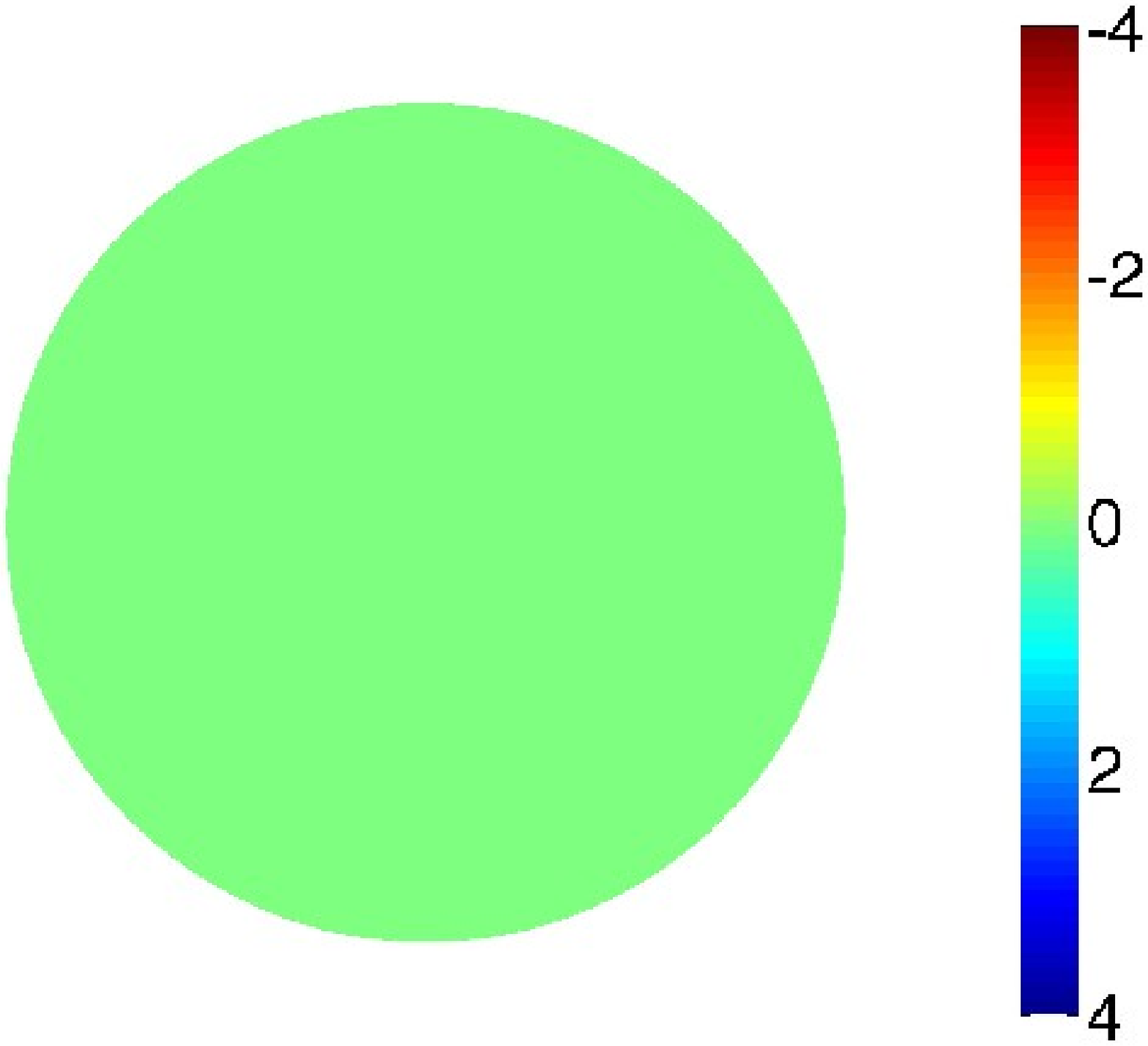}
\includegraphics[width=0.28\textwidth]{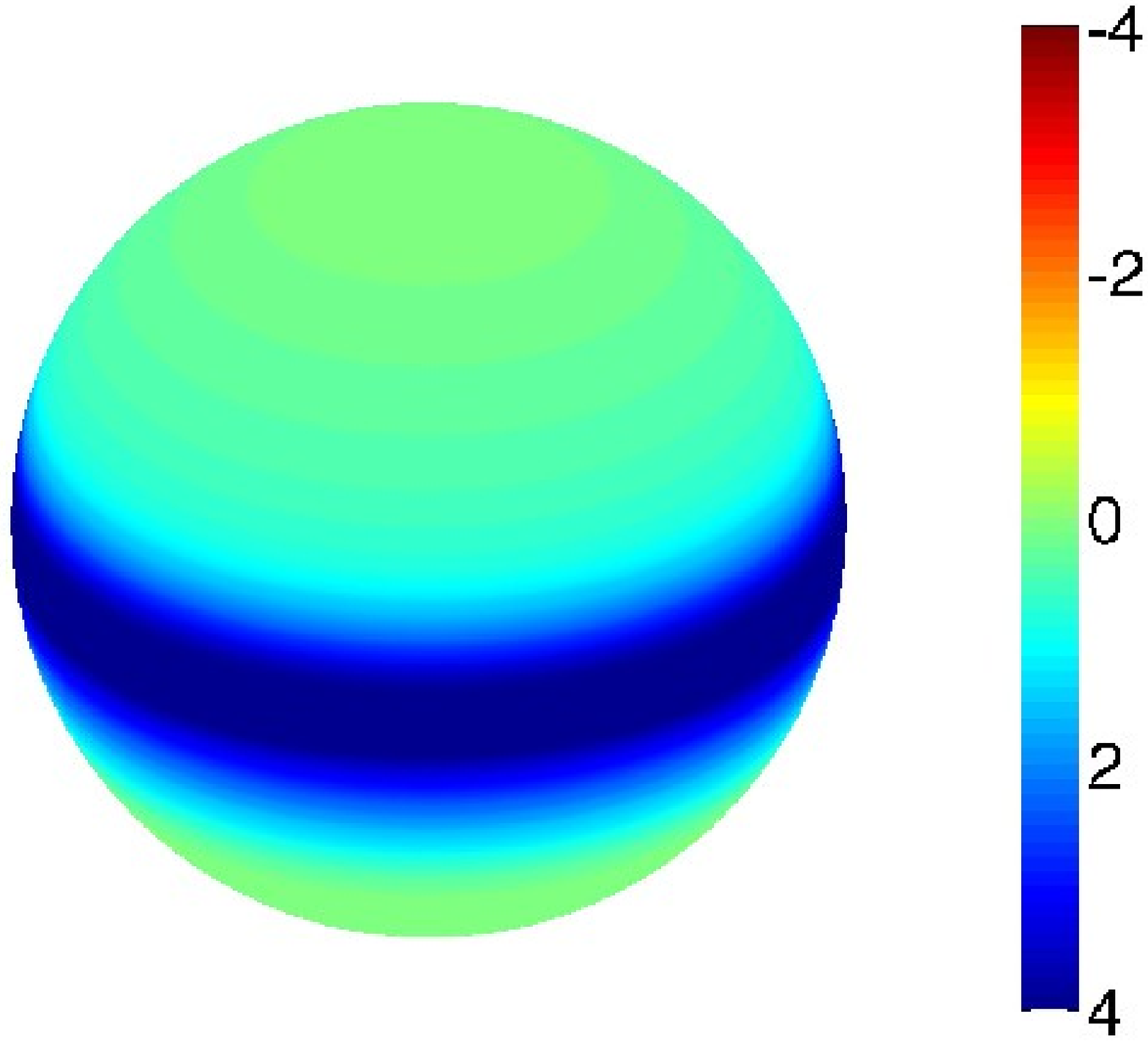}
\caption{\label{fig:horizonedot} Energy flow (first column), torques (second column), and dissipation (third column) on the membrane for our GRMHD simulations with SANE initial conditions.  Rows correspond to different simulations: $a_*=0, 0.7, 0.9, 0.98$ (bottom to top).  In all three columns, the membrane is losing energy at yellow-red regions (jets) and gaining energy at blue regions (accretion disk).}
\end{figure}

\newpage
\begin{figure}
\hspace{1.3in}
{\LARGE $\frac{dM}{dmdA}$}
\hspace{1.4in}
{\LARGE $\Omega_H \frac{dJ}{dmdA}$}
\hspace{1.3in}
{\LARGE $T_H \frac{dS_H}{dmdA}$}
\\
{\LARGE $a_*=0.90$}
\includegraphics[width=0.28\textwidth]{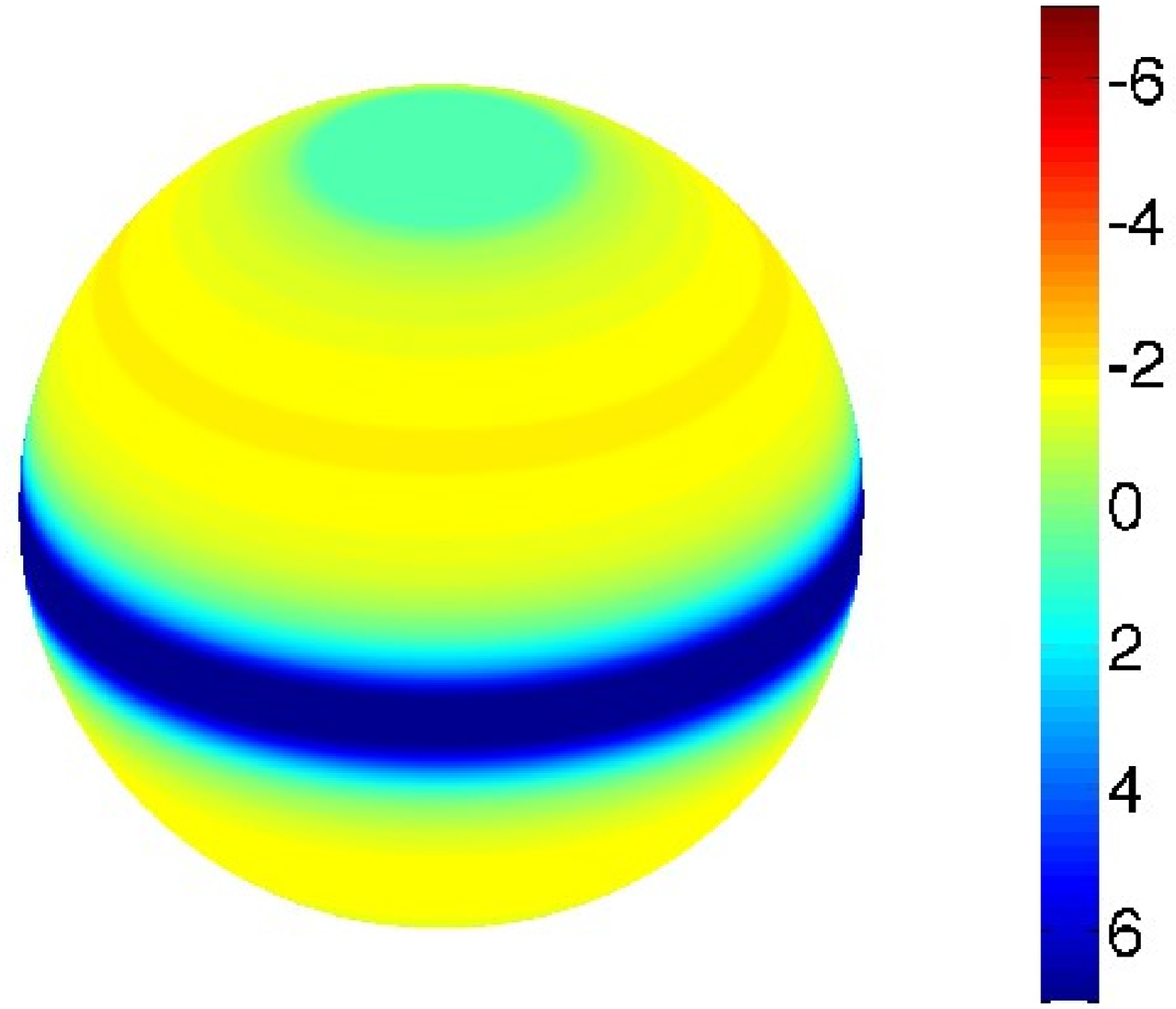}
\includegraphics[width=0.28\textwidth]{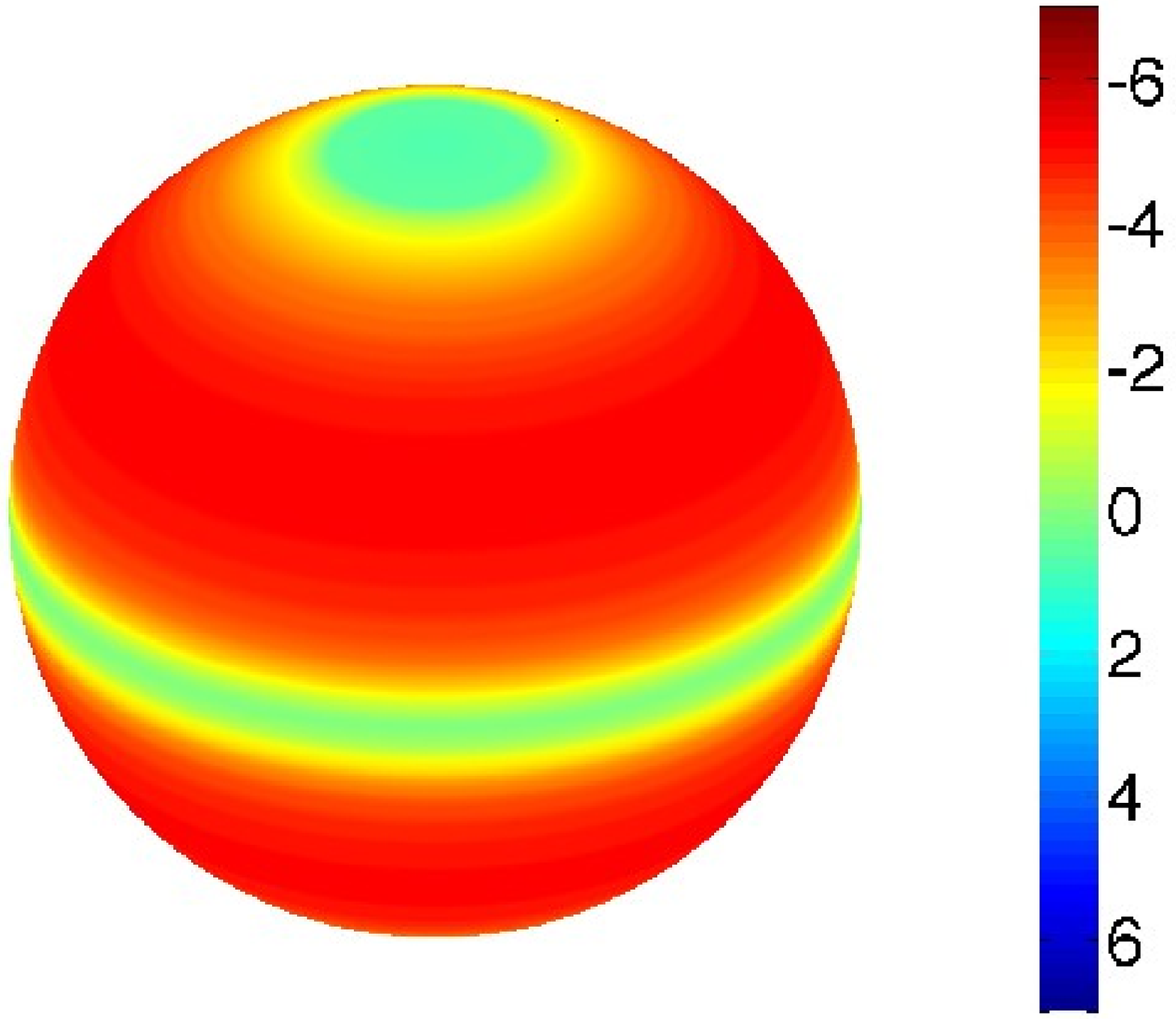}
\includegraphics[width=0.28\textwidth]{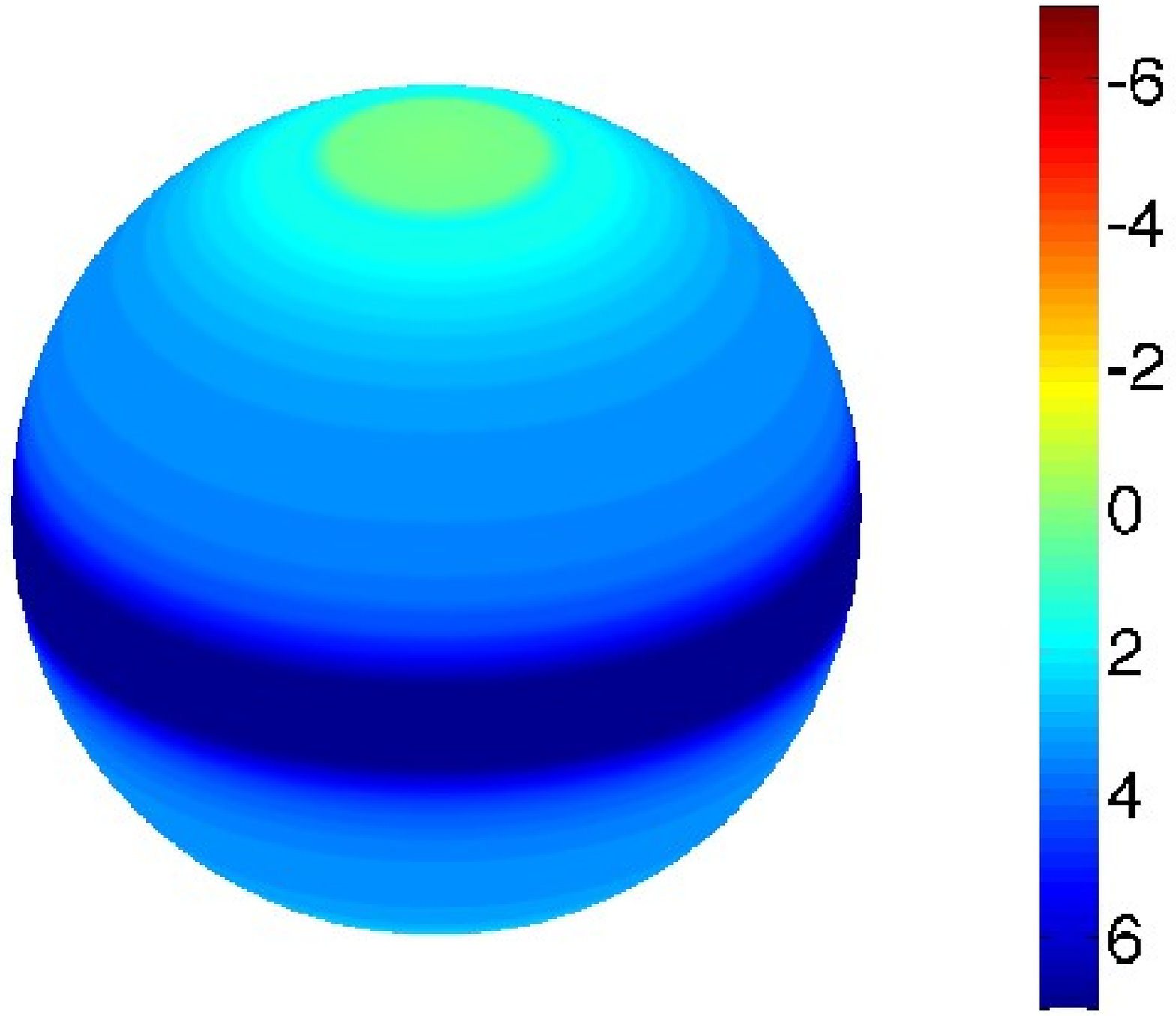}
\\
{\LARGE $a_*=0.70$}
\includegraphics[width=0.28\textwidth]{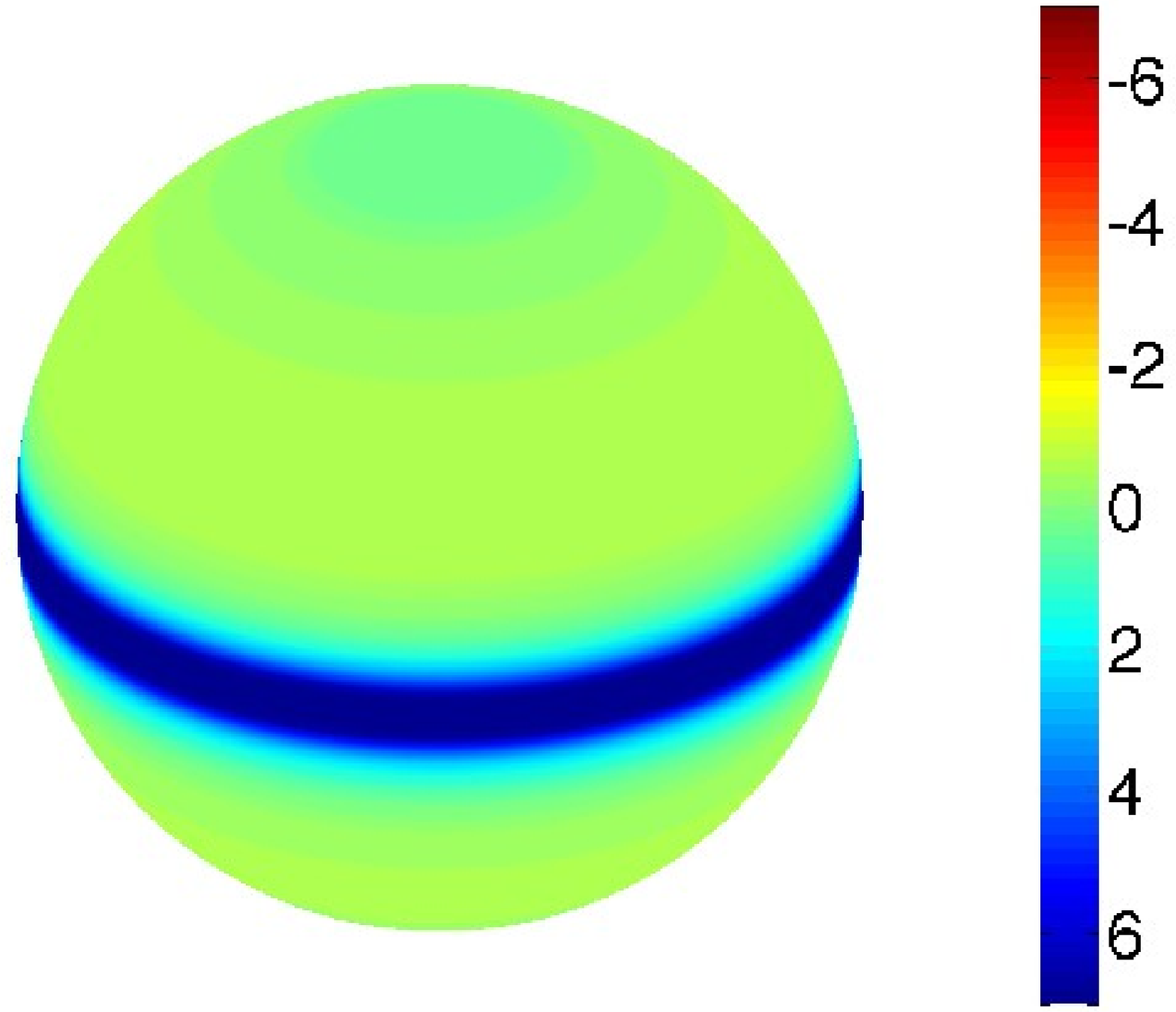}
\includegraphics[width=0.28\textwidth]{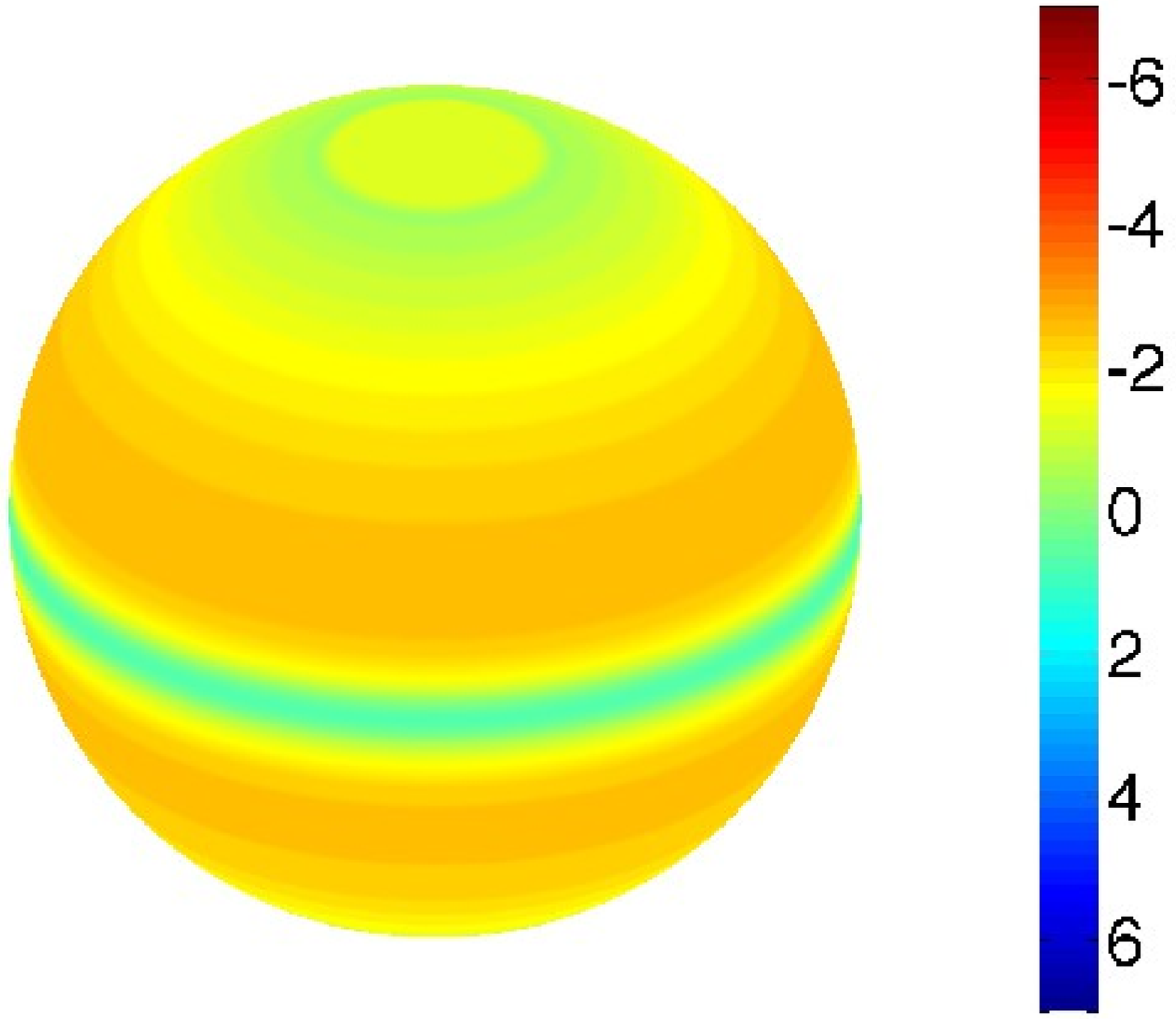}
\includegraphics[width=0.28\textwidth]{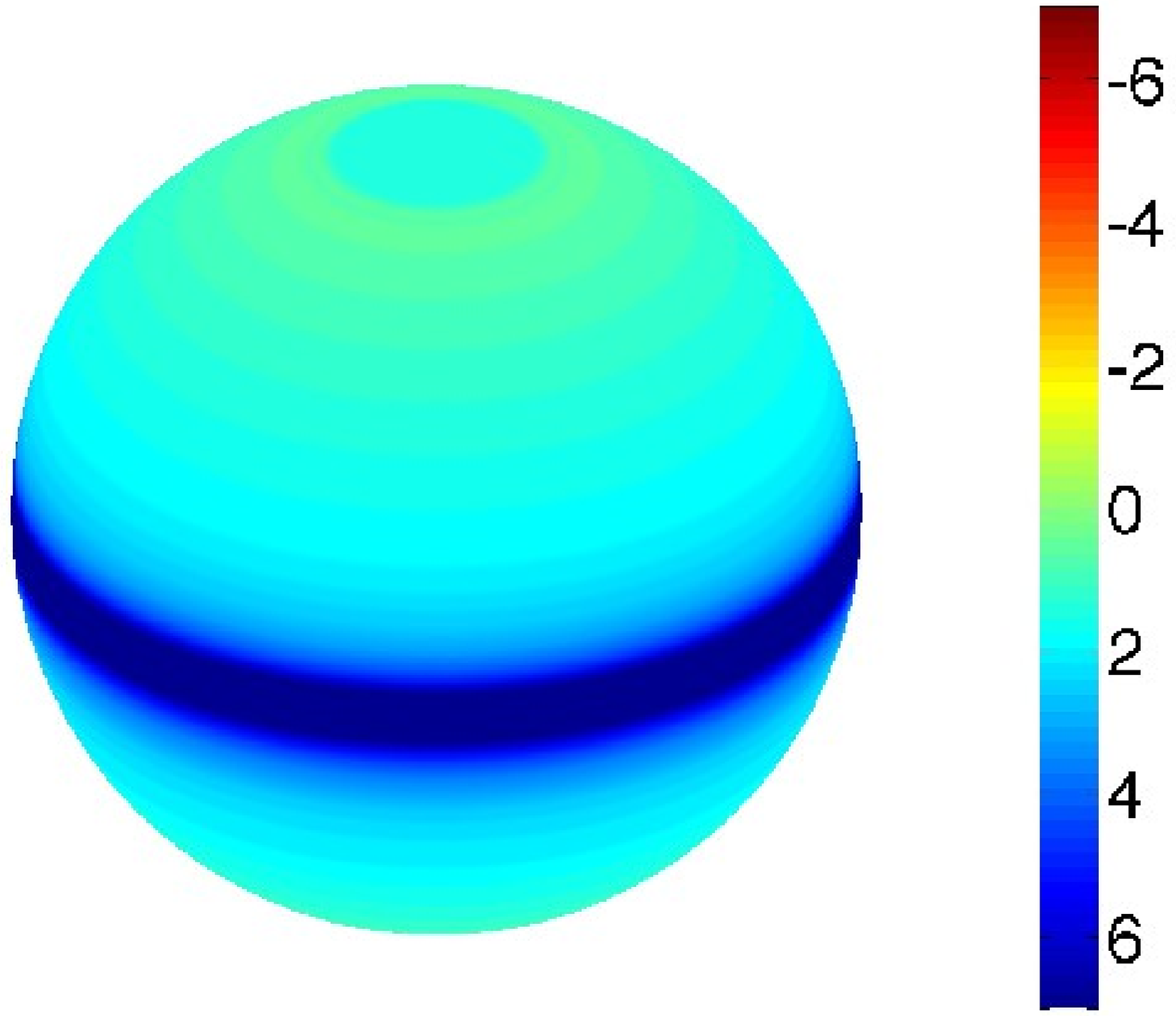}
\\
{\LARGE $a_*=0.00$ }
\includegraphics[width=0.28\textwidth]{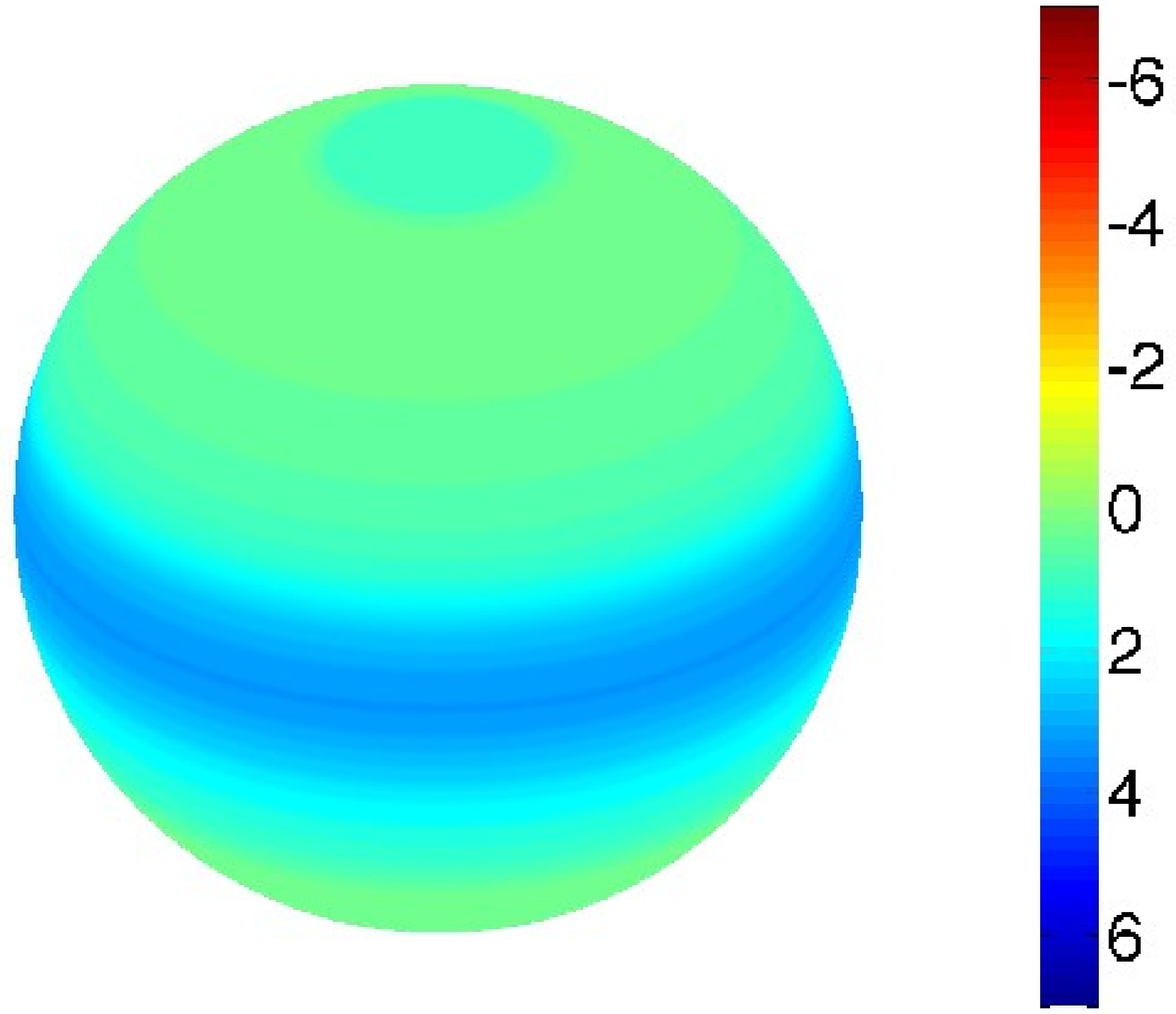}
\includegraphics[width=0.28\textwidth]{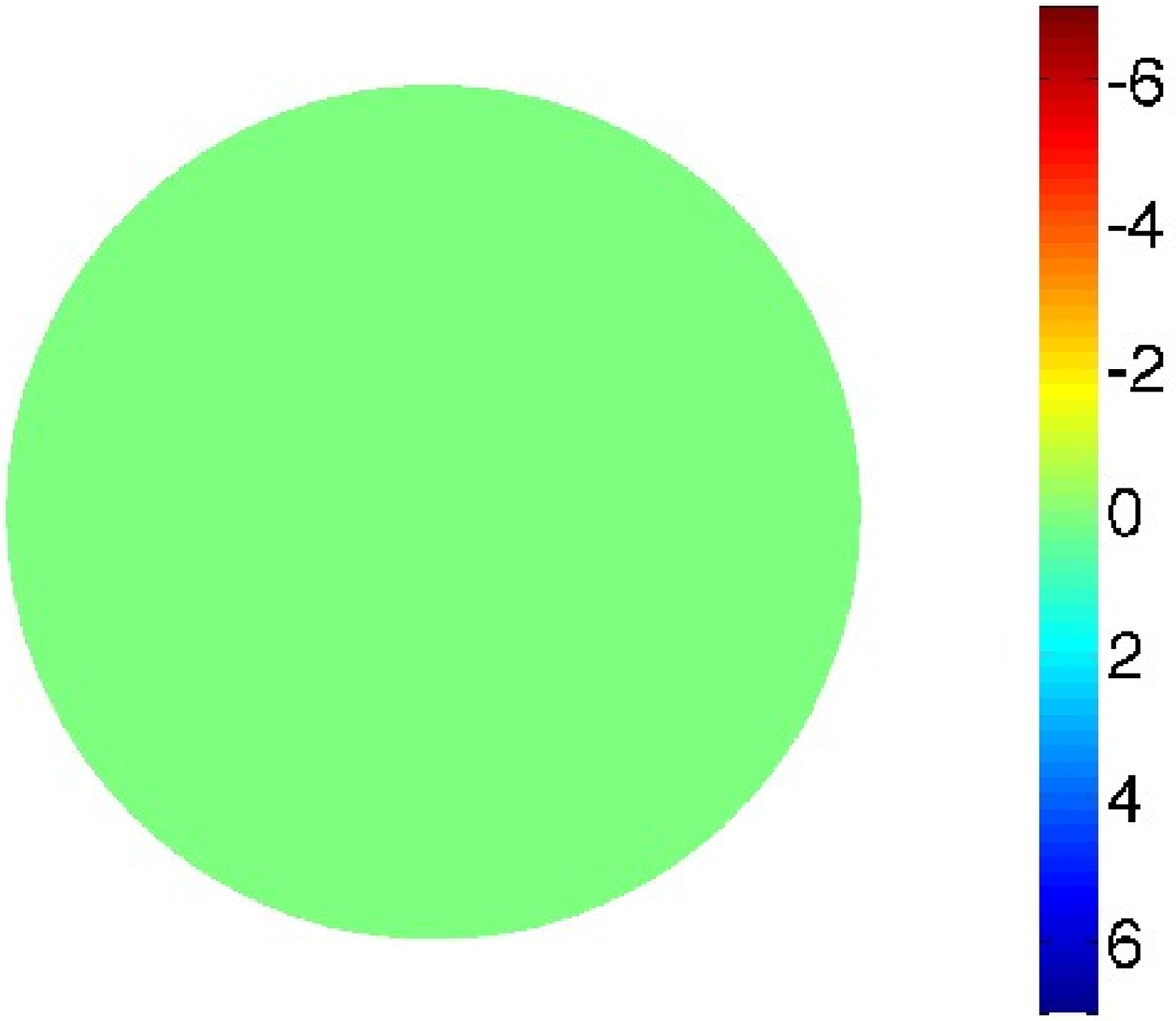}
\includegraphics[width=0.28\textwidth]{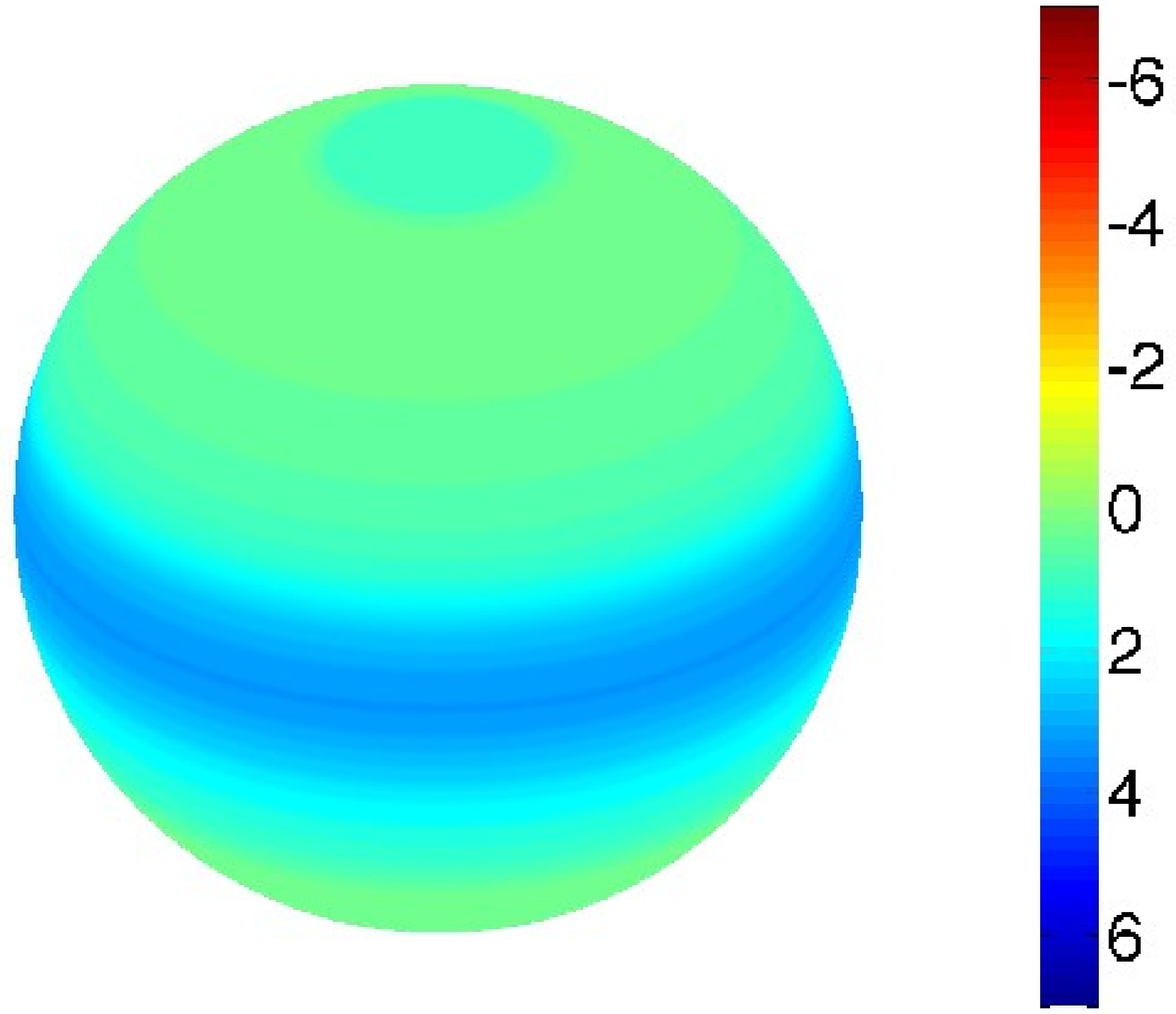}
\caption{\label{fig:horizonedot_mad} Same as Figure \ref{fig:horizonedot} but for MAD runs. }
\end{figure}

%---------- Horizon 1st Law -- Split
  \newpage
\begin{figure}
\hspace{1.3in}
{\LARGE $\frac{dM}{dmdA}$}
\hspace{1.4in}
{\LARGE $\Omega_H \frac{dJ}{dmdA}$}
\hspace{1.3in}
{\LARGE $T_H \frac{dS_H}{dmdA}$}
\\
{\LARGE $a_*=0.98$}
\includegraphics[width=0.28\textwidth]{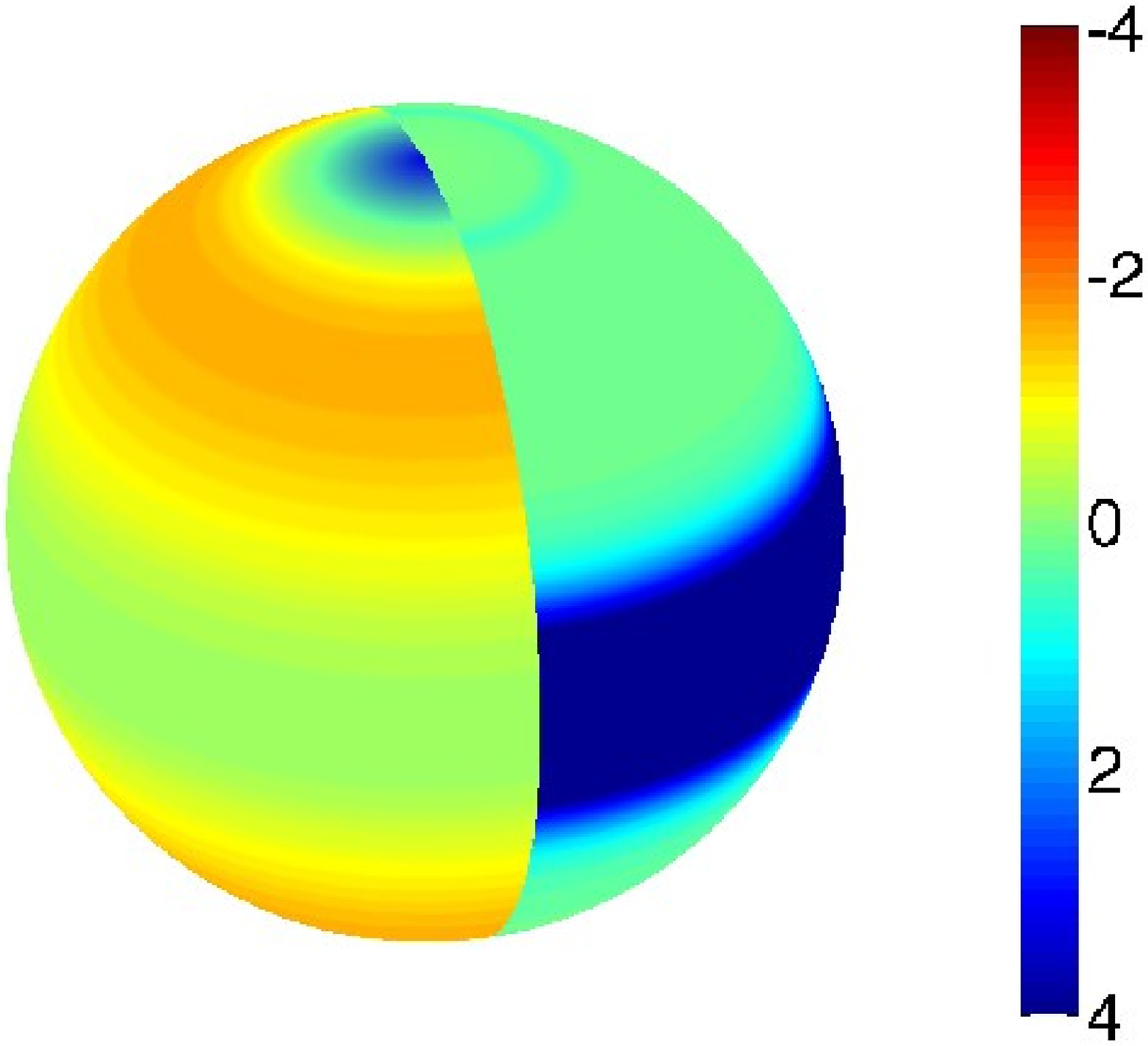}
\includegraphics[width=0.28\textwidth]{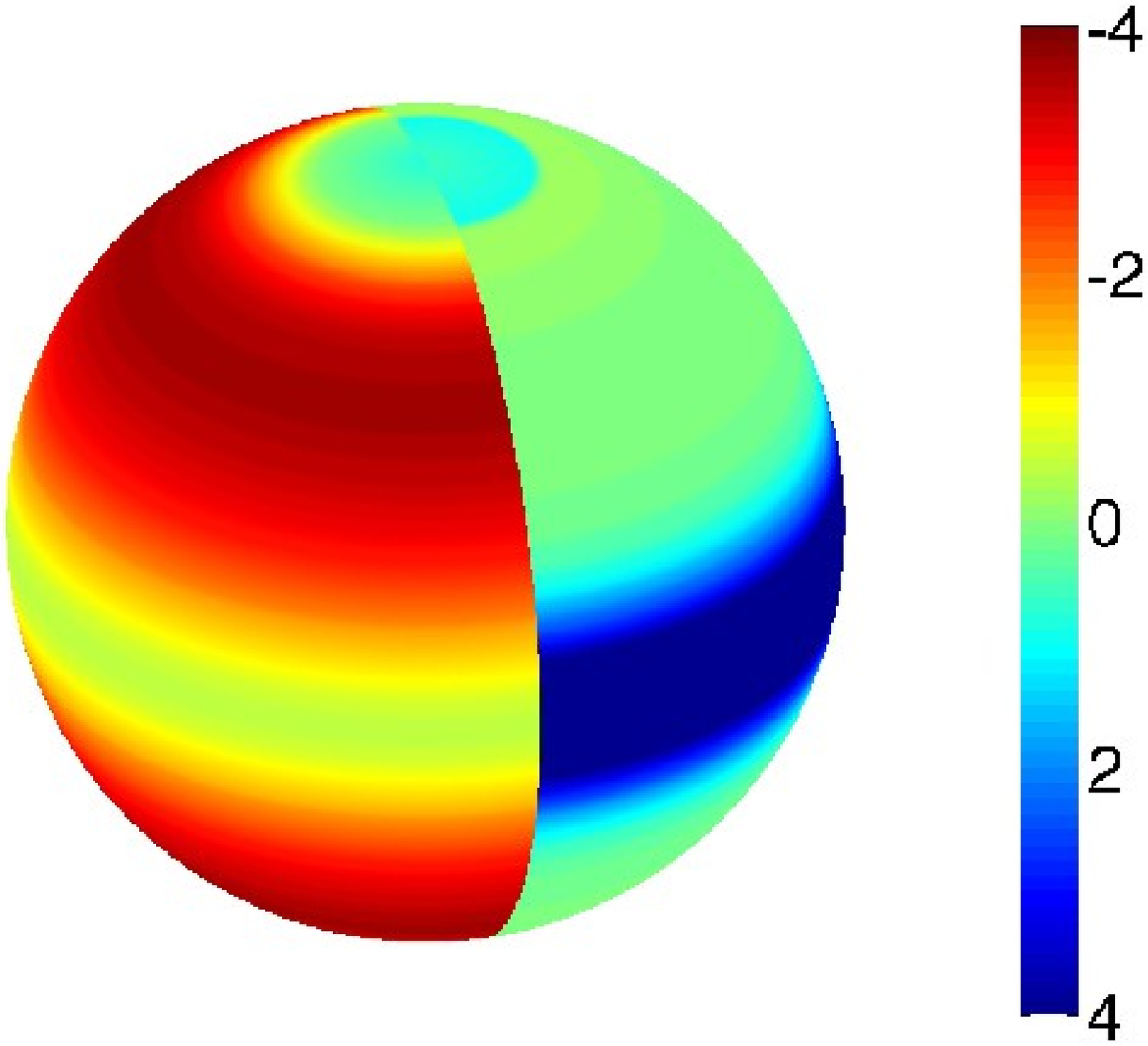}
\includegraphics[width=0.28\textwidth]{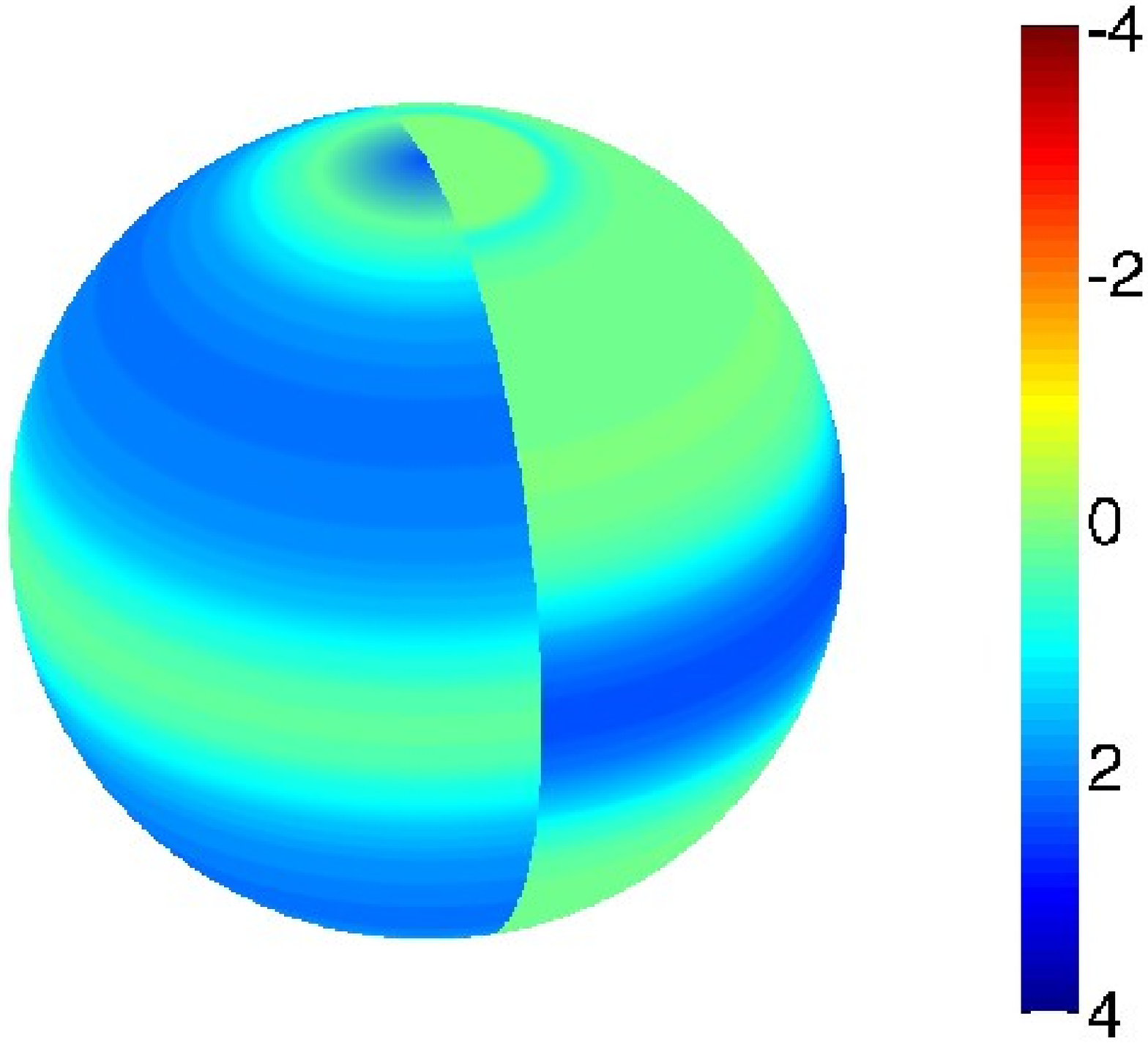}
\\
{\LARGE $a_*=0.90$}
\includegraphics[width=0.28\textwidth]{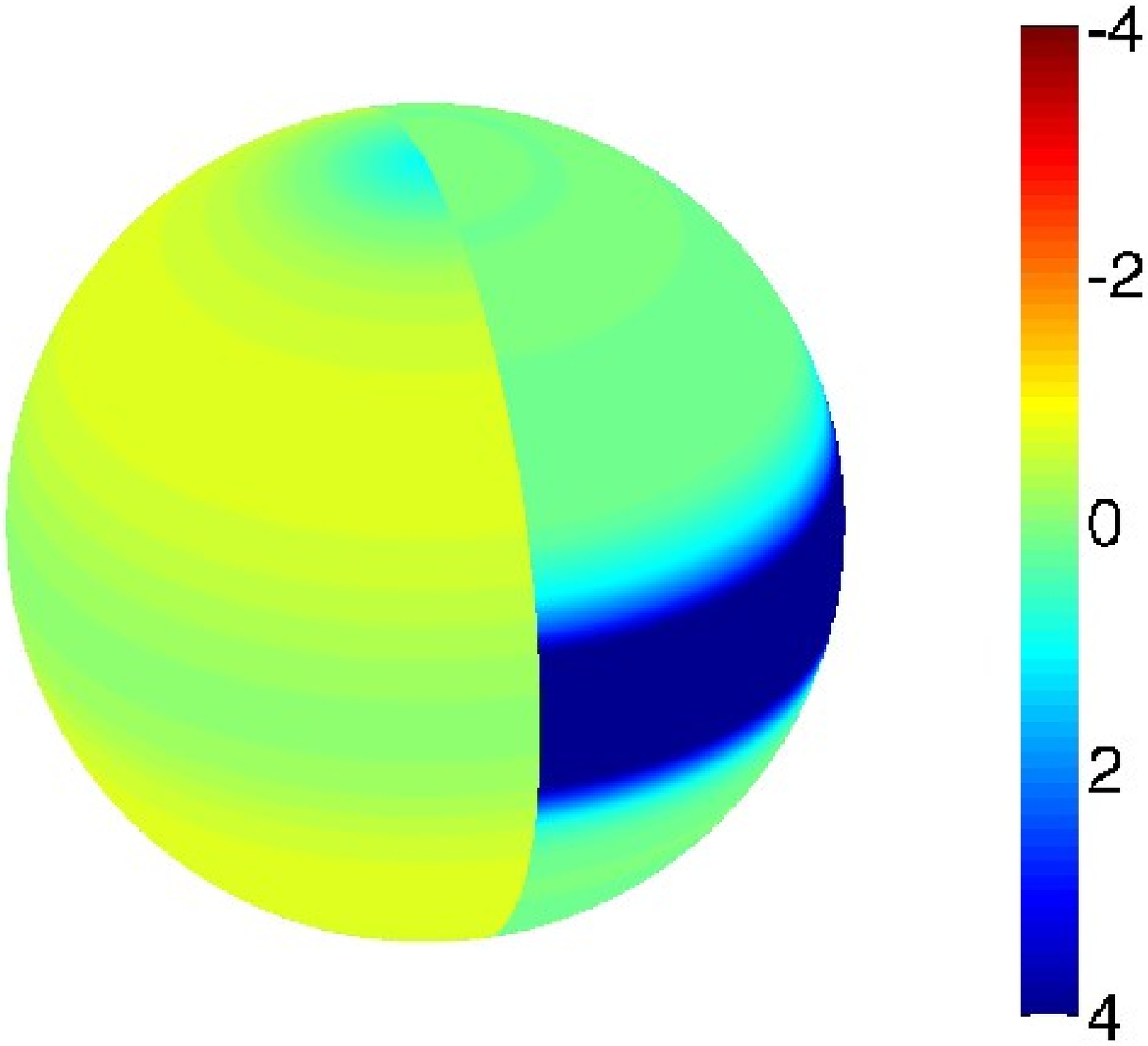}
\includegraphics[width=0.28\textwidth]{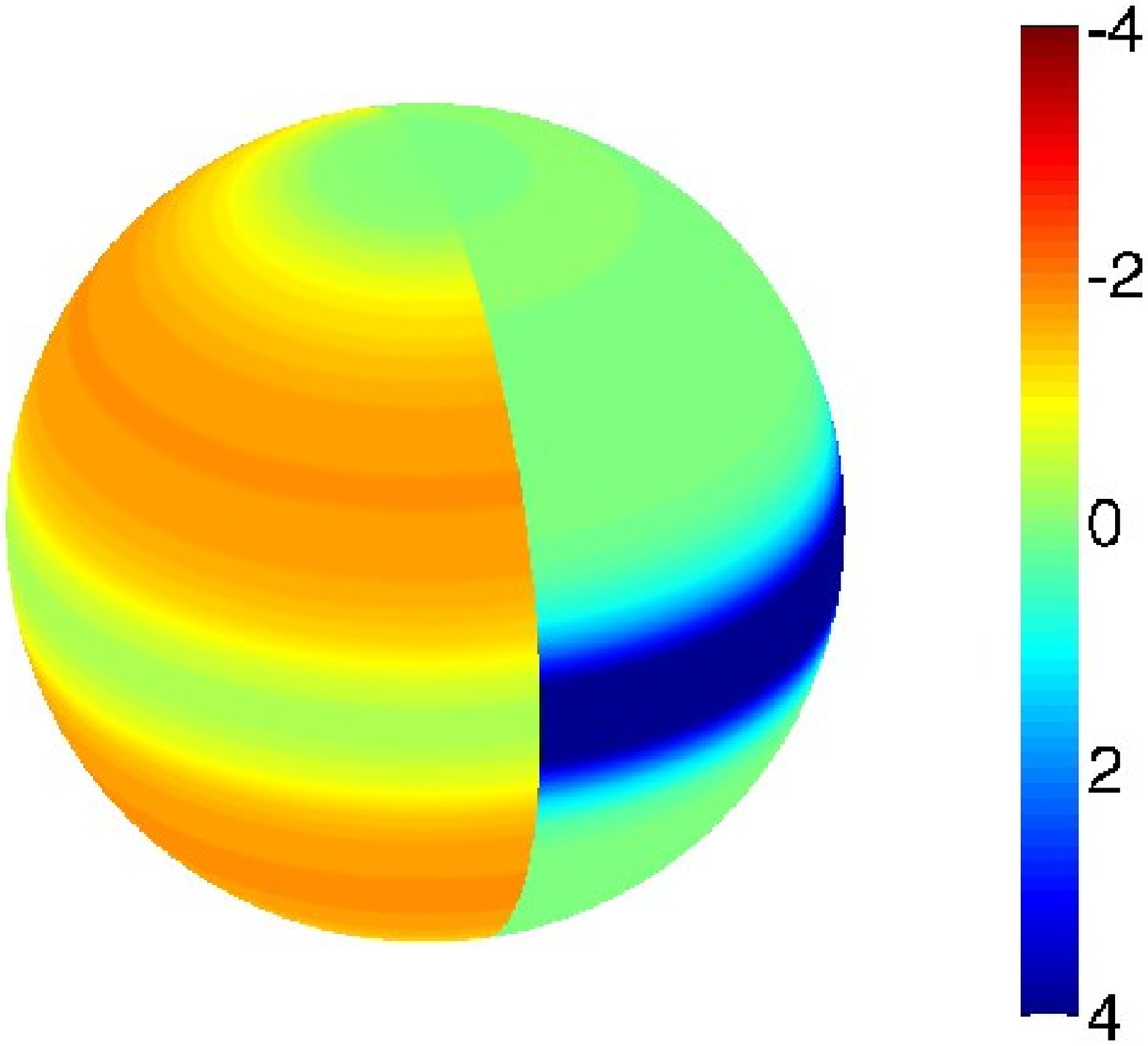}
\includegraphics[width=0.28\textwidth]{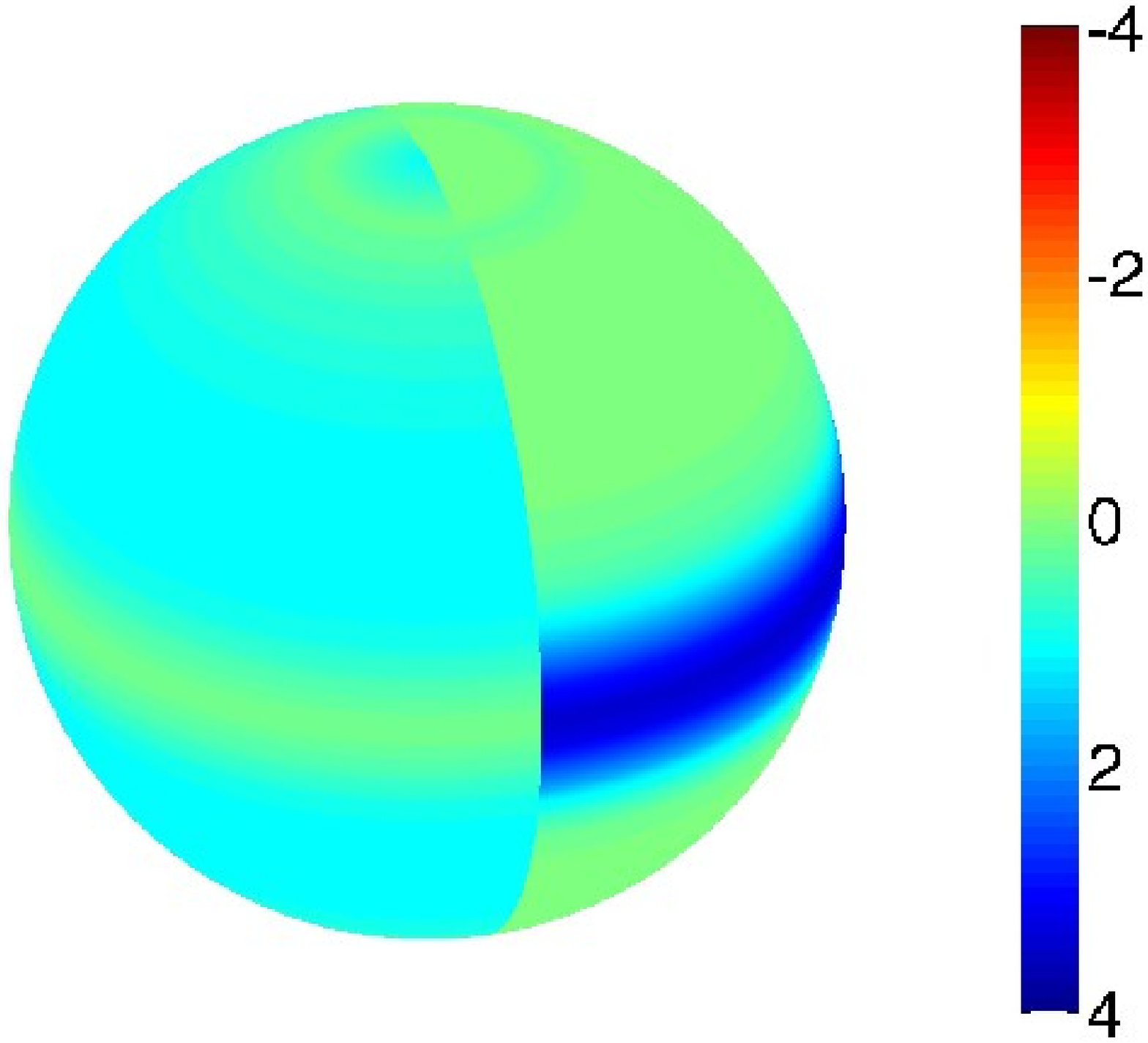}
\\
{\LARGE $a_*=0.70$}
\includegraphics[width=0.28\textwidth]{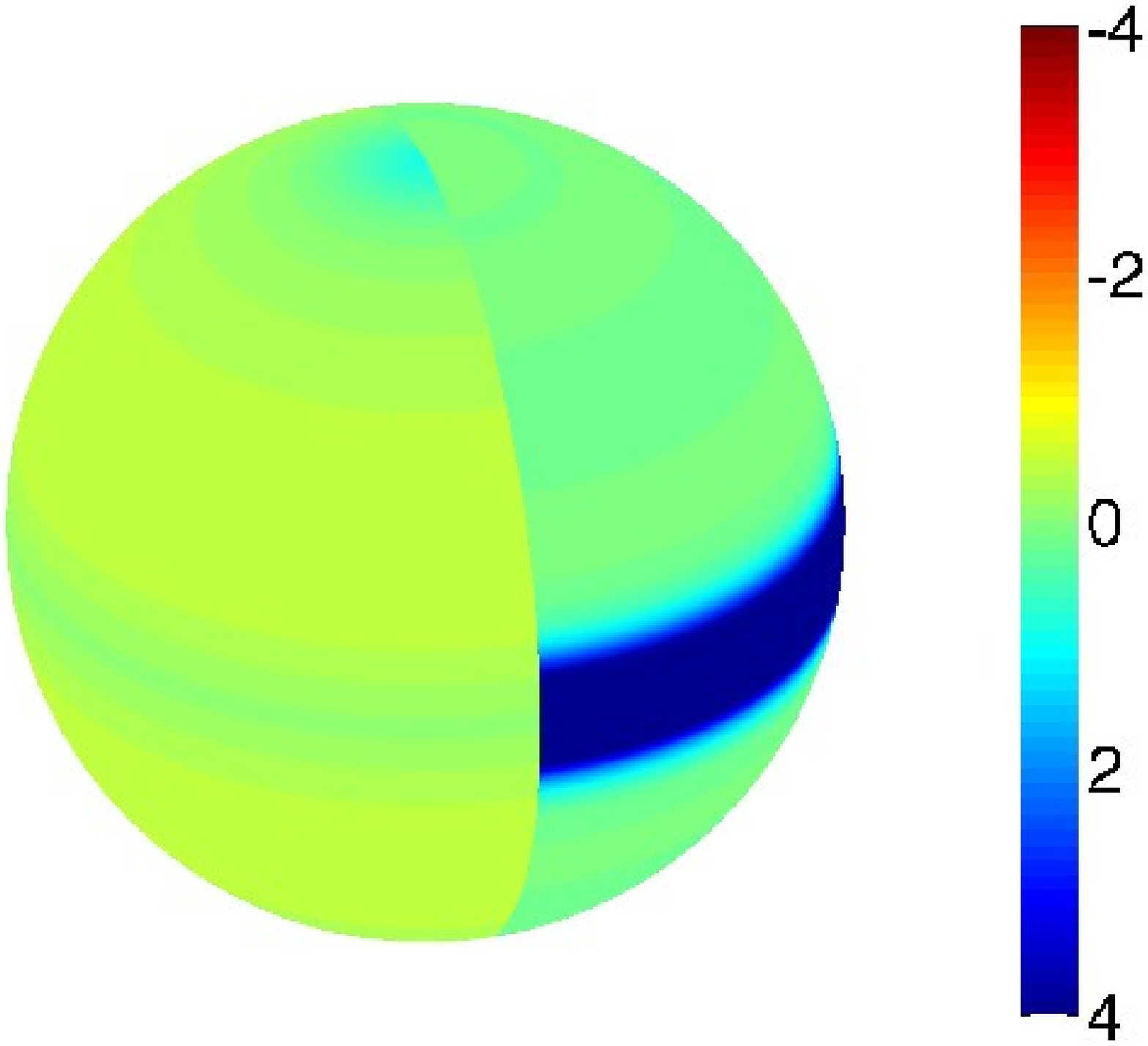}
\includegraphics[width=0.28\textwidth]{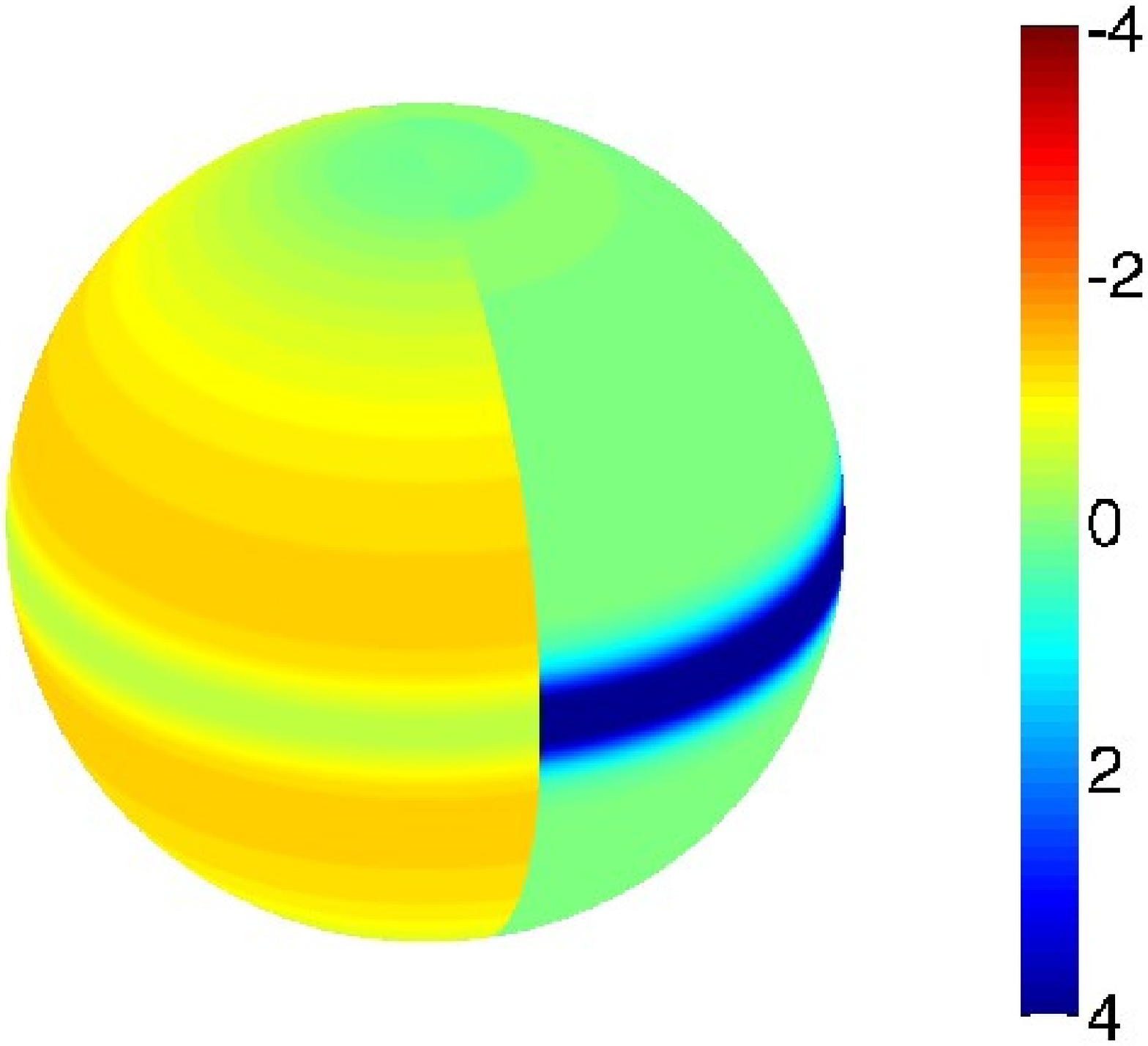}
\includegraphics[width=0.28\textwidth]{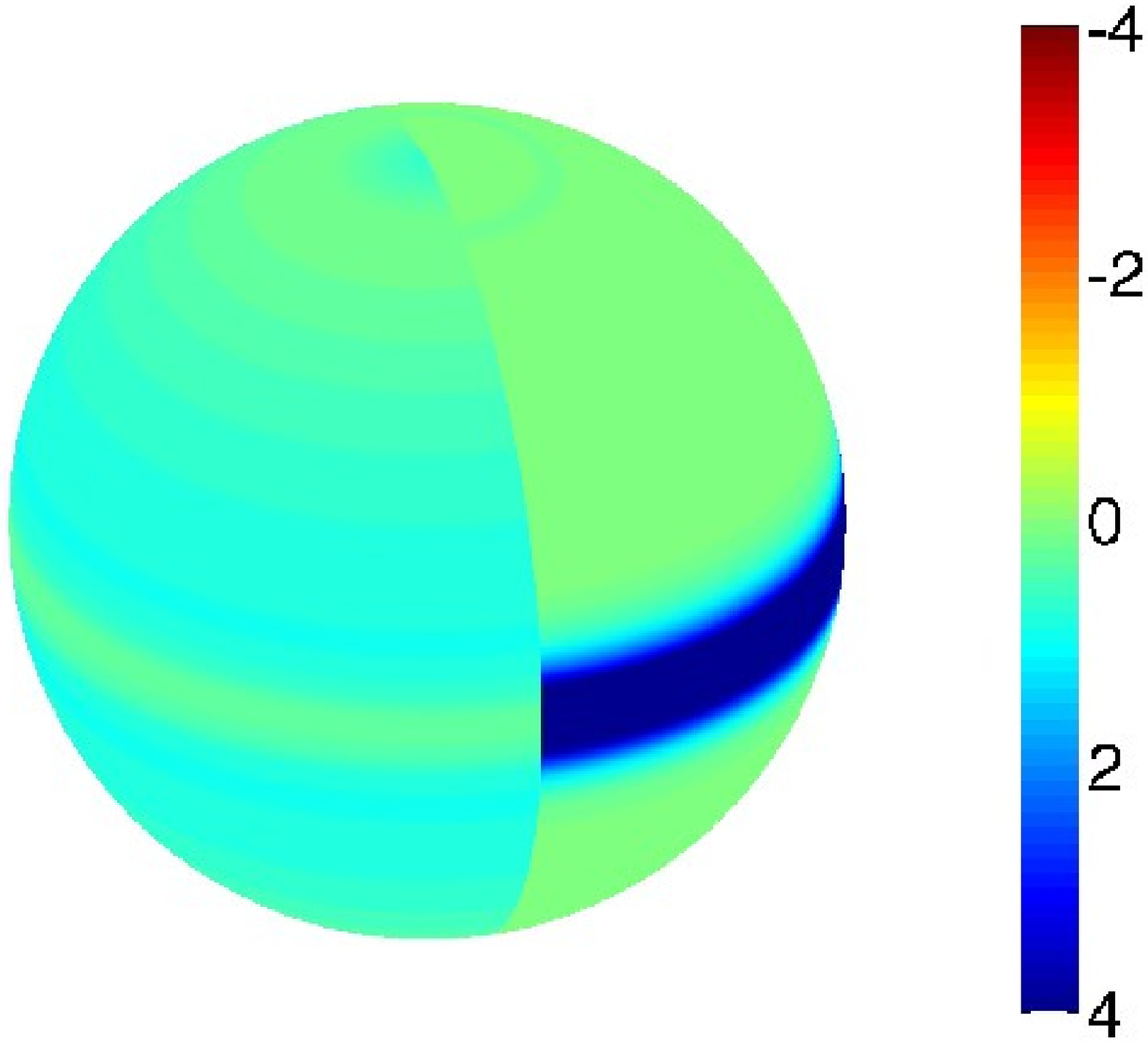}
\\
{\LARGE $a_*=0.00$}
\includegraphics[width=0.28\textwidth]{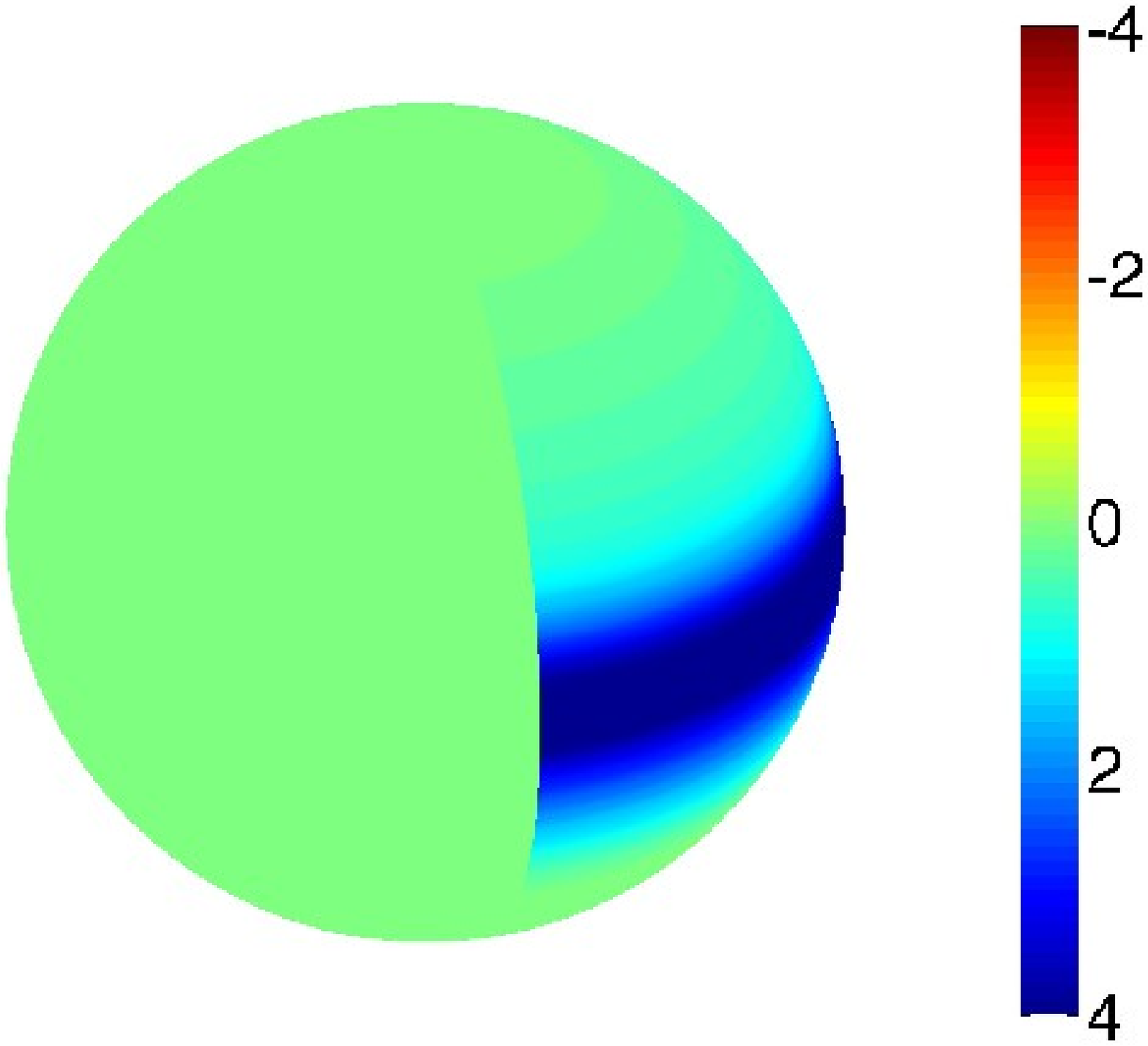}
\includegraphics[width=0.28\textwidth]{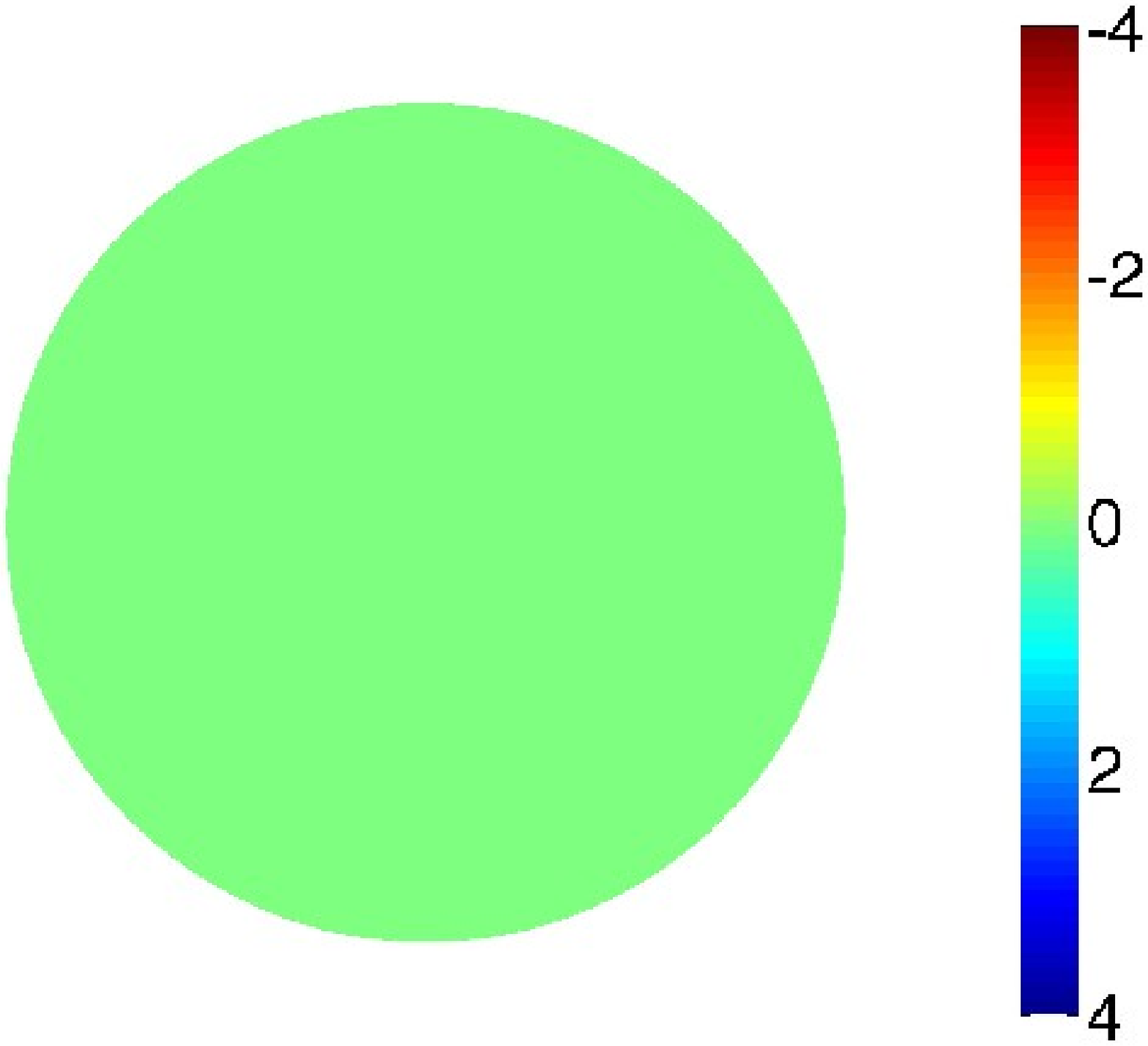}
\includegraphics[width=0.28\textwidth]{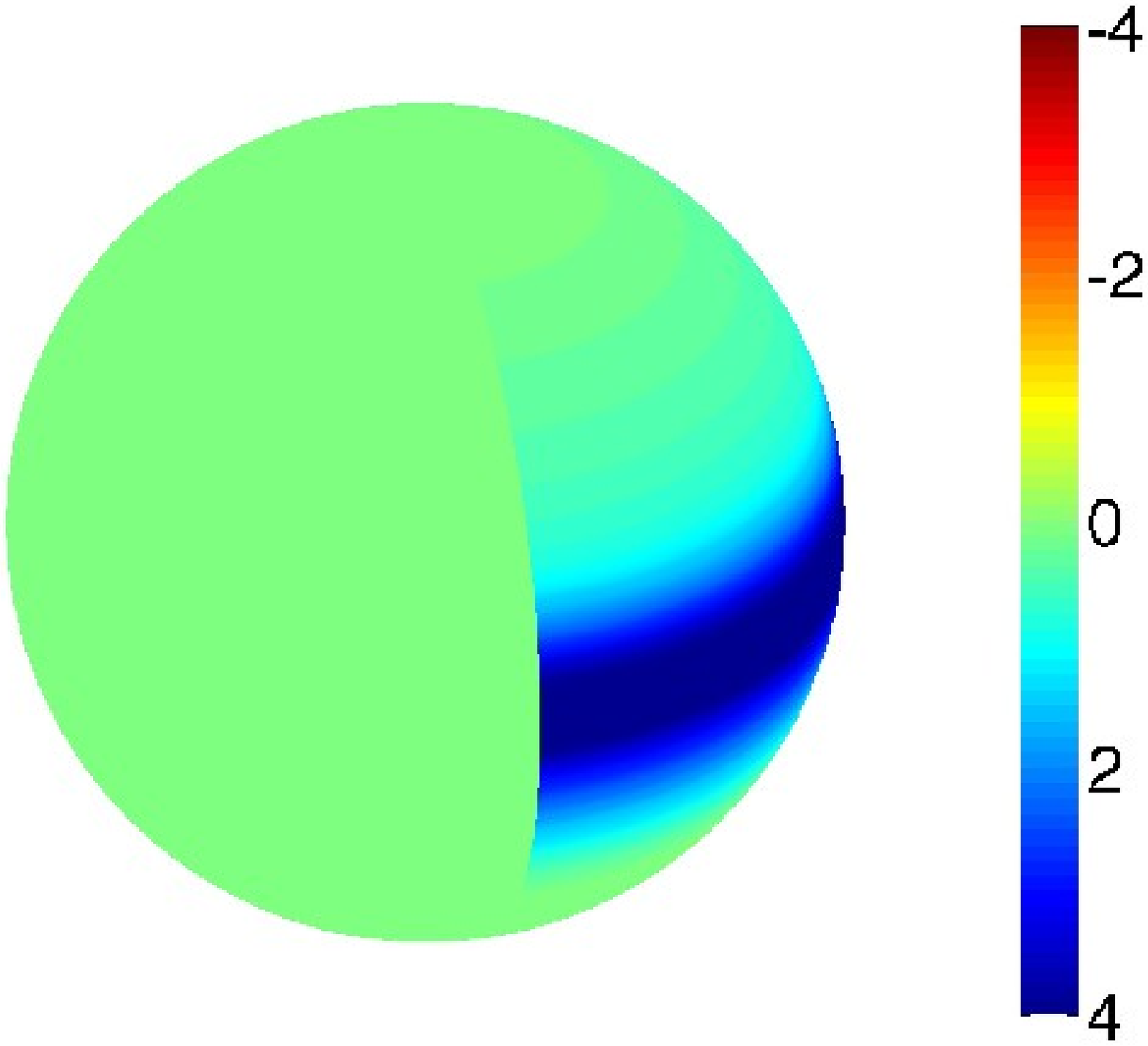}
\caption{\label{fig:horizonsplit} Same as Figure \ref{fig:horizonedot}, except each term is split into its magnetic (left hemisphere) and hydrodynamic (right hemisphere) components.}
\end{figure}

\newpage
\begin{figure}
\hspace{1.3in}
{\LARGE $\frac{dM}{dmdA}$}
\hspace{1.4in}
{\LARGE $\Omega_H \frac{dJ}{dmdA}$}
\hspace{1.3in}
{\LARGE $T_H \frac{dS_H}{dmdA}$}
\\
{\LARGE $a_*=0.90$}
\includegraphics[width=0.28\textwidth]{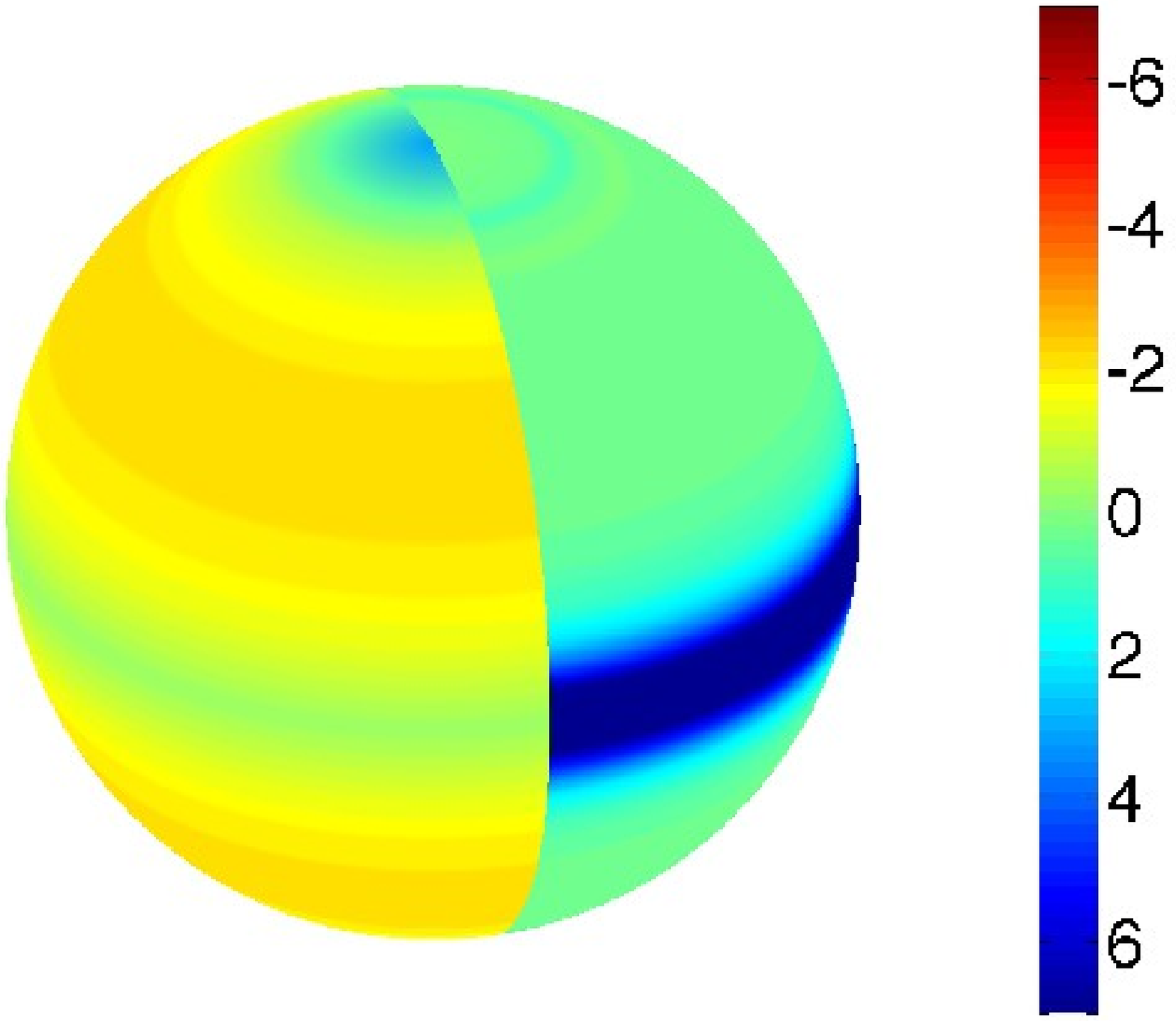}
\includegraphics[width=0.28\textwidth]{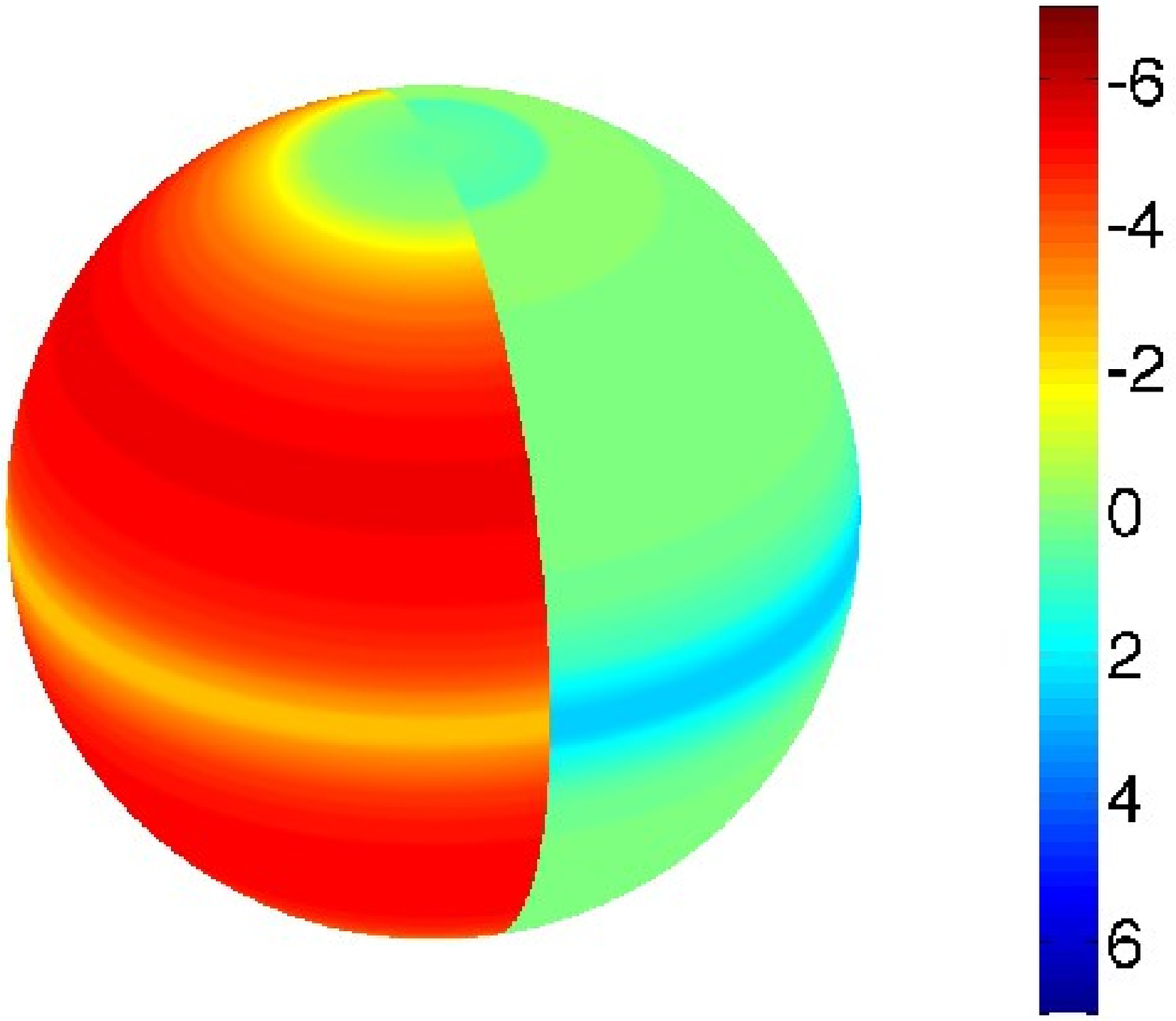}
\includegraphics[width=0.28\textwidth]{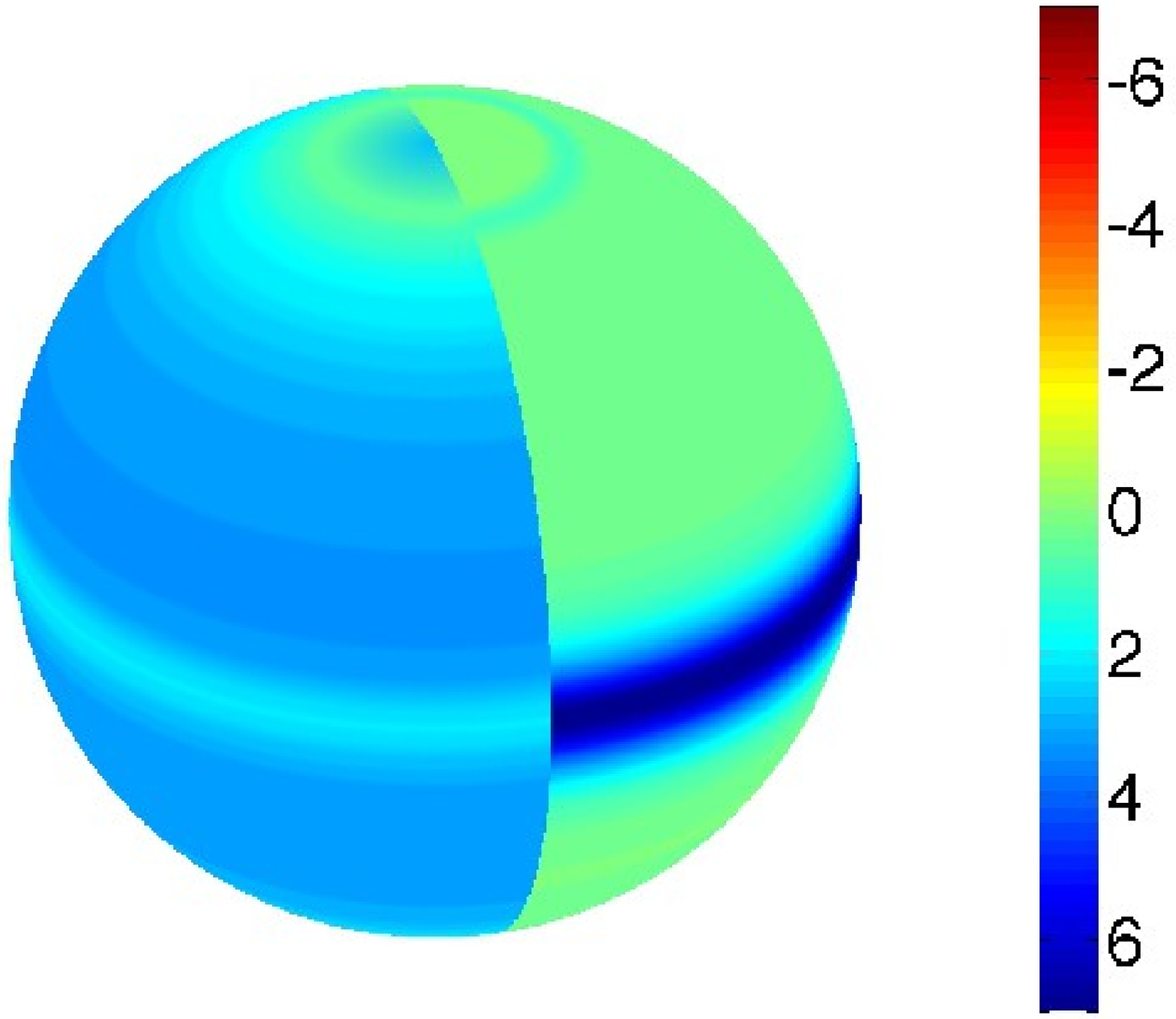}
\\
{\LARGE $a_*=0.70$}
\includegraphics[width=0.28\textwidth]{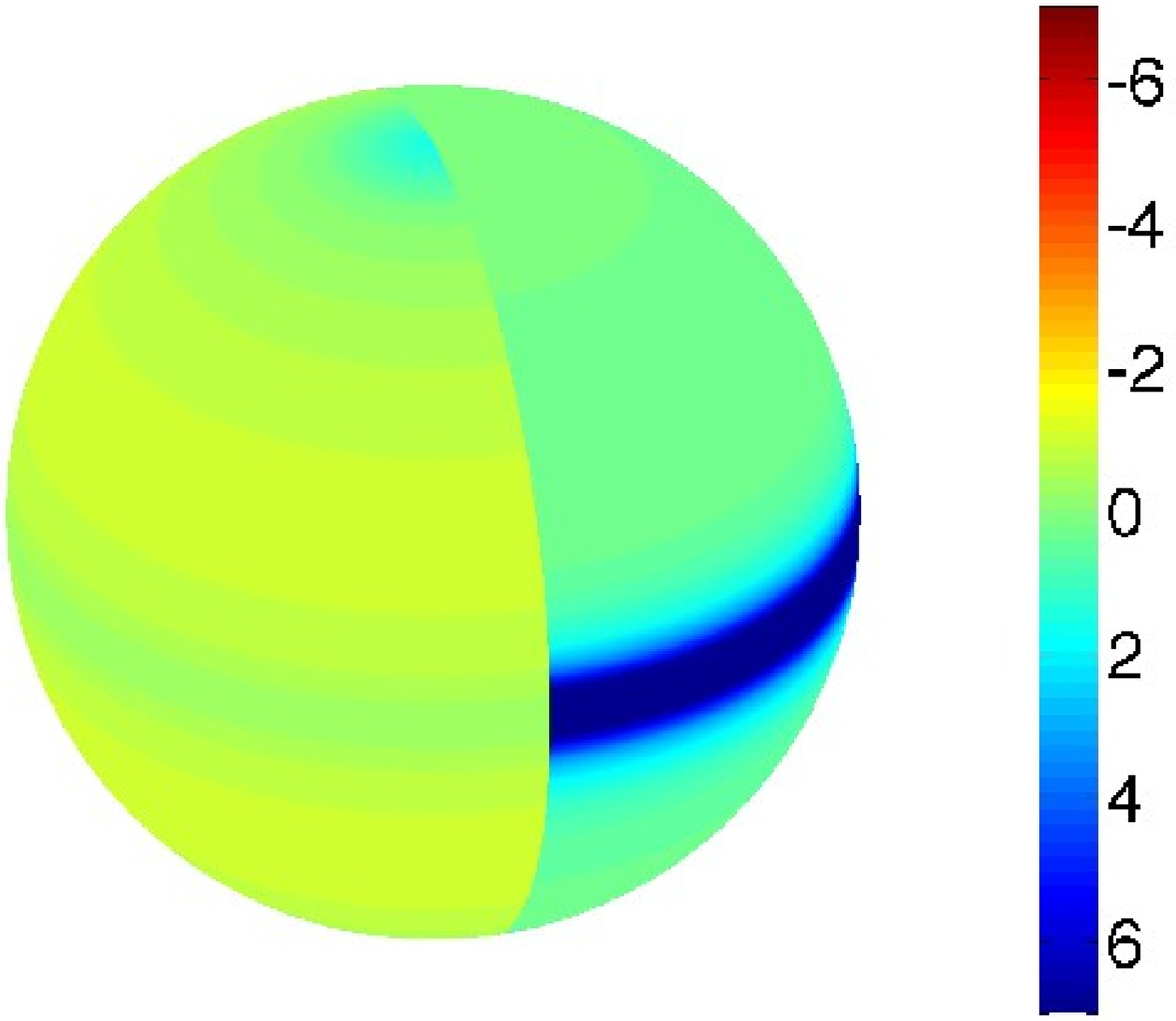}
\includegraphics[width=0.28\textwidth]{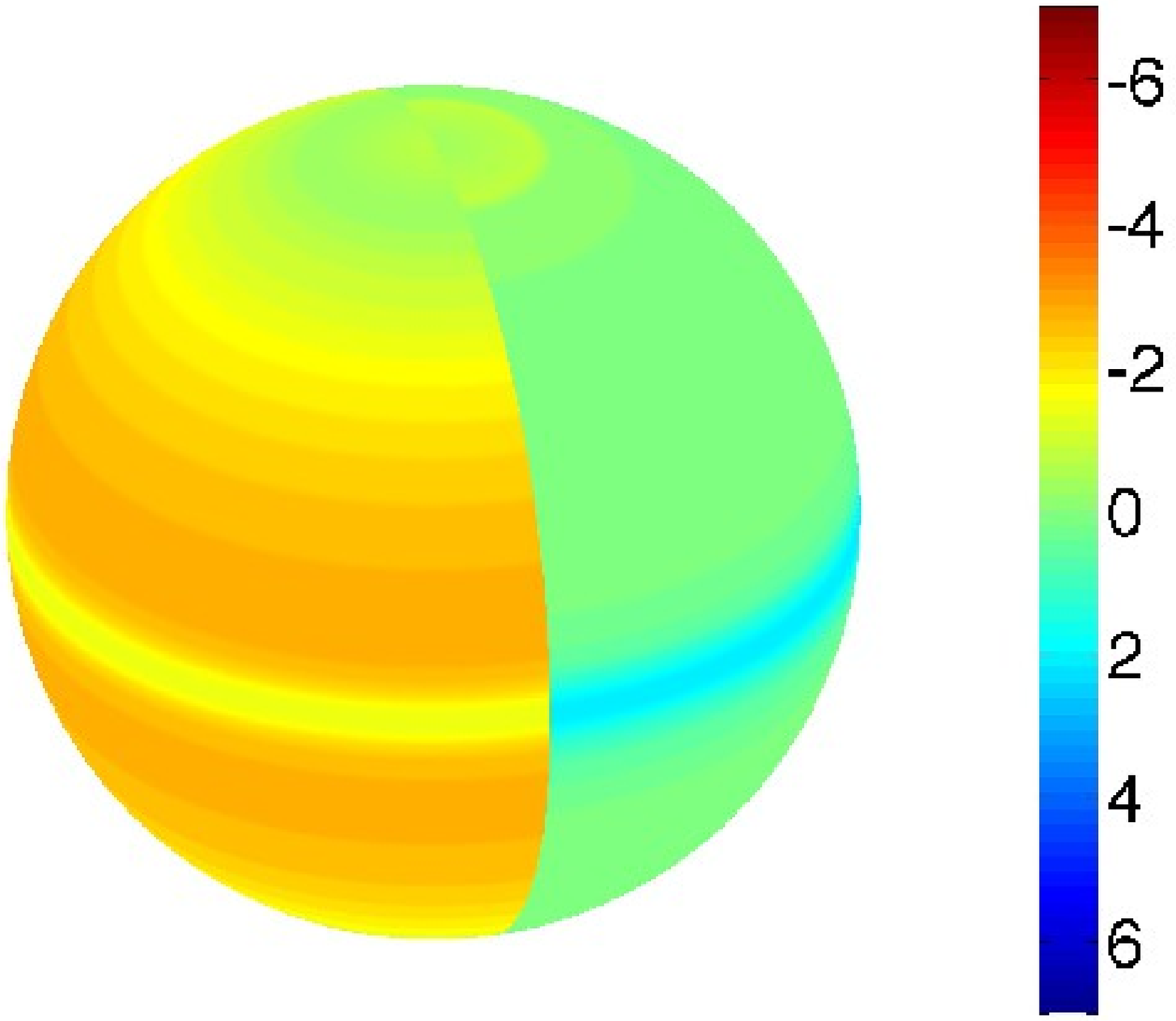}
\includegraphics[width=0.28\textwidth]{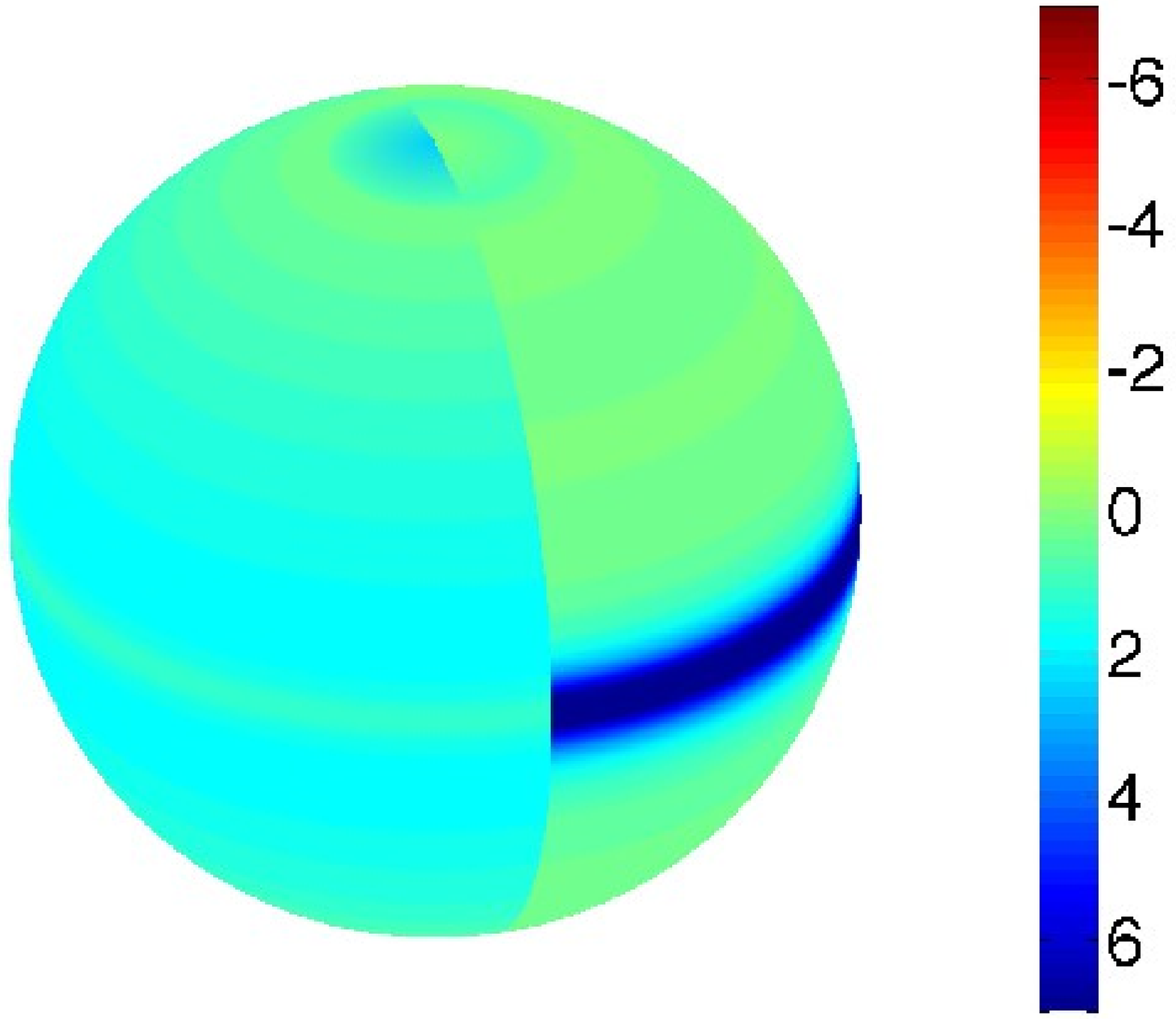}
\\
{\LARGE $a_*=0.00$}
\includegraphics[width=0.28\textwidth]{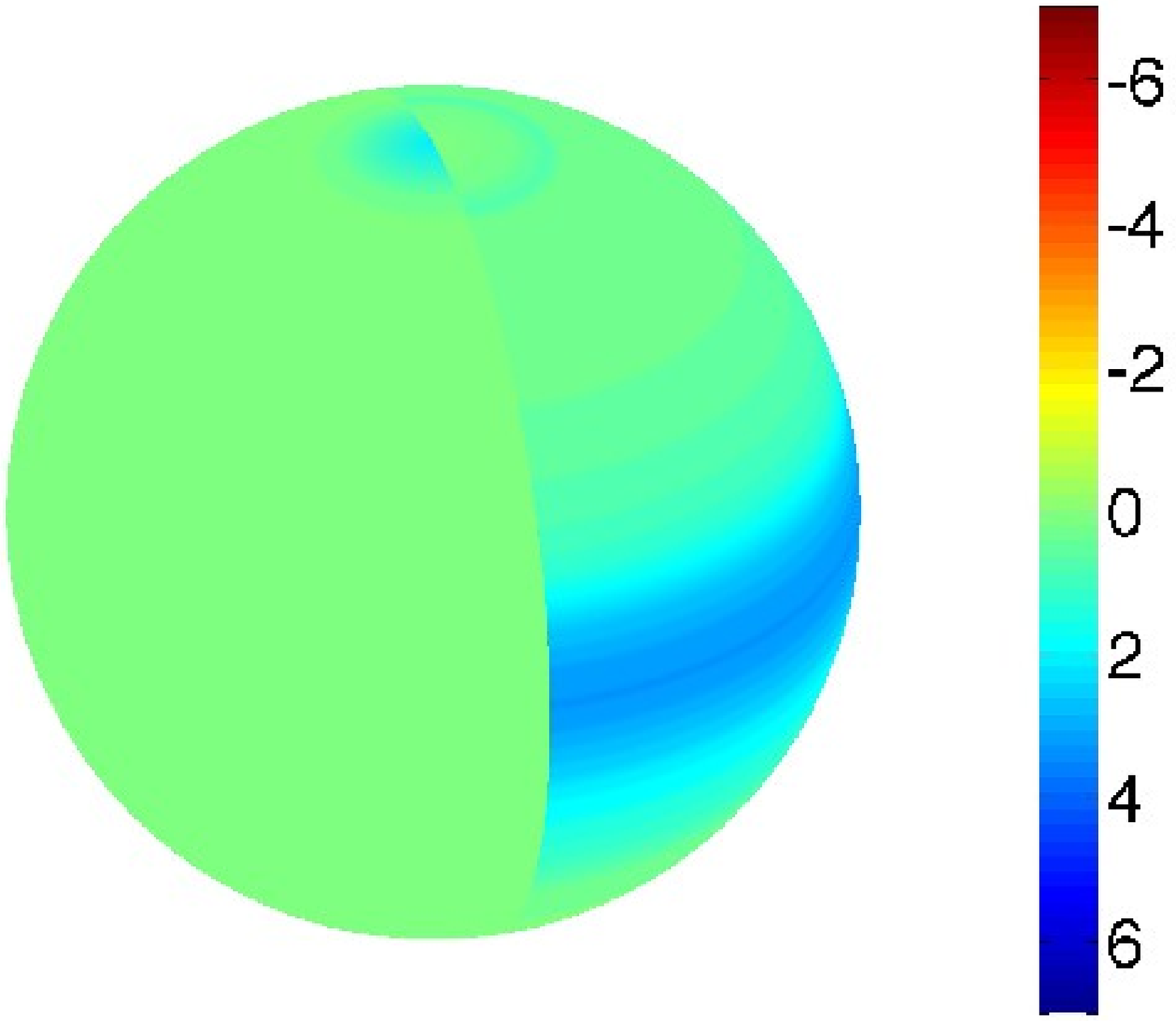}
\includegraphics[width=0.28\textwidth]{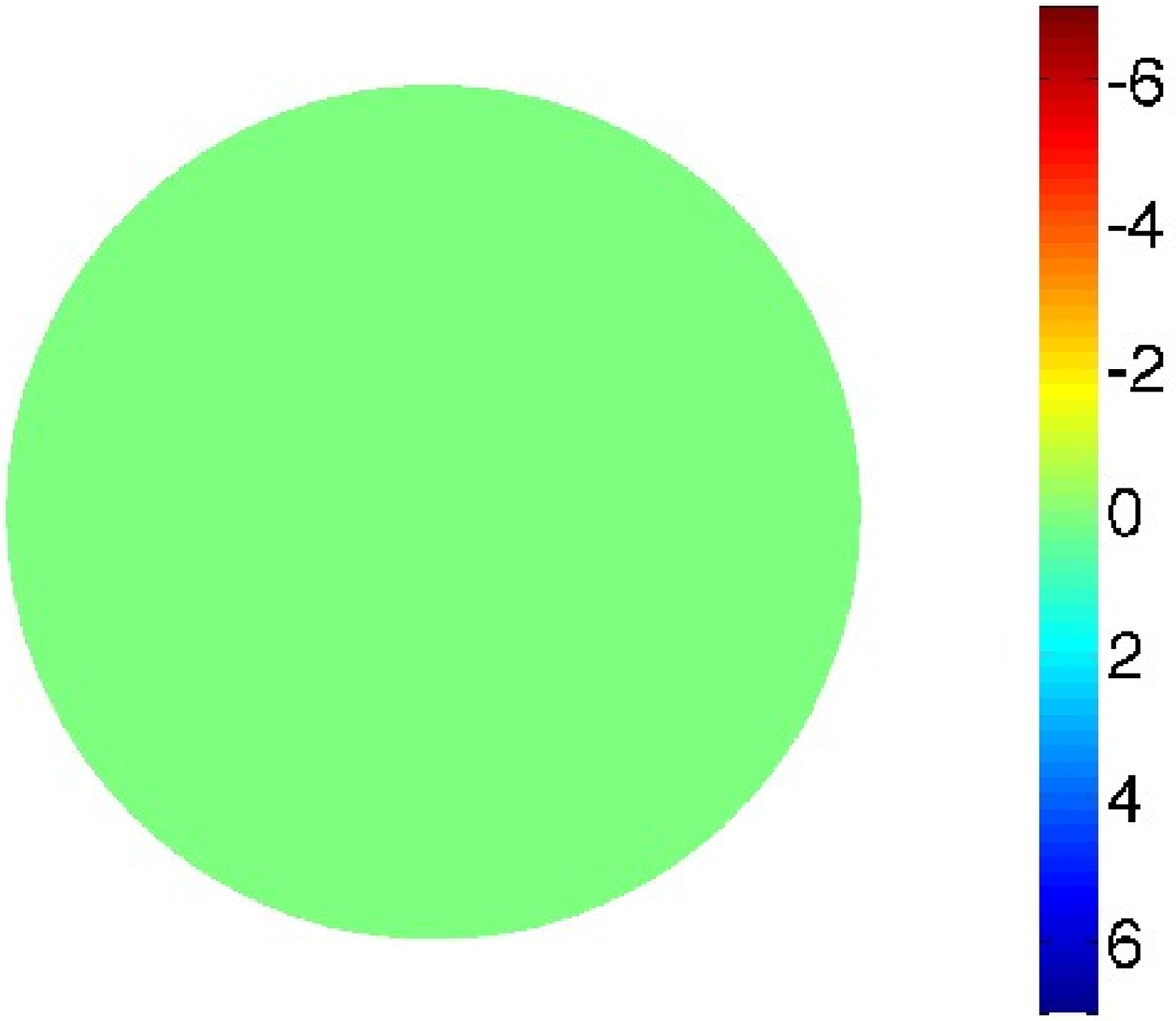}
\includegraphics[width=0.28\textwidth]{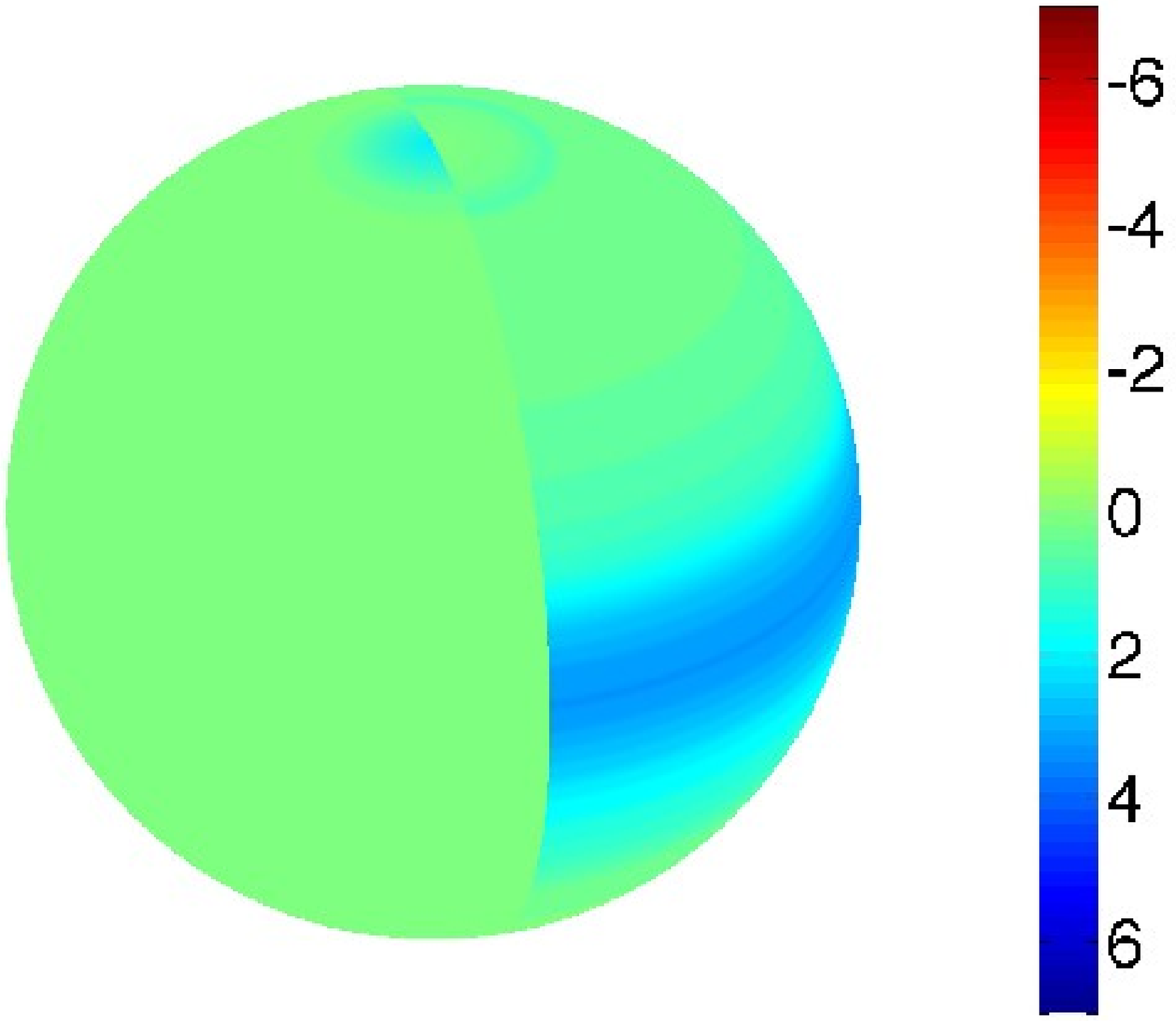}
\caption{\label{fig:horizonsplit_mad} Same as Figure \ref{fig:horizonsplit} but for MAD runs.}
\end{figure}

%------- Horizon Fields --------------
\newpage
\begin{figure}
\hspace{1.2in}
{\LARGE $B_n, \sigma_H$}
\hspace{1.5in}
{\LARGE $\vec{E}_H$}
\hspace{1.7in}
{\LARGE $\vec{B}_H$}
\\
{\LARGE $a_*=0.98$}
\includegraphics[width=0.28\textwidth]{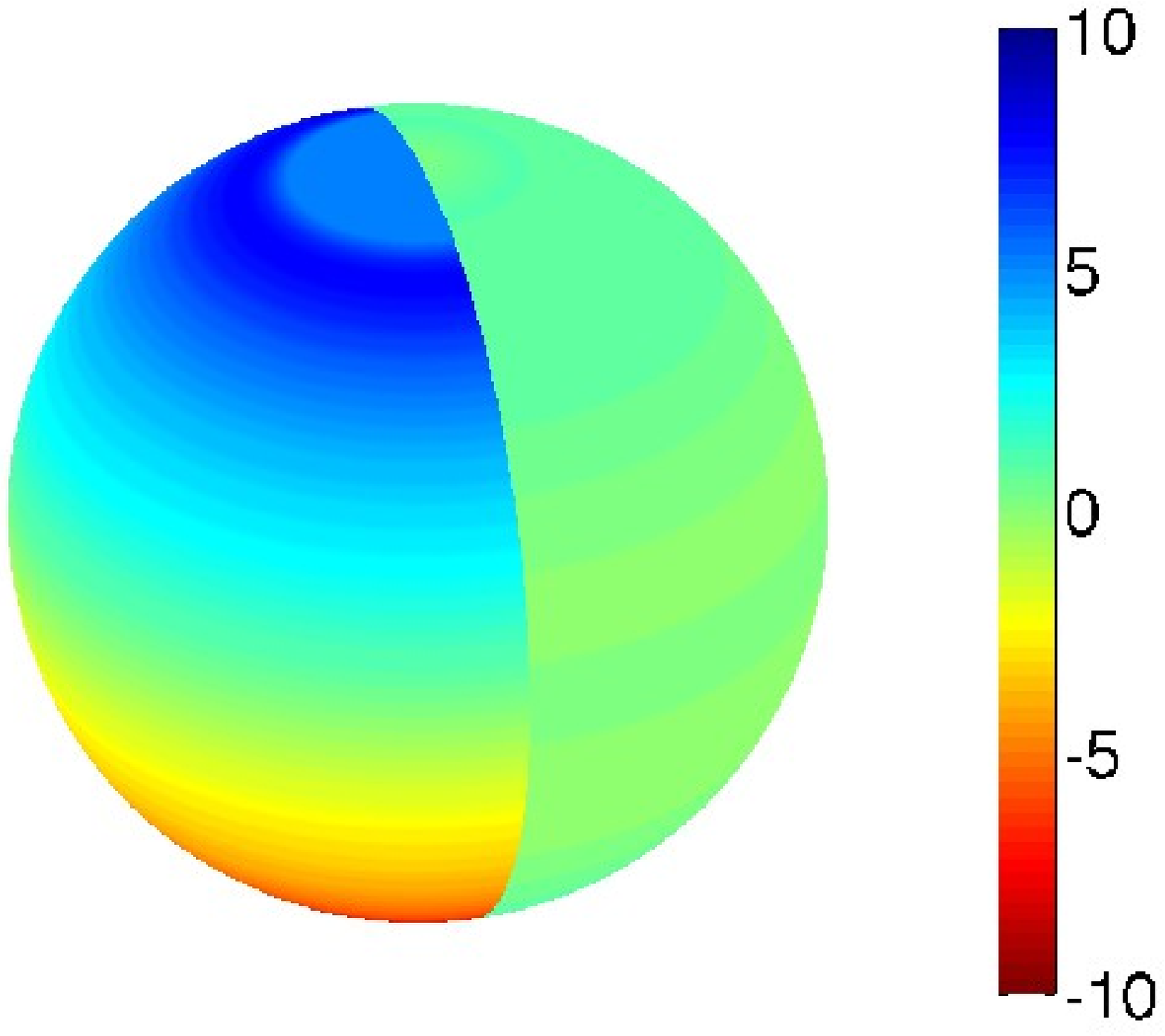}
\includegraphics[width=0.28\textwidth]{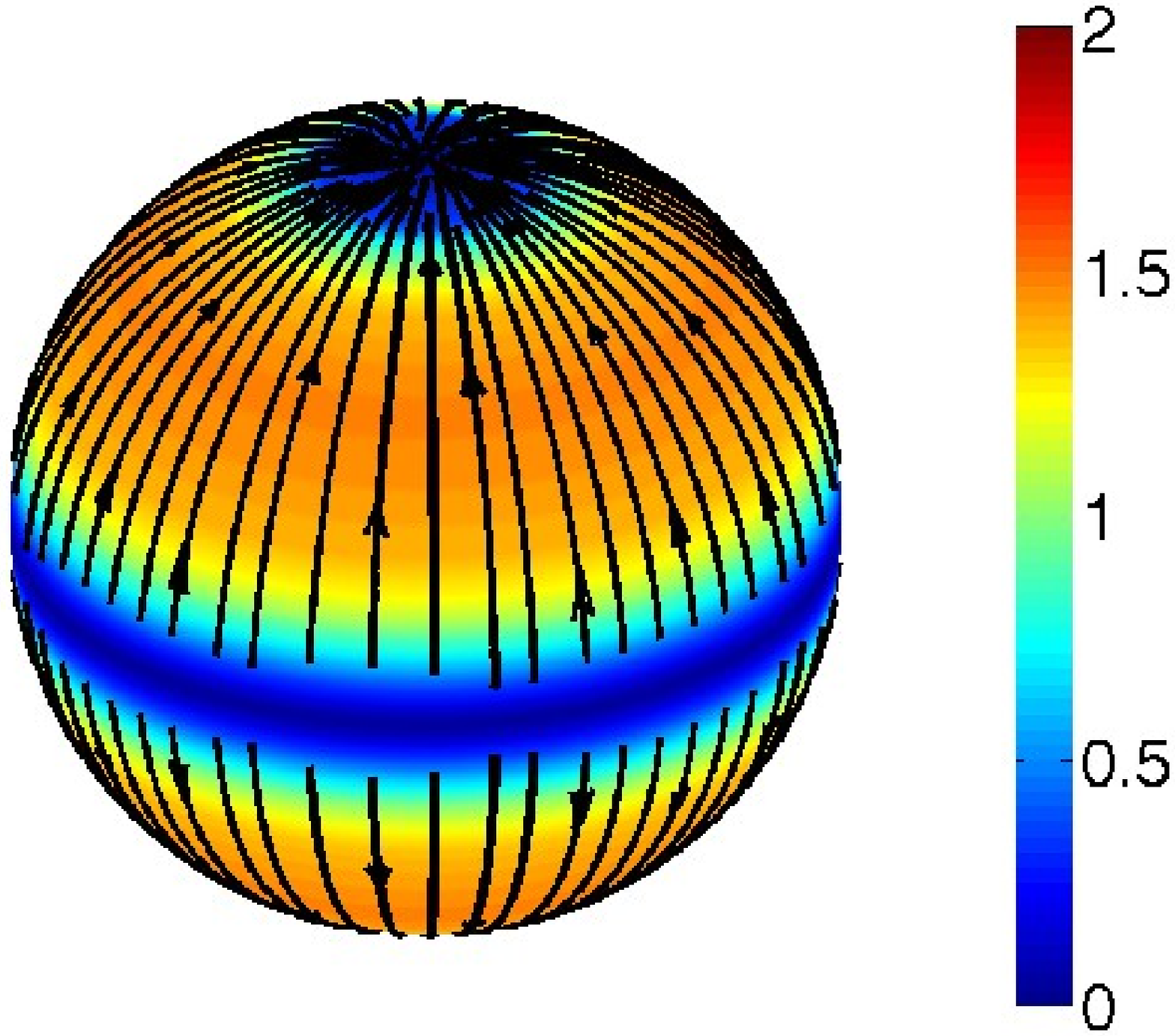}
\includegraphics[width=0.28\textwidth]{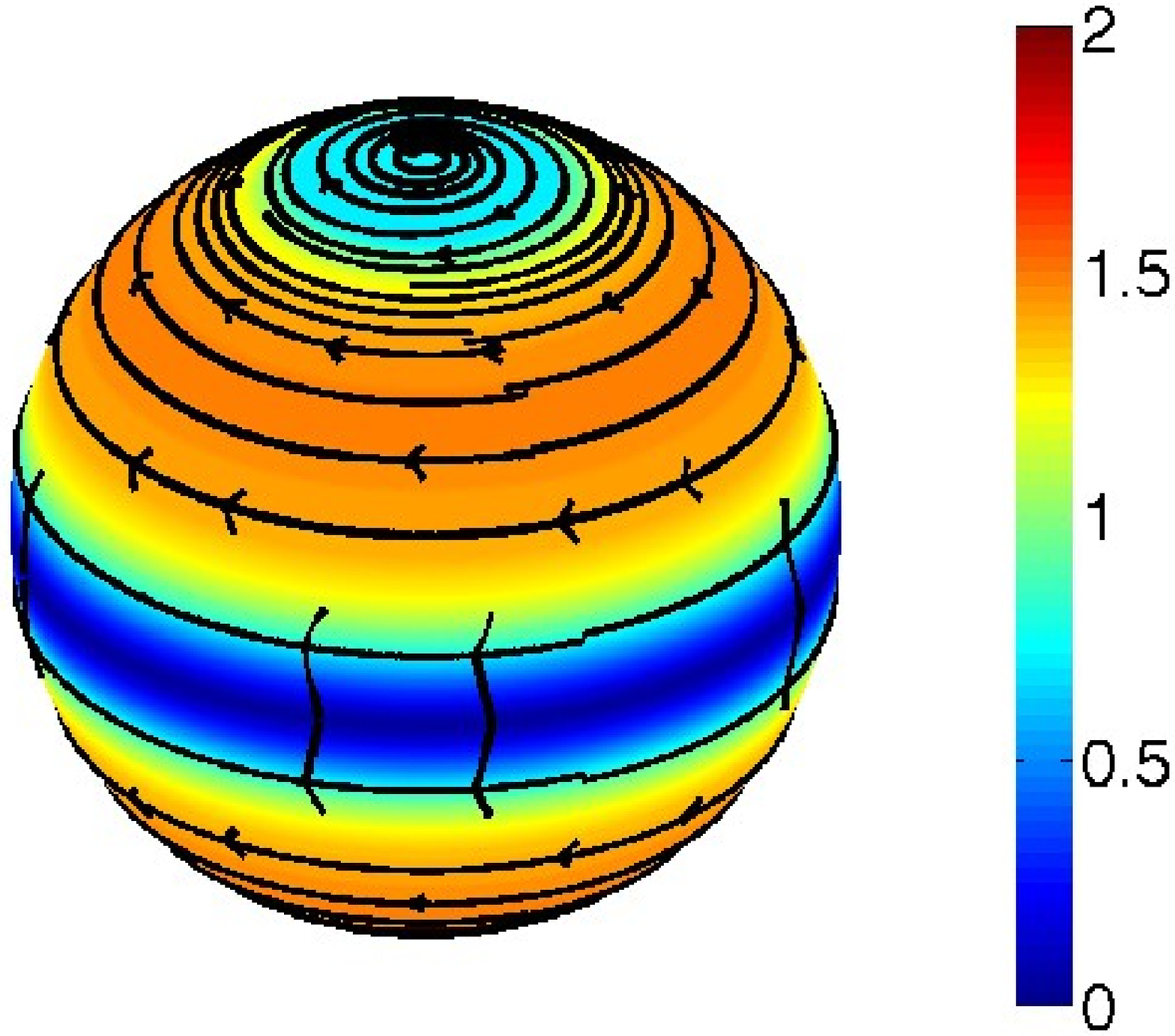}
\\
{\LARGE $a_*=0.90$}
\includegraphics[width=0.28\textwidth]{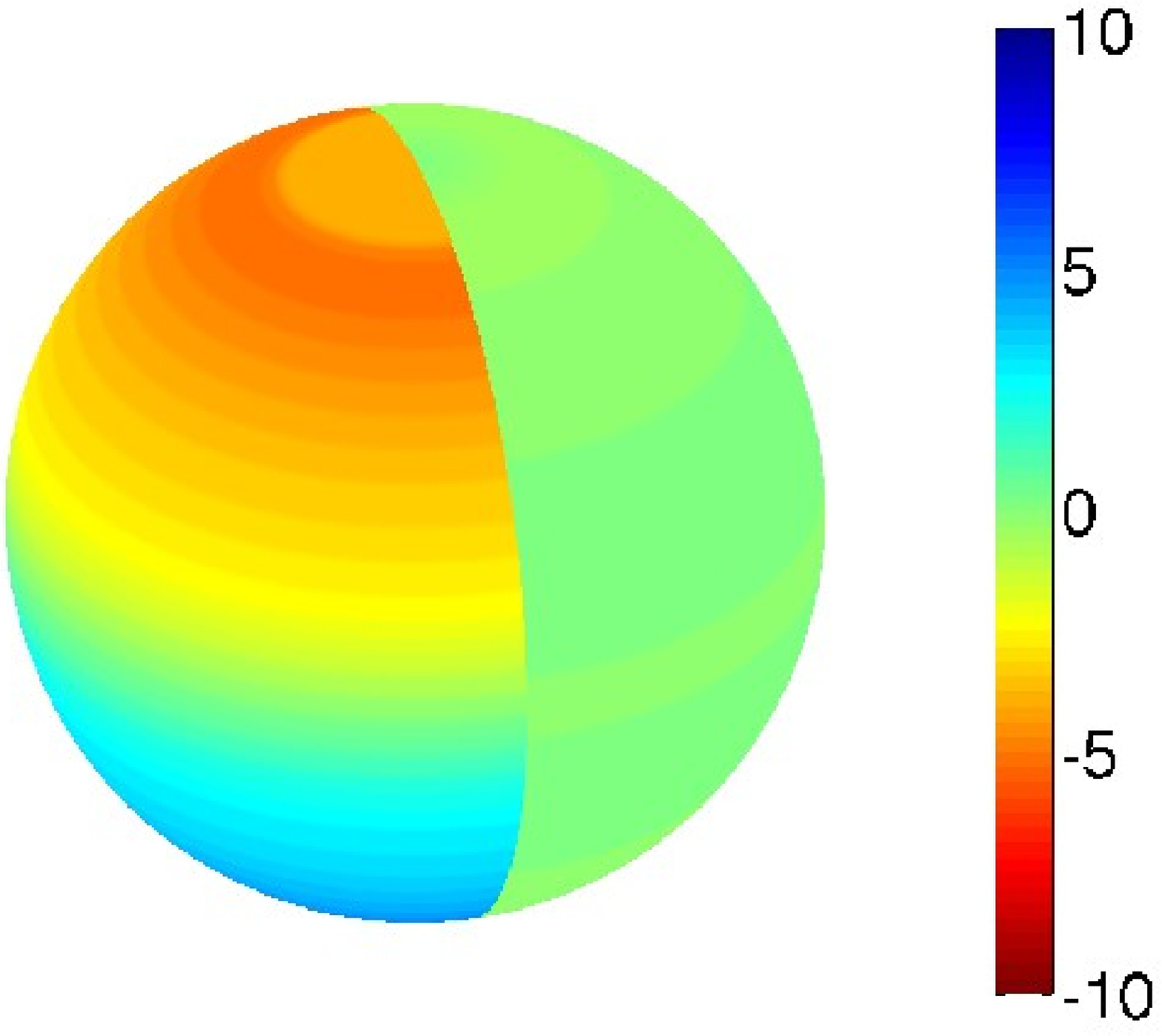}
\includegraphics[width=0.28\textwidth]{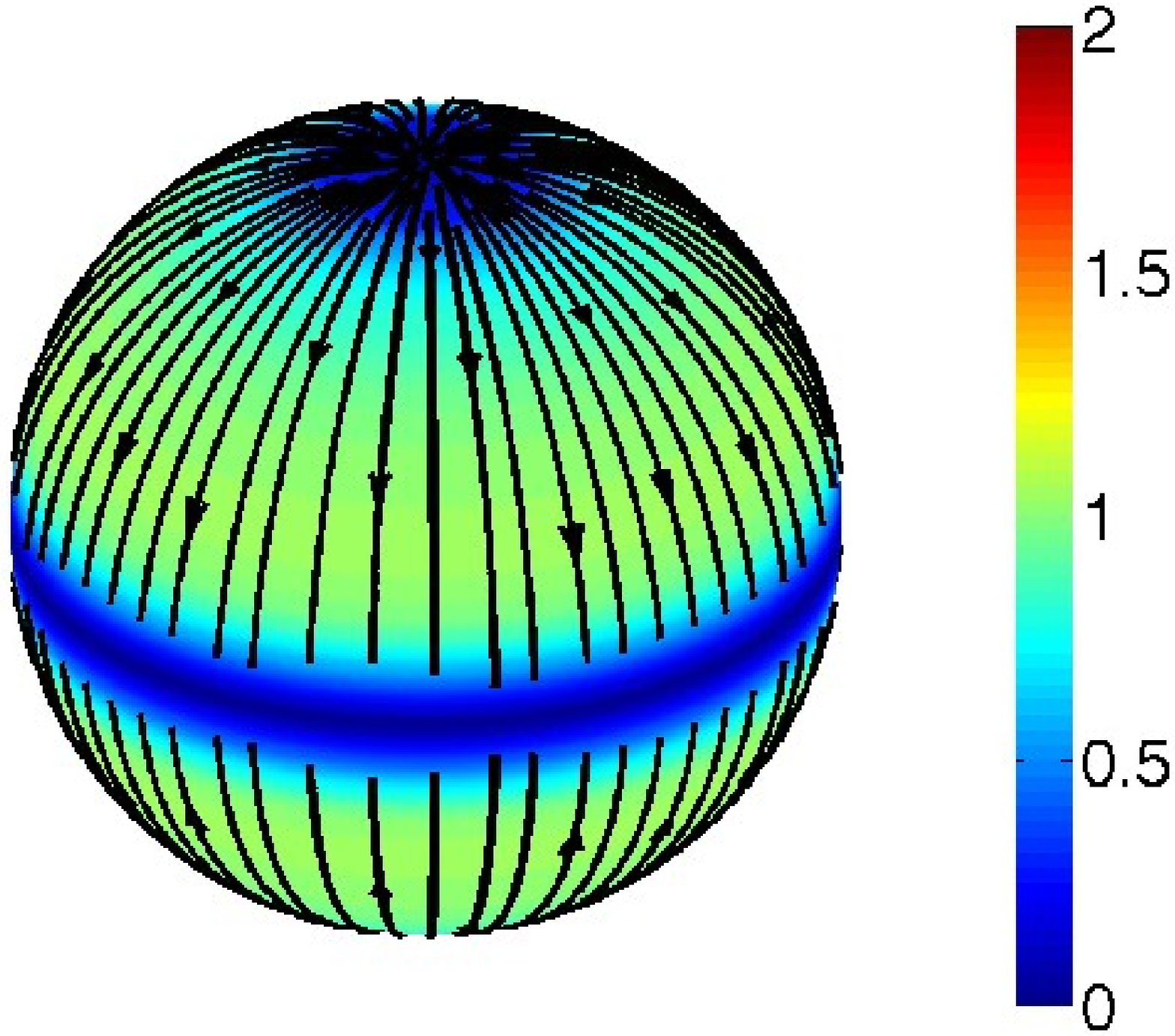}
\includegraphics[width=0.28\textwidth]{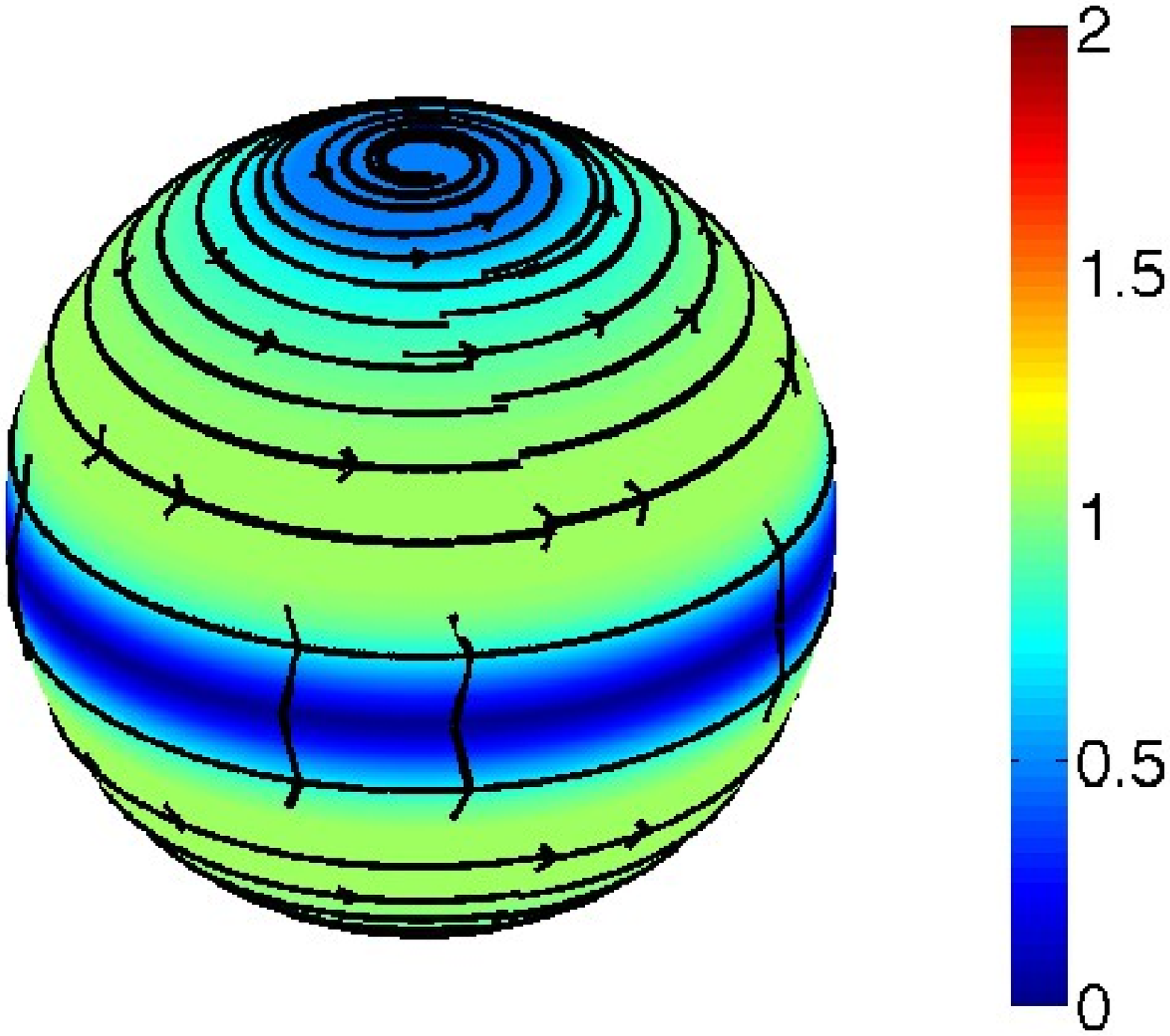}
\\
{\LARGE $a_*=0.70$}
\includegraphics[width=0.28\textwidth]{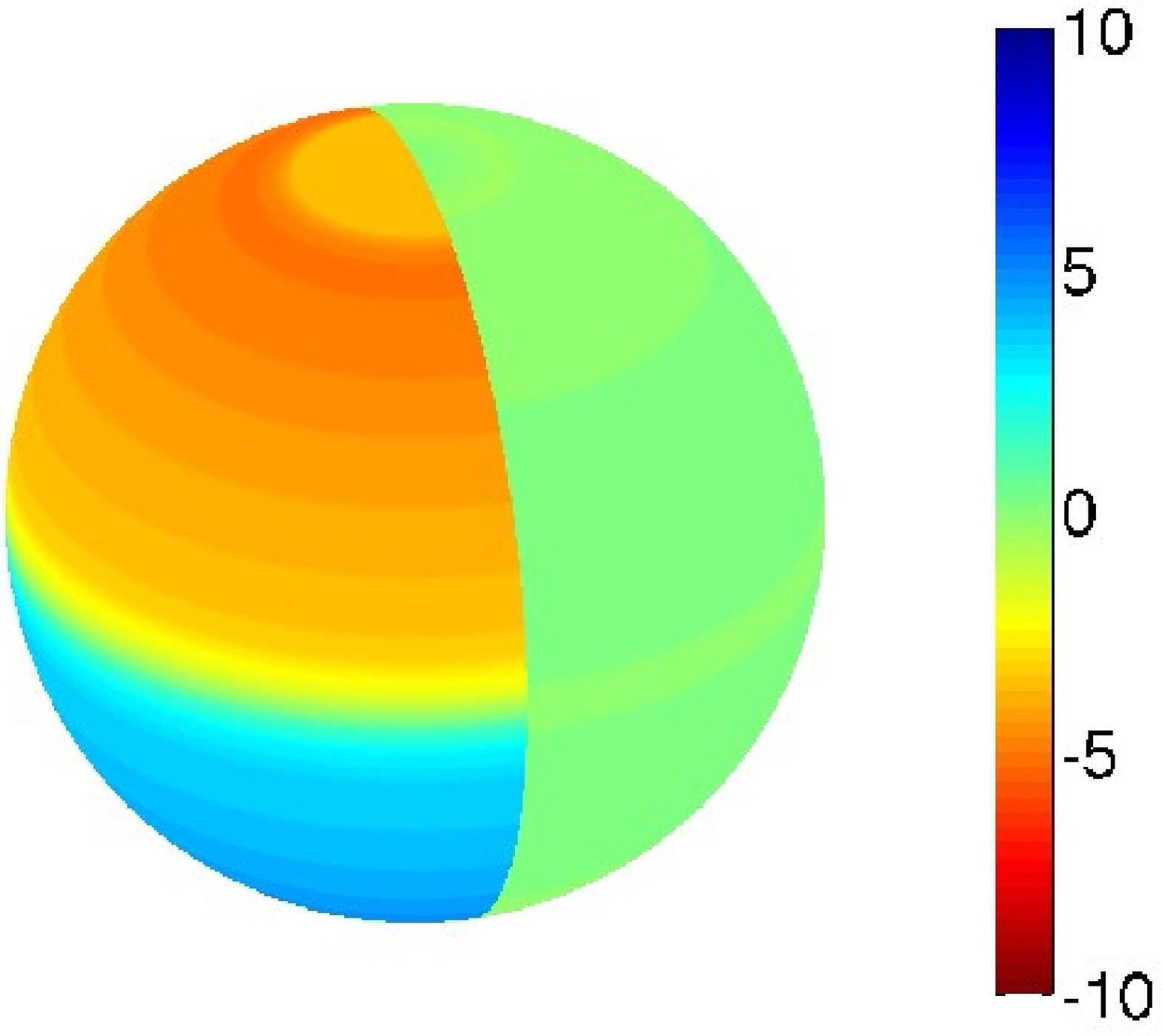}
\includegraphics[width=0.28\textwidth]{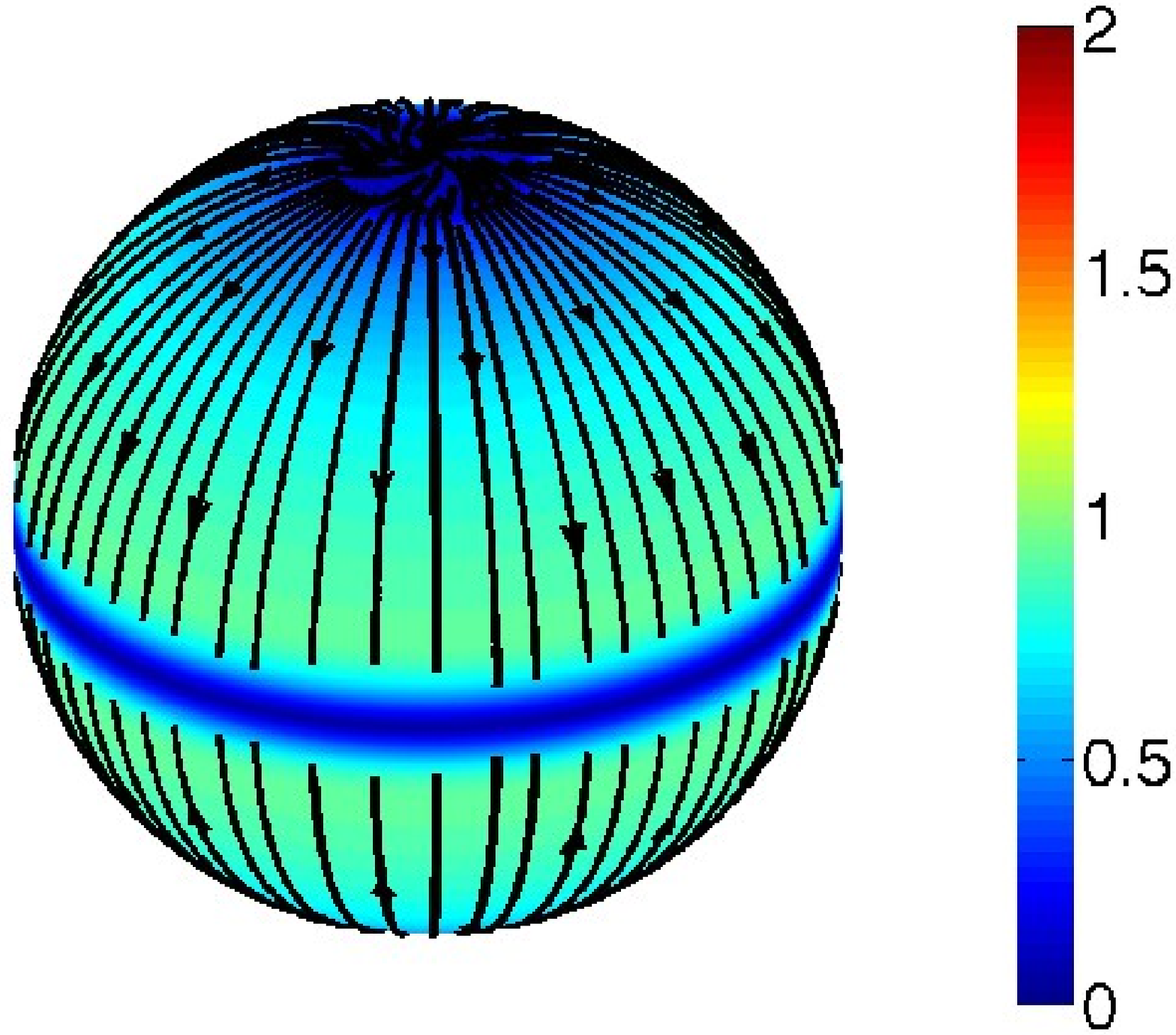}
\includegraphics[width=0.28\textwidth]{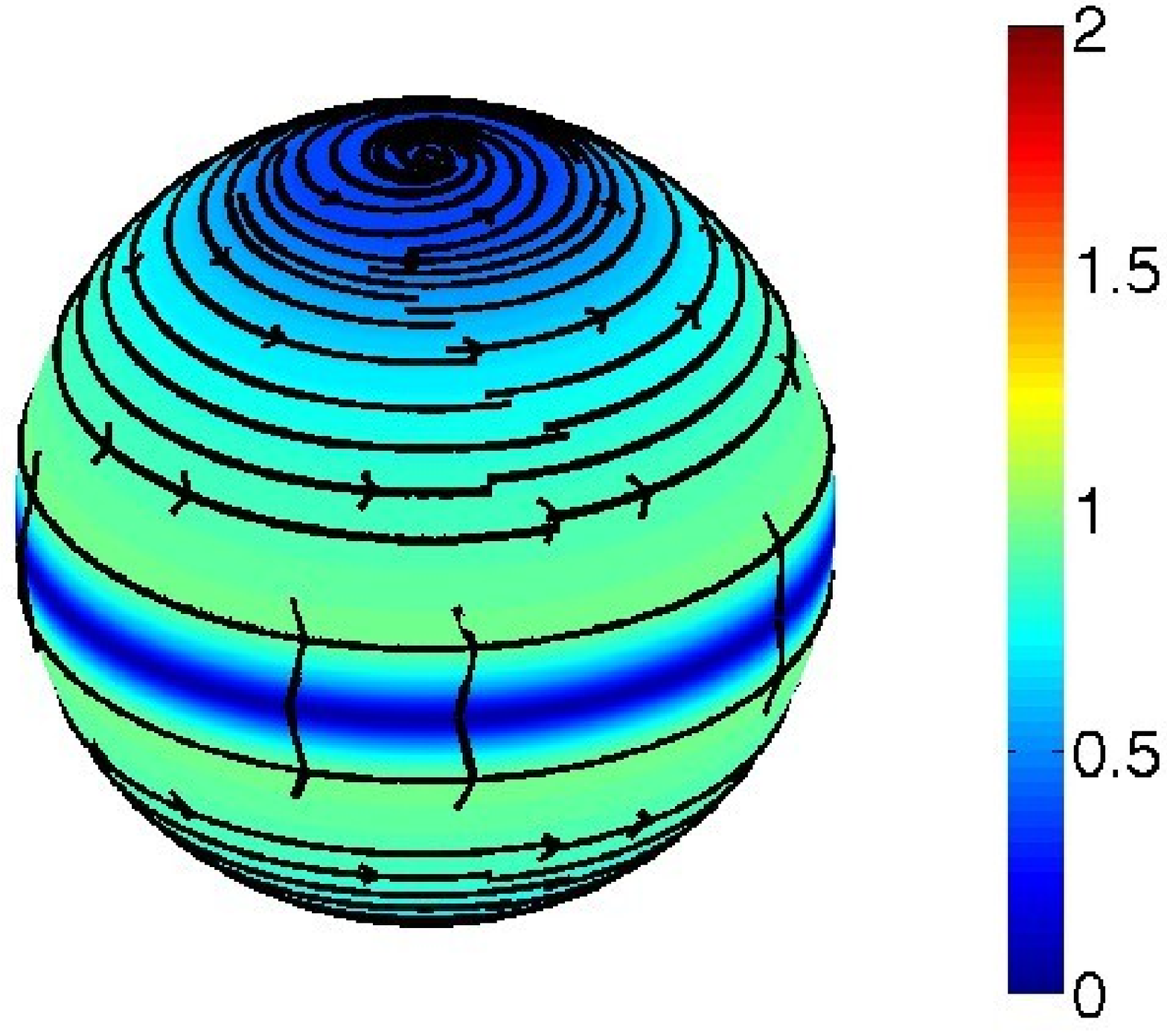}
\\
{\LARGE $a_*=0.00$}
\includegraphics[width=0.28\textwidth]{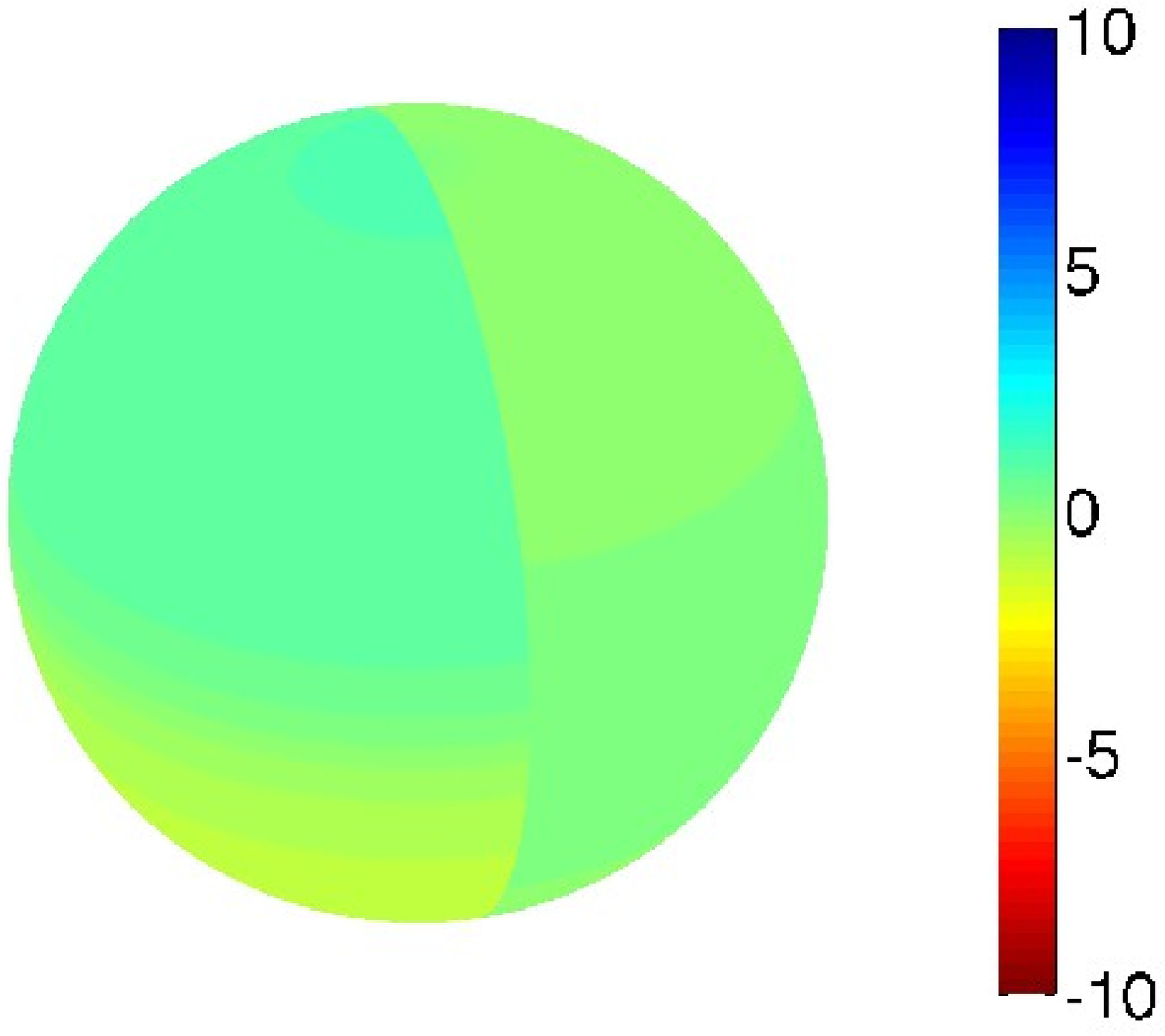}
\includegraphics[width=0.28\textwidth]{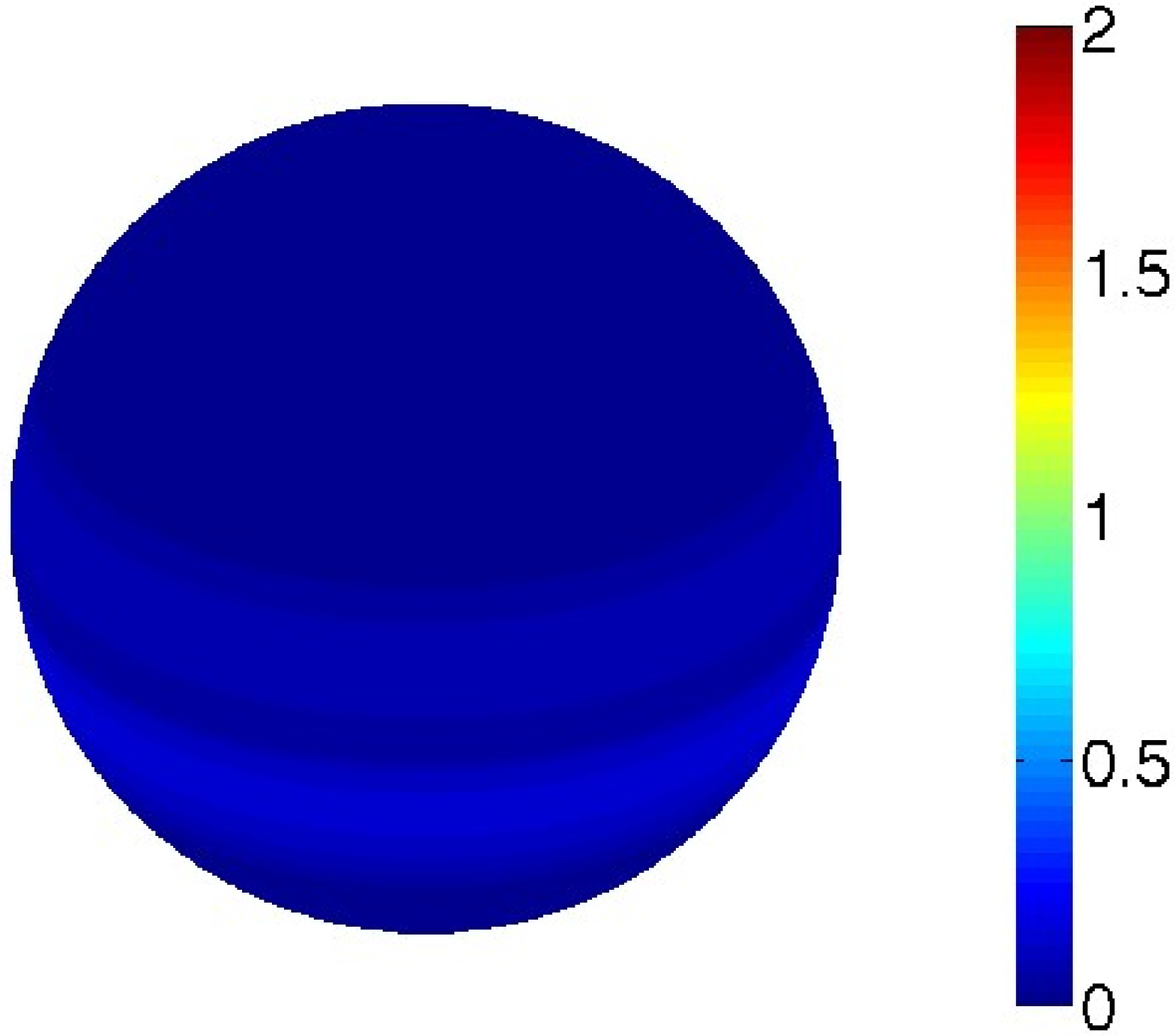}
\includegraphics[width=0.28\textwidth]{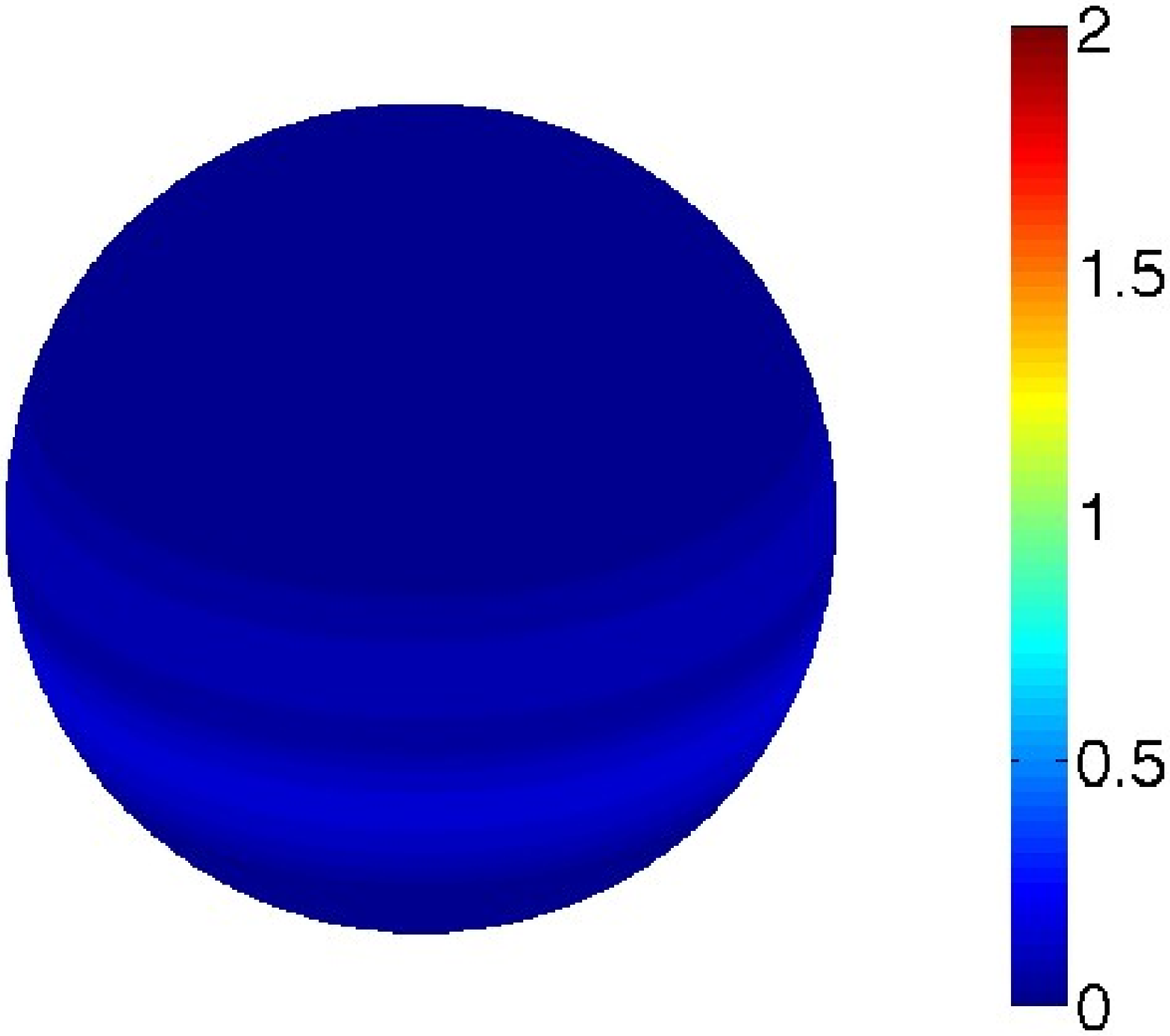}
\caption{\label{fig:horizonfields} Electromagnetic fields at the membrane for our GRMHD simulations with SANE initial conditions.  Membrane $B_n$ (first column, left hemisphere), electric charge density (first column, right hemisphere), electric field (second column), and magnetic field (third column) are shown.  On the membrane $\vec{J}_H=\vec{E}_H$, so the second column is also the membrane's current density.  The GRMHD simulation results shown here are correctly described by the BZ model.}
\end{figure}

\newpage
\begin{figure}
\hspace{1.2in}
{\LARGE $B_n, \sigma_H$}
\hspace{1.5in}
{\LARGE $\vec{E}_H$}
\hspace{1.7in}
{\LARGE $\vec{B}_H$}
\\
{\LARGE $a_*=0.90$}
\includegraphics[width=0.28\textwidth]{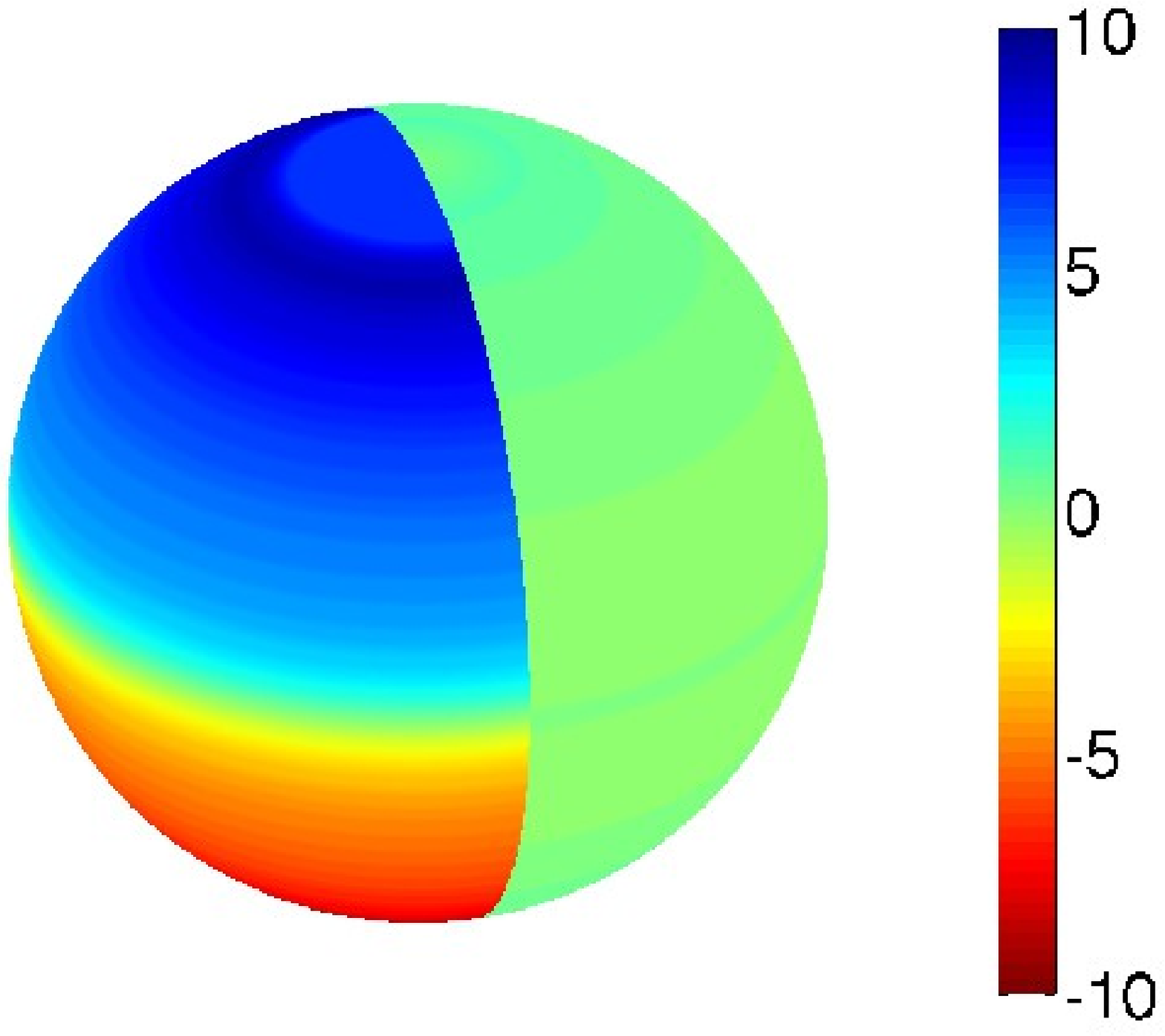}
\includegraphics[width=0.28\textwidth]{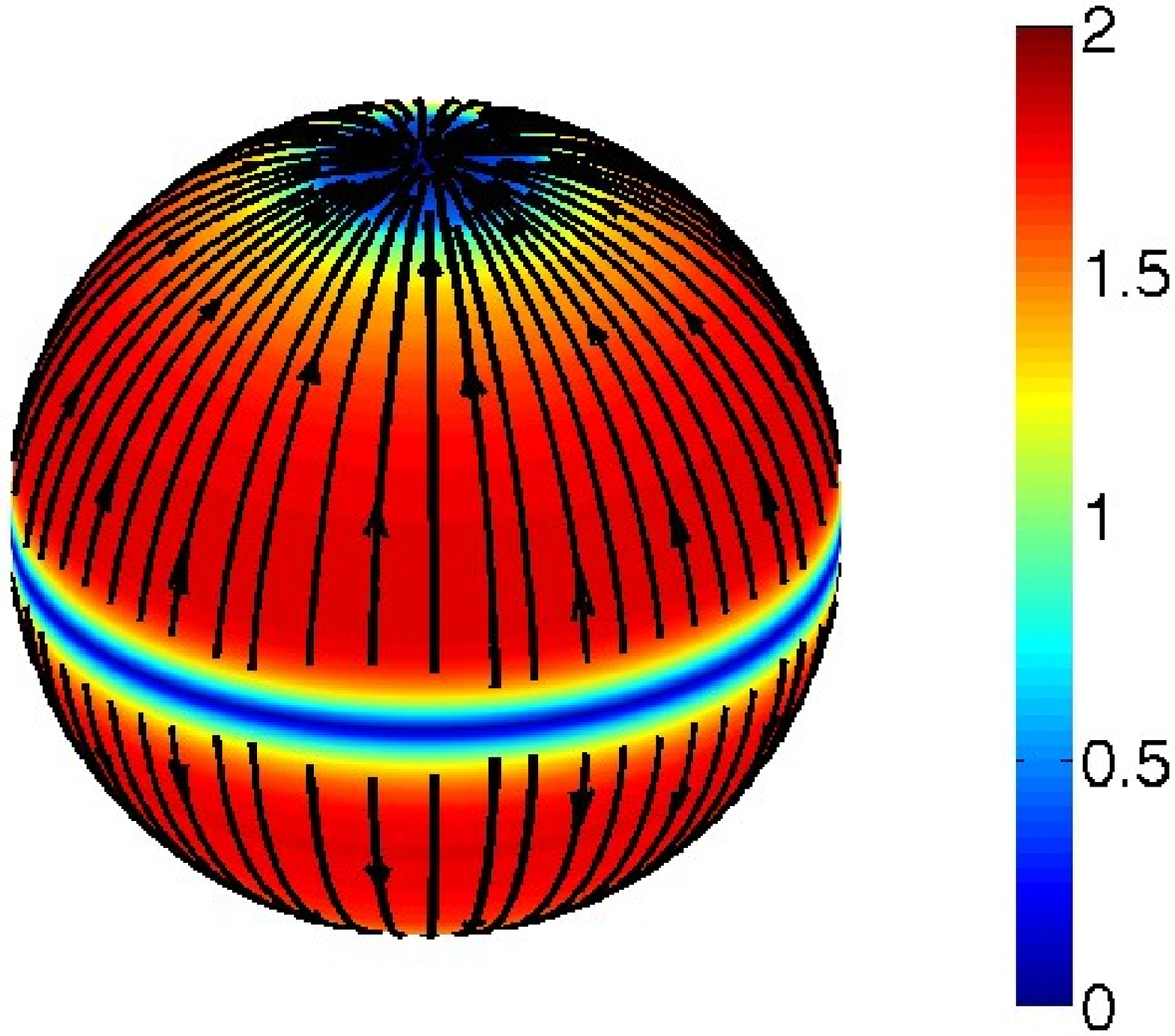}
\includegraphics[width=0.28\textwidth]{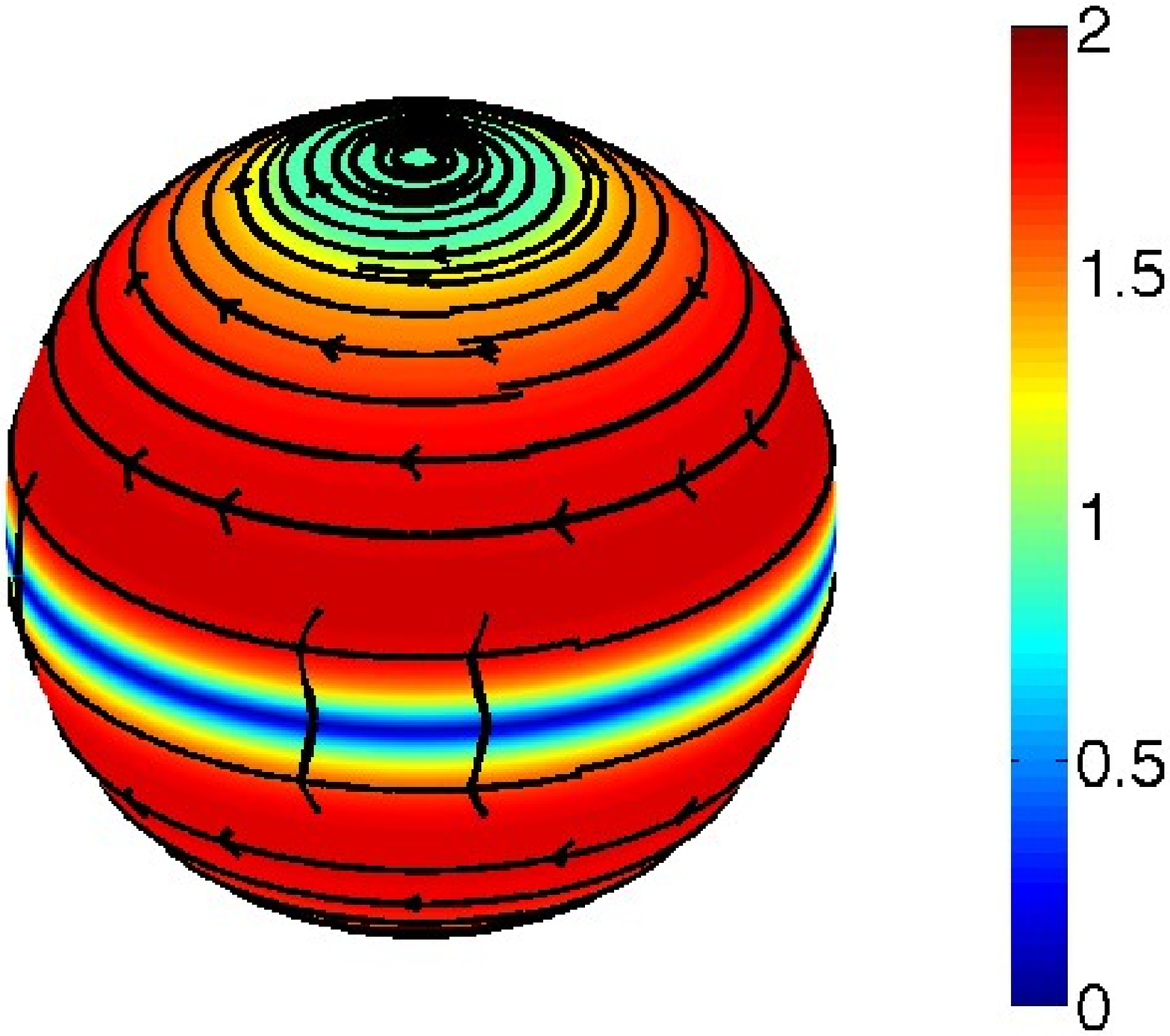}
\\
{\LARGE $a_*=0.70$}
\includegraphics[width=0.28\textwidth]{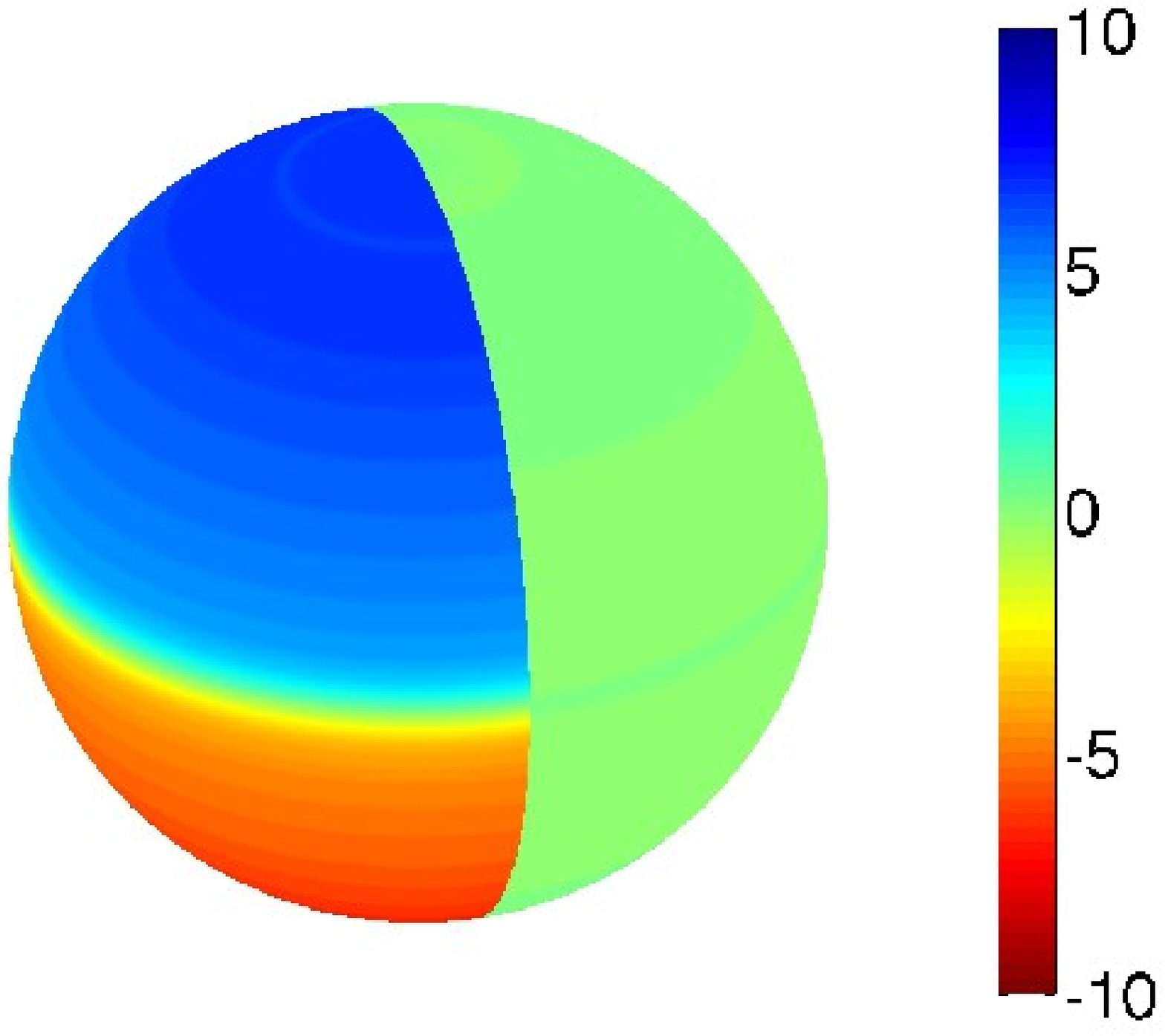}
\includegraphics[width=0.28\textwidth]{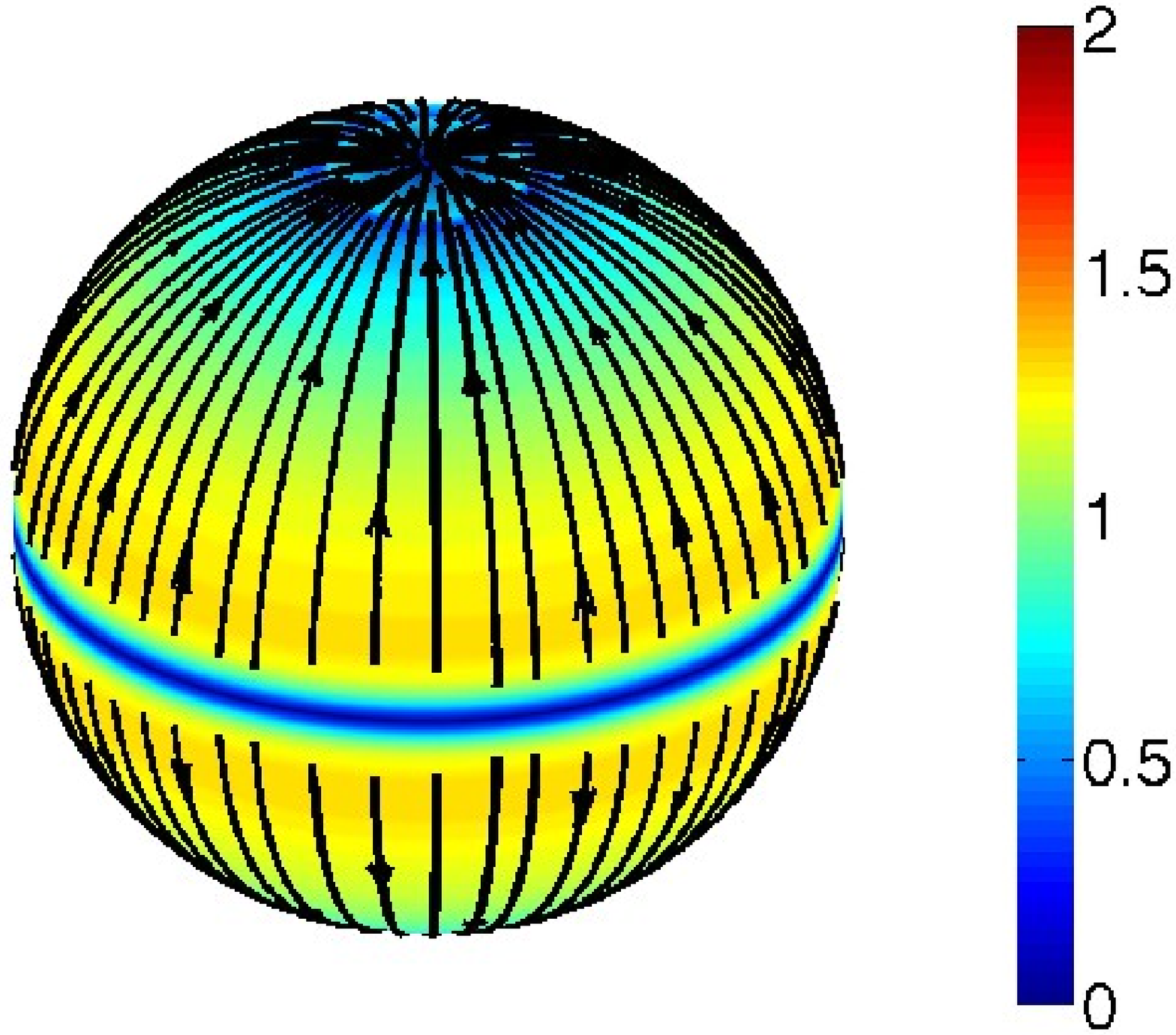}
\includegraphics[width=0.28\textwidth]{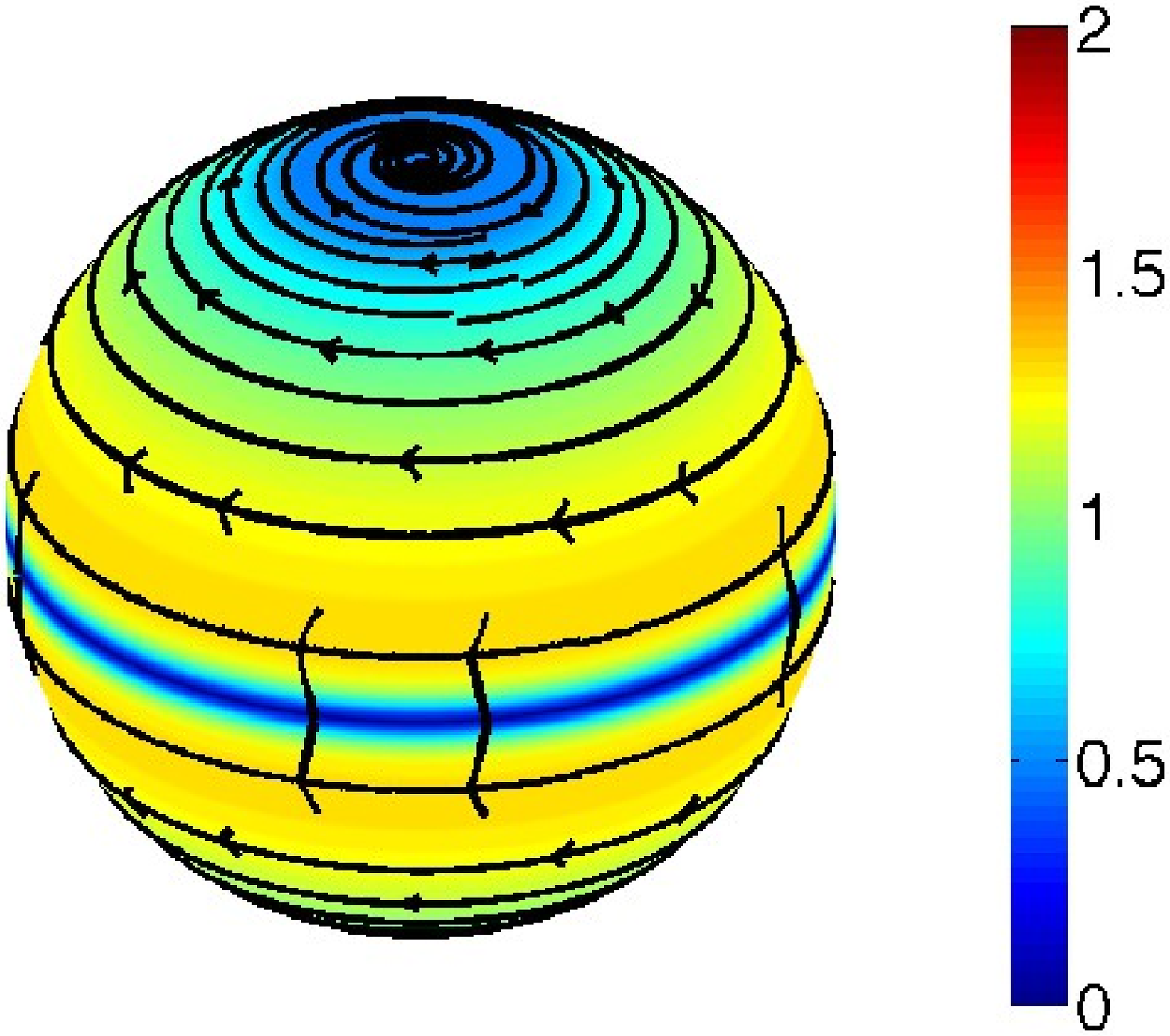}
\\
{\LARGE $a_*=0.00$}
\includegraphics[width=0.28\textwidth]{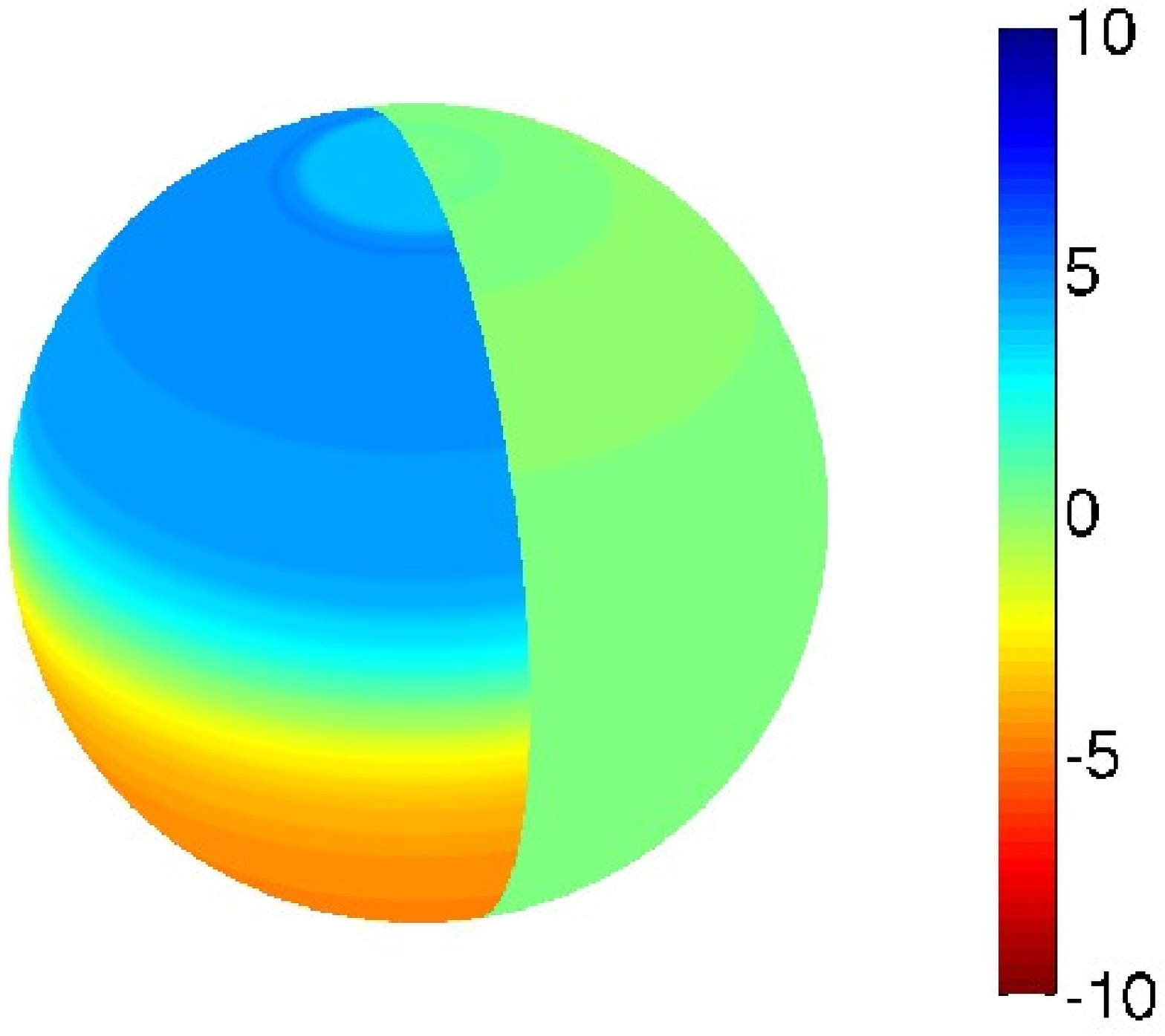}
\includegraphics[width=0.28\textwidth]{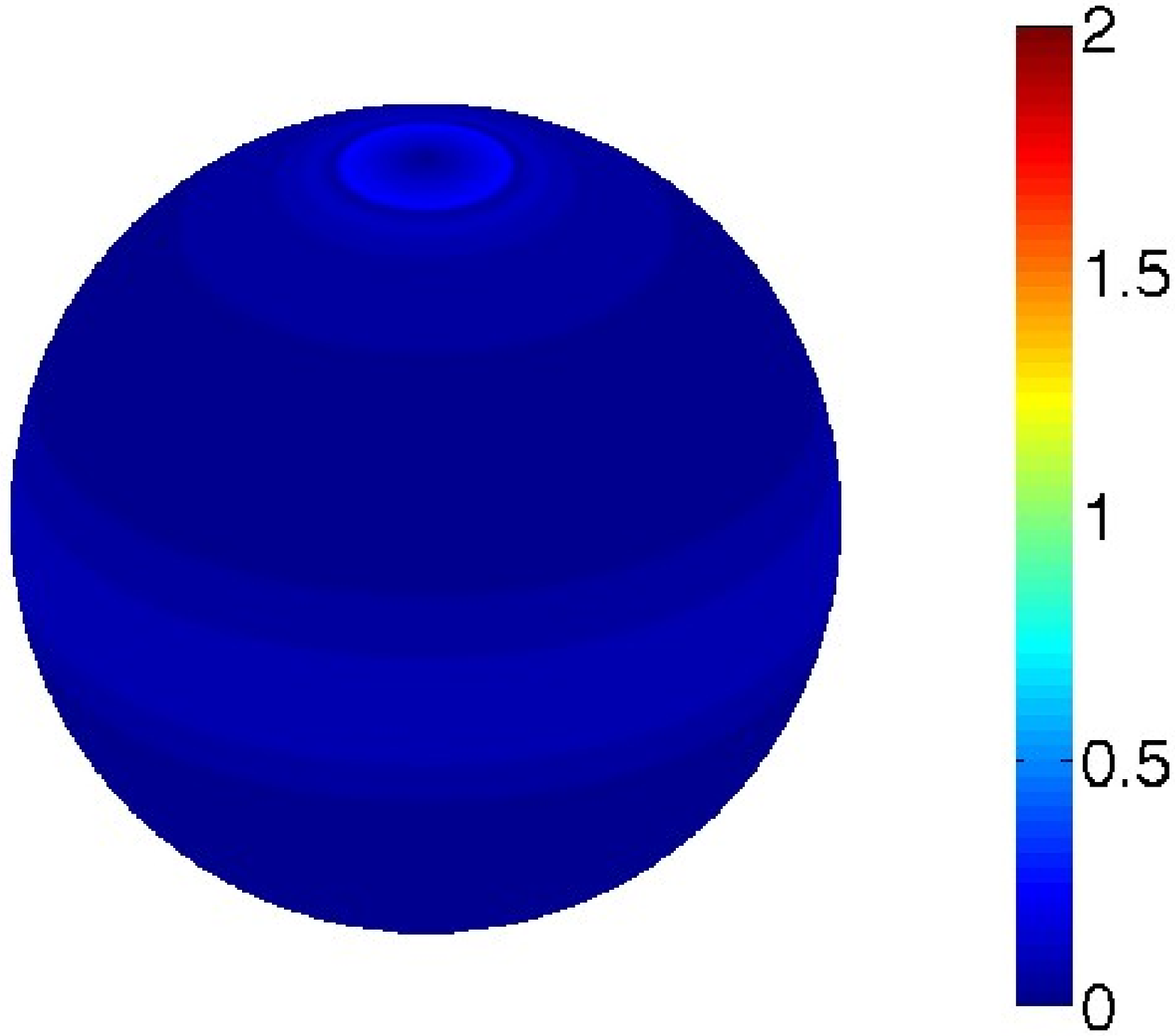}
\includegraphics[width=0.28\textwidth]{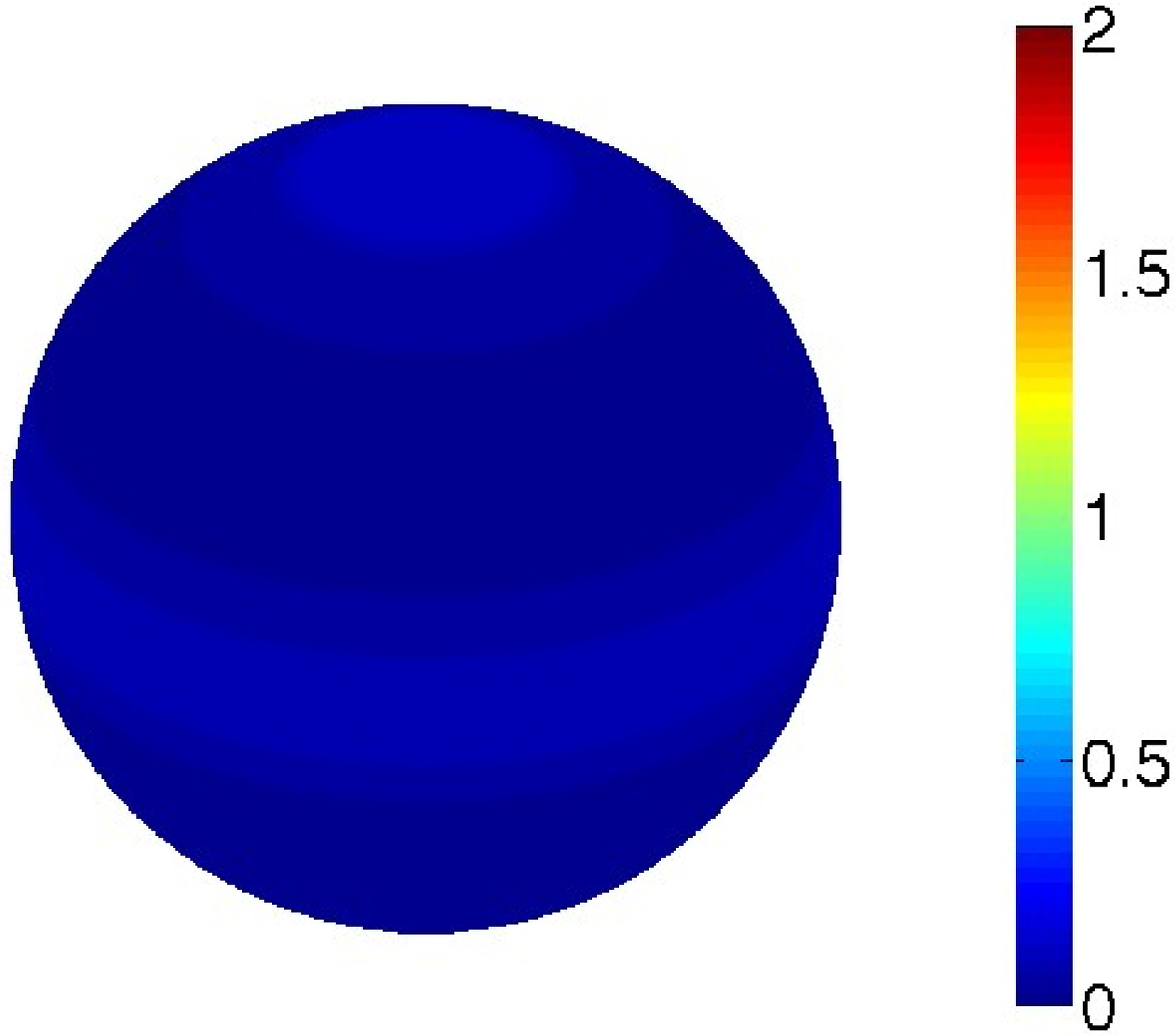}
\caption{\label{fig:horizonfields_mad} Same as Figure \ref{fig:horizonfields} but for MAD runs.}
\end{figure}

\clearpage
\bibliographystyle{mnras}
\bibliography{ms}

\end{document}